\documentclass[prb,aps,twocolumn,showpacs,superscriptaddress]{revtex4-1}
\usepackage{hyperref}
\hypersetup{colorlinks, citecolor=blue, filecolor= black, linkcolor= black, urlcolor= blue}

\usepackage{graphicx}
\graphicspath{ {phase-diag/} {regimes/} }
\usepackage{dcolumn}
\usepackage{bm}
\usepackage{float}
\usepackage{bm}
\usepackage{epsfig}
\usepackage{latexsym}
\usepackage{amsmath}
\usepackage{mathtools}
\usepackage{amsfonts}
\usepackage{color}
\usepackage{array}
\usepackage{enumitem}
\usepackage{hyperref}
\hypersetup{
    colorlinks=true,
    linkcolor=blue,
    filecolor=magenta,    
    urlcolor=cyan,
}
\usepackage{lipsum}
\usepackage{amssymb, amsmath}

\DeclarePairedDelimiterX\braket[2]{\langle}{\rangle}{#1 \delimsize\vert #2}

\renewcommand{\Im}{\text{Im}}
\renewcommand{\Re}{\text{Re}}

\newcommand{\<}{\langle}

\renewcommand{\>}{\rangle}
\renewcommand{\<}{\langle}
\renewcommand{\(}{\left(}
\renewcommand{\)}{\right)}

\renewcommand{\d}{\partial}

\begin{document}
	\preprint{Preparing}
	
	\title{Weak-coupling theory for partial condensation of mobile excitons}
	
	\author{Igor V. Blinov}
	\email{blinov@utexas.edu}
	\affiliation{
		On leave from Department of Physics, The University of Texas at Austin, Austin, Texas 78712, USA}
		
	\date{April 15, 2024}
	
	\begin{abstract}
		\noindent
We studied formation of charge density wave between valleys in a system with double-well-like dispersive valence band which captures some of the features of rhombohedral graphene trilayer. In a regime with 2 Fermi surfaces: electron- (with radius $p_i$) and hole-like ($p_o$) -- an instability in particle-hole channel appears at $q=q_c+\delta q$, where $q_c=p_o-p_i$. In a weak coupling regime ($x/\epsilon_F\ll 1$) presence of an additional energy scale $\propto m q_c\delta q$ gives rise to several regimes with distinct spectrum and transport properties: in a regime with small order parameter $x\lessapprox m \<p_F\> \delta q$ Fermi arcs show up and change conductance qualitatively. At larger values of the order parameter Fermi arcs are gapped out. Regimes are also distinguished by different effective exponents $\zeta$ in conductance correction $\sigma\propto \tau_D^\zeta$ where $\tau_D$ is scattering time off disorder and $1\leq\zeta\leq 2$.
  \end{abstract}

	\maketitle
	\newpage
	

\section{Introduction}
When electrical current fixed by the external conditions, dissipated heat goes as $\propto j^2/\sigma$, so that in order to minimize external influence \cite{landau2013statistical}, most of the things at low temperatures conduct well. It is then reasonable to expect finding unusual physics in areas where highly resistive phases are present.\\ Insulators ($\sigma=0$) usually appear because of very severe restrictions, such as statistics\cite{mott1968metal}, particle conservation \cite{mermin1968crystalline} or charge of particles effectively being zero \cite{frenkel1931transformation}. \\ 
Besides true insulators, highly resistive phases can be of interest (known as failed insulators \cite{cao2020tunable}). We may expect resistance per quasiparticle being high in systems where two interacting subsystems present: one with charged carriers, and the other without such. One of the examples of such phase is high-temperature superconductors, where both spinons and chargons are present \cite{lee2006doping}.\\
Another system of this kind would consist of excitons (compound particles with zero charge) and holes/electrons. \\
When a system consists of several layers with energy hybridisation $\gamma$ doped away from the charge neutrality, it is reasonable to expect that at some displacement field $D>\gamma$, but not $D\gg \gamma$ a hole-like Fermi surface will be localized in the top-most layer, and the electron-like Fermi surface will be largely in a bottom-most layer, given the field direction is positive. ABC-stacked (or rhomboedral) graphene has three layers and in sufficiently strong  electrical field  (of order 100 mV/nm) perpendicular to the plane seem to satisfy a condition with two types of carriers present. In this regime, two Fermi surfaces\cite{zhang2010band}, electron- and hole-like for each spin and valley are present.\\
There, a number of correlated phases were observed\cite{zhou2021half}.
A resistive phase, called partially isospin polarized (PIP) is mostly consistent with an intervalley coherence \cite{zhou2021half, chatterjee2021inter}. Similarly to twisted graphene bilayer, adjacent phase is superconducting \cite{zhou2021superconductivity} and appears in a regime with two Fermi surfaces.  Besides the $C_6$-symmetric intervalley coherent phase \cite{chatterjee2021inter}, a phase with additionally broken rotational symmetry \cite{huang2022spin} was suggested to explain the experiment. However, it is known that Stoner transition in a system with quadratically dispersive bands and screened interactions in 2D does not change shape of the bands and therefore does not change the resistance according to Drude formula. Then there are several alternatives:
\begin{enumerate}
	\item Drude formula is inaccurate due to large quantum corrections;
	\item long-range nature of interaction is important;
	\item a phase realized in nature is not intervalley coherent.
\end{enumerate}
Mean-free path in the sample $l_D\approx 1 \mu m$ \cite{zhou2021superconductivity}, which makes the first alternative less plausible, since quantum correction decays as a power of $l_D p_F$. And, while the band curvature induced by the momentum-dependence of interaction could be important, it is known that in many cases most of the relevant physics can be modeled with a properly chosen value of the contact interaction in the limit of small density \cite{lifshitz2013statistical} or small screening length. As a consequence, in this, as well as in the previous paper\cite{blinov2024partial} we have explored the last path.
\\Namely, we suggest as a candidate a phase with valley and spatial symmetries broken simultaneously: e.g. an intervalley coherence established at a vector $Q=K-K'+q$, where $q\approx p_o-p_i\ll Q$, which would correspond to formation of an incommensurate large-period crystal lattice in the second order of the order parameter or, another words, to an incommensurate charge density wave. In the language of the band theory, the intervalley potential flattens bands, and as a result, the effective mass changes. This phase explains qualitatively suppression of conductivity at low temperatures. In this case, as long as the state is metallic, correction should go as $\delta \sigma\propto \tau_D$, where $\tau_D$ is the initial scattering time.  \\
However, because of presence of two Fermi surfaces in the system: inner, electron-like, with a smaller Fermi momentum $p_i$ and outer, hole-like, with a large Fermi momentum $p_o$, local gaps may open and, consequently, system may enter non-metallic regime of conductance.  As a result, correction to the conductance contradicts Drude formula. Conductance correction depends on $q$ and has, as a function of it, few regimes, with a leading term going as some power of $\tau_D$ (the scattering time due to disorder), and $m_e$ is the effective mass of electrons:
\begin{equation}
	\delta \sigma\propto \(\frac{x_q}{m_e}\)^\alpha (m_e\tau_D)^\zeta ,
\end{equation}
where effective $\zeta$ is from $1$ to $2$ (in the experimentally relevant regime it has a value around $3/2$).
To shed some light on these regimes, in this paper we start from a simple fermionic model (Sec \ref{sec:band-model}) that can be of relevance both for ABC-graphene\cite{zhou2022isospin,dong2021superconductivity, bernevig2025berry}, and systems with Rashba electrons\cite{nagaev2020two}, then build an effective bosonic theory valid in the low-temperature regime, evaluating all the relevant coefficients (Sec \ref{sec:low-energy-theory}) microscopically, and calculate the correction to the conductance due to formation of the intervalley order in Sec \ref{sec:resistance} both in the second order in $x_q$ and, extending perturbative in $x_q \tau_D$ expansion, obtain a general formula.  
\\ To conclude, we comment on the difference with related theories \cite{larkin1965inhomogeneous, brazovskii1975phase}, problems with the current theory and its application to transport calculations, as well as future prospects.

\section{Order parameter}
To clearly define an intervalley order at finite $q$, we first revise a definition of the intervalley coherence. A knowledgeable reader may consider skipping this section. \footnote{In short, it is charge density wave at a momentum close but not equal to the momentum connecting two valleys. An author received a lot of repetitive questions during seminars that showed that some sort of clarification of the latter is necessary. A reader familiar with graphene may consider skipping this section.}
Charge density order (CDW) is defined through the Fourier components of electron density at non-zero $q$:
\begin{equation}
    \rho_q=\sum_{k,\sigma}\<c^\dagger_\sigma(k+q)c_\sigma(k)\>,
\end{equation}
where summation over quasimomentum $k$ goes over the whole Brillouin zone. For single layer graphene, effective low-energy description involves 4 flavors of fermions: valley and spin, with valley being an area in k-space close to one of energy-extrema at $K$ and $K'$ points: $K/K'=(\pm 4\pi/3\sqrt{3},0)/a_0$. Then Fermi creation/annihilation operators can be represented as a sum of two operators defined in different subregions of $k$-space: 
$c_\sigma(k)=c_{\sigma,K}(k)\delta_{k,K}+c_{\sigma,K'}(k)\delta_{k,K'}$, $\delta_{k,K}=\theta(-|k-K|+k_\delta)$ are the Heaviside functions with cutoff $k_\delta<|K-K'|/2$. Introducing additionally $p=k-K^{'}/K$ and Pauli matrices $\tau^{0,x,y,z}$ acting in the valley space, one can rewrite the CDW-order in the form 
\begin{multline}
    \rho_q=\sum_{p,\sigma}\<c^\dagger_{\sigma,a}(p+q)c_{\sigma,b}(p)\>\delta_{q,K-K'}(\tau^0_{ab}+\tau^z_{ab})/2
    \\+
    \sum_{p,\sigma}\<c^\dagger_{\sigma,a}(p+q)c_{\sigma,b}(p)\>(1-\delta_{q,K-K'})(\tau^x_{ab}+i\tau^y_{ab})/2,
\end{multline}
with arguments of creation/annihilation operators $c_{\sigma,a}(p)/c^\dagger_{\sigma,a}(p)$ being by modulus of $K$. Or, finally, introducing pseudomagnetization vector with projections $x,y,z$ at a momentum $q$ in analogy to the ferromagnetism labels, have it in the form:
\begin{equation}
    \rho_q=\rho^0_q/2+z_q/2+x_q/2+iy_q/2,
\end{equation}
where the last two correspond to an order parameter that we call intervalley coherence at finite $q$. Note here that each component of pseudospin has both $K-K'+q$ and $-K+K'+q$ terms, which will give rise to charge modulation at momentum $q$. Assume that intervalley coherence at a finite $q$  established, say in $x$-direction: $x_q\neq0$ and there are at least 2 reciprocal lattice vectors $q$. In real space, expectation value of the density is:
\begin{equation}
    \rho(r)=\sum_{\sigma}\<c^\dagger_\sigma(r)c_\sigma(r)\>
   =\sum_{\sigma,k,q}\<c^\dagger_\sigma(k+q)c_\sigma(k)\>e^{iq \cdot r},
\end{equation}
meaning that in the second order in $x_q$ there will be long-range variations of density with wavevector $K-K'+q-(K-K'-q)=2q\ll|K-K'|$ with large period $\pi /|\kappa|$, with $\kappa=min_{i,j}(q_i+q_j)$ which is not generically commensurate with the period of the density wave appearing in the first order, which is roughly equal to $2\pi/|K-K'|$. 
In what follows, I denote a SO(3) order parameter as $M_q=(x_q,y_q,z_q)=(\bar{X}_q,z_q)$. 
\\
Displacement field breaks $z\to -z$ symmetry and opens a gap. When the gap is sufficiently large in comparison to other energy scales present in the model (kinetic, coupling between the layers, interaction), the low energy description may involve a single band. Sufficiently strong displacement field $D\| \hat{z}$ distributes electrons between the layers. At very small momentum valence band of the trilayer is mainly located within the conduction band of the B-sublattice of the bottom layer, while at larger momentum it restores the original form, which is delocalized in the sublattice space. As a result, Fermi surface has a property of having positive mass for small momentum and negative mass for large momentum. Hence a regime with 2 Fermi surfaces (annular Fermi surface) can be established, with bands being electron-like for inner circle, and hole-like for outer. \\\\
Because of presence of 2 different Fermi surfaces within each electron flavor, the order parameter $M_q$ at sufficiently large momentum will have two components qualitatively different from each other: $M_q=M_{m,q}+M_{ex,q}$, where $M_{m,q}$ is a metallic component and corresponds to coherence established between alike Fermi surface (inner-inner, outer-outer). We call it metallic for it does not lead to a gap opening.  $M_{ex,q}$ is an excitonic component that establishes coherence between different Fermi surfaces (inner-outer, outer-inner). Having in mind  presence of both component may help to understand both the critical behavior and electromagnetic response.  We will see later that the excitonic component is related to Fermi arcs.

\section{Band model} \label{sec:band-model}
Within the tight-binding model \cite{ashcroft1976solid, wallace1947band} electronic states in graphene trilayer are described by $2\times 3$ spinors in each (valley and spin) flavor. For ABC-stacked graphene, the largest interlayer hybridization ($\gamma_1$) is between $B_1$ and $A_2$, and  $A_3$ and $B_2$. Then in the regime $\gamma_1\gg v_D p_F\approx D\gg \gamma_{2}\gg\gamma_{3...6}$, where 
($\gamma_{2...6}$ are the remaining tunneling constants) a valence band can be described by double-well dispersive band \cite{blinov2024partial} with trigonal warping (a term that breaks continous rotational symmetry down to $C_3$):
\begin{equation}\label{order:dispersion}
    \epsilon(p)=(mp^2+\lambda p^4)+\Delta p^3\cos(3\theta_p),
\end{equation}
where $\lambda<0$ and $m>0$. The role of the last term is to lift the $SU(2)$ degeneracy in the valley space, choosing valley coherent state ($x-y$) instead of valley polarized ($z$). However, when anisotropy is small $\Delta/m_e\ll 1$, we may safely make $\Delta=0$ (corresponds to the next-nearest vertical neighbor tunneling, coupling between $A_1$ and $B_3$, $\gamma_{2}\to 0$) and postulate $x$ ordering. In this work, we fix $\Delta=0$ for simplicity, to make $\epsilon(p)$ a function of $p^2$ only. We claim, however, that most of the conclusions of the paper will still survive finite $\Delta$. \\
Then Fermi momentums for inner- (i) and outer-Fermi (o) surfaces are given by $p_{i/o}^2=-m/(2\lambda)\pm\sqrt{(m/2\lambda)^2+\mu/\lambda }$. Presence of 2 Fermi surfaces with Fermi momentums $p_o$ and $p_i$ implies that an instability at $q_c\approx p_o-p_i$ should be present. 

\section{Response} \label{sec:response}
In a trilayer, phase transition happens as a function of hole density at a certain critical $n_c$. Usually it can be attributed to increase of the density of states $\nu$ at the Fermi level. Usual epistemology of phase transitions says that since electron susceptibility to an inhomogenous pseudomagnetization $x_q$ is a function of $q$, a phase at $q$ that minimizes non-interacting susceptibility is realized at $n_c$. 
\\\\
Within RPA \cite{altland2010condensed} for contact interaction, susceptibility (or response) can be expressed through components of non-interacting polarization operator $\Pi^a=\text{Tr}(\hat{\Pi}\sigma_a)/2$: $\Pi^a_q=\Pi^a_q/(1+\lambda \Pi^a_q)$. In negligence of the trigonal warping the model is rotationally invariant in the pseudospin space, and hence $\Pi^a_q=\Pi_q$. \\
At low temperatures, only points close to the Fermi energy should contribute to the response. We then divide $\Pi^x$ into 4 parts: between alike Fermi surfaces in different valleys $\Pi_{ii/oo}$ (homo-part), and between different Fermi $\Pi_{io/oi}$ (hetero-part). Such division is legitimate for $q^2<(p_o^2-p_i^2)/2$.
A component of the polarization function can be evaluated through:
\begin{equation}
    \Pi_{aa'}(q)=\sum_p \frac{n_a(p+q)-n_{a'}(p)}{\xi_a(p+q)-\xi_{a'}(p)+i\delta}= I_{a}(q)+I_{a'}(q),
\end{equation}
where $n_a(p+q)\equiv n(\xi_a(p+q))$, $\xi_a(p+q)\equiv\frac{\d^2 }{\d p^2}\epsilon(p_a)(p^2-p_a^2)$, $p_a$ characterizes Fermi surface ($a=i,o$) and $\Pi_{a(a')}(q)$ simply to denote terms with single density $n_a(p)$ in the numerator.  
\\\\

\subsection{Zero temperature response}\label{sec:response:sub:response:0T}
For $q<2p_{i/o}$ response of approximately quadratically dispersive particles in 2D changes slowly – for $q<p_f$:
\begin{equation}\label{instability:polarization:homo:q-small}
	\Pi_{hom}=-\frac{1}{2|m_e|},
\end{equation}
where an effective inverse mass of the quasiparticles $\frac{\d^2 }{\d p^2}\epsilon(p_a)\equiv \pm m_e=\pm\sqrt{m^2+4\lambda \mu}$ for inner/ outer Fermi surfaces. When $q$ no longer can connect two points on Fermi surface,
\begin{equation}\label{instability:polarization:homo:q-large}
	\Pi_{hom}=-\left(1-\frac{1}{2}\sqrt{q^2-(2p_i^2)}\right)\frac{1}{2m_e}.
\end{equation}
Because $p_i^2=(m_e-m)/2\lambda$ the system enters the second regime for electrons located at the inner Fermi surface for  $\mu<\mu_u=(1-(4/5)^2)m^2/4\lambda$.
\\\\
For hetero-processes, static response vanishes below $q_{bound}=q_c-\delta q_{b}$, with $\delta q_{b}\ll q_c\equiv p_o-p_i$, since it will be impossible to satisfy energy conservation condition. \\\\
To calculate the hetero-response, we note that quadratic mass approximation is valid whenever the quartic part is small: $-\delta<\xi(p+q)<\delta$ and $-\delta<\xi(p)<\delta$, so that the integration limits $-\delta/m_e+p_{s}^2<p^2<\delta/m_e+p_{s}^2$ and 
$-2\delta/m_e+p_{s'}^2-p_{s}^2-q^2<2p_{s}q\cos(\theta)<2\delta/ m_e-p_{s}^2+p_{s'}^2-q^2$, and $\delta=m_e(m_e/\lambda)\delta_d$ with $\delta_d\ll 1$, which we estimate later in this section.\\

Dividing the hetero-response into the inner- and outer-parts, we get an integral with $\epsilon_o(p+q)-\epsilon(p)$ in the denominator. Each difference defines angle $\theta_{c,i/o}$ for which energy conservation is satisfied together with electron/hole being close to the Fermi surface.  Clearly, such integral is maximized when $\theta_{c,i}$ equal to one of the integration limits, so that the integrand does not change sign. Not surprisingly, we will get an expression analogous to response to the Peierls instability for quasi-one-dimensional electrons constrained to angle $\theta_{c,i/o}$ \footnote{Alternatively, we could have talked about 1D particles with large inverse effective mass $m_{eff}=m_e/\theta_{c,i}$.}:
\begin{equation}
	I_a=\frac{\theta_{c,i}}{2m_e\pi }\frac{p_i}{p_i+p_o}\log\(\frac{\epsilon_{c,i}}{\<\epsilon_{F}\>}\),
\end{equation} 
with cutoffs dictated by the kinetic energy of relative motion of a hole and an electron. As such, $\<\epsilon_{F}\>$ is the maximum kinetic energy of relative motion that satisfies energy conservation: $\<\epsilon_{F}\>=m_e (p_i+p_o)^2/^2$, $\epsilon_{c,i}$ plays the role of the low-energy cutoff: $\epsilon_{c,i}=m_eq_c^2\theta_{c,i}^2$, and an analogous expression for $\Pi_o$\footnote{See Appendix B}.  For small $\delta q$, 
\begin{equation}
	\theta_{c,i/o}^2\equiv \(\frac{2\delta q p_{o/i}}{p_{i/o} q_c}\pm\frac{\delta q^2(p_o^2+p_i^2)}{q_c p_{i/o}(p_o+p_i)^2}\).
\end{equation}\\
Angle $\theta_f$ depends on cutoff $\delta$. Then small momentum offset $\delta q$ also is a function of the cutoff, and the equality of two angles is what determines $\delta q$. Clearly, good choice of $\delta$ will make the sum $\Pi_{aa'}$  closest to the exact response with quartic dispersion. One source of inaccuracy is due to the incompleteness of the integration range over energies, and goes as $\Delta P_{range}\propto (\delta-\delta_{1/2})/(m_e \delta)$ where $\delta_{1/2}$ is such that $(p_o^2-p_i^2)/2=\delta_{1/2}/m_e$. The other source of mistake is due to the approximation to the quartic dispersion
$\Delta P_{app}\propto 2\lambda^2 \delta^4/m_e^3 \delta^2 $, which is minimized at $\delta=\delta_{1/2}|m_e/2\lambda|=|m_e/2\lambda|^2\ll 1$, and then:
\begin{equation}
	\delta q^2=\frac{|m_e/\lambda|^2}{2\(\frac{p_o+p_i}{q_c}+\frac{p_o^2+p_i^2}{(p_o+p_i)^2}\)},
\end{equation}
We now choose the interaction constant in order to match the density of the first transition $n_{c}\approx 1.2\times 10^{12} cm^{-2}$ (which for single band model corresponds to $\mu_c=7\times 10^{-4}m$):
\begin{equation}
 -1+\frac{1}{2\pi }\<\theta_c\>\log\(\frac{1}{e^2}\(\frac{\epsilon_{c,i}\epsilon_{c,o}}{\epsilon_{F,i}\epsilon_{F,o}}\)^{1/2}\)\approx -m_e V^{-1}_c,
\end{equation}
with $\<\theta_c\>=(\theta_{c,i}+\theta_{c,o})/2$, so that $V_c\approx 0.45 m$.  Note that the hetero-part of the response does not diverge and of the same order that the homo-part. Therefore it gives significant increase of critical temperature as well as value of the interaction at which the phase can be stabilized.

\subsection{Critical temperature}

\begin{figure*}[htb]
\includegraphics[width=2\columnwidth]{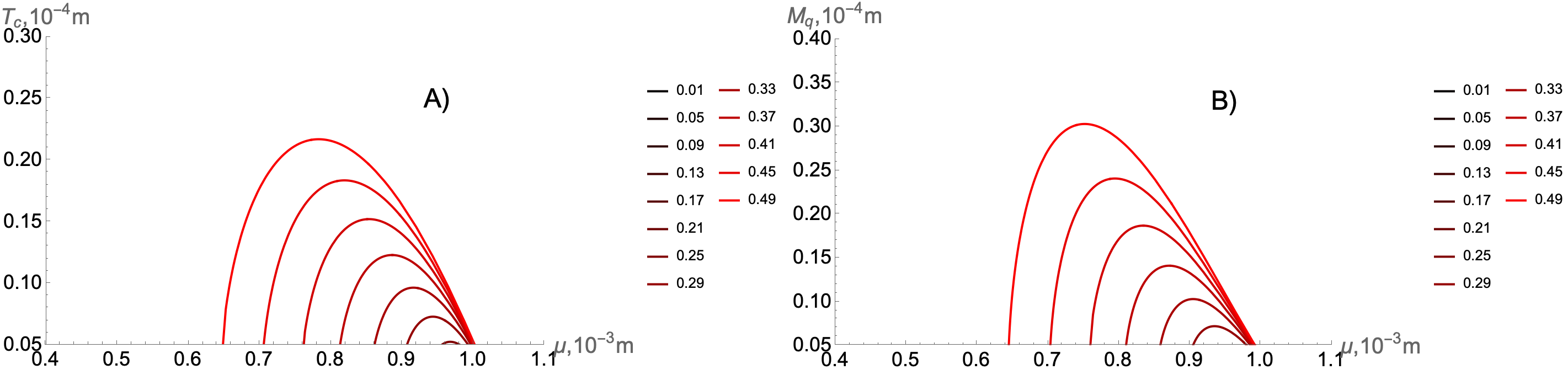}
\caption{Behavior of the critical temperature (A) as a function of the chemical potential and an order parameter (B) for infrared cutoff $\Omega=T_L$ for different values of the interaction constant $V=0.17,...,0.49$. Here, parameter $\lambda=-240 m$. Plot $A$ gives an estimate of the critical temperature in the range 0.1-0.5 K.} 
\label{fig:crit-temp:TcMq}
\end{figure*}
Normally, in an electron liquid the only relevant temperature scale is the Fermi energy $\mu=p_f^2/2m$. For low temperatures static linear responses of Fermi liquid at finite $q$ are at best quadratic functions  of $(T/\mu)^2$ and hence can be considered nearly temperature independent for $T\ll \mu$. We thus look at the hetero-Fermi response only. There will be one additional characteristic energy for each $\Pi_{i/o}$\\
\begin{equation}\label{response:critical:characteristic-temperature:M}
	T_{L,i/o}=\frac{m_e p_i q_c\theta_{c,i/o}^2}{1+\frac{q_c}{2p_{i/o}}},
\end{equation}
\begin{equation}\label{response:critical:characteristic-temperature:F}
	T_{F,i/o}=2m_e p_i^2
\end{equation}
The former energy is to represent transversal energy scale: energy increases as we increase angle of the momentum from $0$ to $\theta_{c,i}$ quadratically for fixed $q=q_c$.
At vicinity of zero temperature, response has a usual for Fermi gases \cite{landau2013statistical} quadratic temperature-dependence. In negligence of the temperature-dependence of the homo-part,
\begin{equation} \label{response:critical:response-T-dependence:hetero:low-T}
    \Pi_{low,i/o}(T)\approx\Pi_{i/o}+ \frac{\pi^2 \theta_{c,i/o}}{12 m_e (1+\frac{q}{2p_{i/o}})}\frac{T^2}{T_{L,i/o}^2},
\end{equation}
In the regime of interest, small characteristic temperature $T_{L,i/o}\approx 10^{-4}-10^{-5} m\approx 10^{-2}-10^{-1} meV$ is an order of magnitude smaller than the Fermi temperature. 
 If the  phase transition happens within this regime, the critical temperature is
\begin{equation} \label{response:T-dependence:Tc-quadratic}
    T_{c,L}=T_{L}(-V \Pi_0-1)^{1/2}\(\frac{m_e}{V}\)^{1/2},
\end{equation}
where \begin{equation}
T_{L}^{-1}\equiv\frac{\pi}{12^{1/2}}\sqrt{\frac{\theta_{c,i}}{ (1+q/2p_i)}T_{L,i}^{-2}+\frac{\theta_{c,o}}{ (1+q/2p_o)}T_{L,o}^{-2}} 	
 \end{equation}
A prefactor $-m_e\Pi_0-m_eV_c^{-1}\propto   \d (m_e \Pi)/\d \mu (\mu-\mu_c)$ should be less than $1$ to satisfy $T_{c,L}<T_{L,i/o}$ condition. Since $\d (m_e \Pi)/\d \mu (\mu=\mu_c) \approx 10^3$ the condition is satisfied in the range $\mu-\mu_c\approx 10^{-3} meV $, which lies within the experimentally relevant regime.  See Fig. \ref{fig:crit-temp:TcMq} A for plot of the low-temperature regime. With $T_{L,i/o}\propto m_e p_{i/o}q\theta_{c,i/o}^2\propto m_e \delta q$ we conclude that $T_L\propto \theta_c^{-1/2}\propto q_c^{1/4}$, so that $T_c$ non-analytically vanishes at $q_c=0$ point.
\\
At higher temperatures, $T\gg T_1$, momentum difference between two Fermi surfaces becomes smeared: difference between $p_i$ and $p_o$ can be neglected. Another words, the system becomes effectively indistinguishable from a system with an electron-like and a hole-like Fermi surfaces with the same Fermi momentum (e.g. BCS or excitonic condensate). As a consequence, for $T_L<T<T_{F,i/o}$, direct calculation shows that the temperature-dependence follows the logarithmic law:
\begin{multline}
   \label{response:T-dependence:T-high}
	\Delta \Pi_{high,i/o}(q,\Omega,T)
	\approx
	\frac{T_{L,i/o}(1+\Gamma_0(1))}{2m_e^2 p_i q\sin(\theta_{c,i})}\\+
	\frac{T_{L,i/o}}{2m_e^2 p_i q \sin(\theta_i)}
	\(\log\bigg(\frac{T}{T_{L,i/o}}\bigg)-\frac{1}{2}\(1-\frac{T_{L,i/o}}{T}\)\),
\end{multline}
which using \eqref{response:critical:characteristic-temperature:M} could be interpreted as temperature-dependent part of the polarization operator of light electrons with inverse mass $m_{eff}=m_e/\theta_{c,i}$ and 
 agrees with earlier numerical studies \cite{blinov2024partial}.
Note here that because $T_{F,i/o}$ and $T_{L,i/o}$ are different by a factor of 10-100,  temperature dependence in this regime is nearly linear. 
In this regime, the critical temperature goes as 
\begin{equation}\label{response:T-dependence:Tc-log}
   T_{c,H}=\<T_L\>_g e^{-(\Pi_o+V^{-1})\frac{m_e^2 q}{\<T\>_a}}e^{1+f\frac{m_e^2 q}{\<T\>_a}}\approx \<T_L\>_g e^{-(\Pi_o+V^{-1})\frac{m_e }{\theta_{c}}}e,
\end{equation}
where $\<T_L\>_g=\sqrt{T_{L,i}T_{L,o}}$ is the geometric average of two characteristic temperatures, $\<T\>_a=T_{L,i}/(2p_i \theta_{i})+T_{L,o}/(2p_o \theta_{o})$, $\Delta \<T\>_a=-T_{L,i}/(2p_i \theta_{i})+T_{L,o}/(2p_o \theta_{o})$, and $f=\Delta \<T\>_a/(m_e^2 q)\log(T_{L,o}/T_{L,i})\approx 0$ \footnote{While deriving this formula, we neglected $\Gamma_0(1)\approx 0.21$. It is possible, in fact, to prove, that this term is a result of the approximation.}. 

\section{Low energy theory}\label{sec:low-energy-theory}
We now construct a low-energy theory through a Hubbard-Stratonovich transformation. We proceed through usual steps\cite{altland2010condensed}: decouple interaction in particle-particle channel, then integrate out fermion modes, expand the logarithm in effective action in powers of $M_i$. To justify the next step, the order parameter ($M_i$) must be small in comparison to an inverse mass of a fermion ($\propto \nu^{-1}$, where $\nu$ is the density of states) or, alternatively, temperature. Truncating the expansion at the 4-th order, we obtain a standard Landau-Ginzburg\cite{landau1950} (LG) theory:
\begin{multline}\label{low-energy-theory:free-energy}
    f=
    \sum_i 
    M_i(q)
    \(\chi(q)+\lambda^{-1}\)M_i(q)
    \\+
    \sum_i U(q_i,q_i+q_j,q_i+q_j+q_k)M_l(q_i)M_l(q_j)M_n(q_k)M_n(q_l),
\end{multline} 
The energy difference between the valley-polarized state and the valley-coherent state goes as $\Delta^2x_q^2/\mu^3\propto \Delta^2/\mu$, which justifies $\Delta=0$ limit discussed earlier for $\Delta^2/\mu\ll 1$. Without loss of generality then, we take $M_i(q)=(x_q,0,0)$, where $|q|$ is a momentum that we treat as a parameter of the theory. As we discussed before\cite{blinov2024partial}, for very weak interaction $\lambda\to 0$ the state is fully $SU(2)$ symmetric $x_q=0$. However, when $\lambda$ increases fully symmetric state becomes unstable towards formation of the intervalley coherence at a finite $q= p_o-p_i+\delta q$.

\subsection{Higher order terms}
To find the order parameter behavior, we calculate the higher-order terms in the expansion of the free-energy \eqref{low-energy-theory:free-energy} $U(q_i,q_i+q_j,q_i+q_j+q_k)$, which are the functions of 3 2-dimensional momentums as well as frequencies.  It is one of the coefficients that distinguish the partial excitonic condensation from LO-phase and lead to the formation of the $C_6$ rotationally invariant intervalley crystal. After the summation over the Matsubara frequencies $U$ can be represented as a sum of 4 terms $U_1+U_2+U_3+U_4$, each with 3 energy differences in the denominator in the form $\Delta \epsilon_i=\epsilon(p+q_i)-\epsilon(p)$. \\
Integrands are most peaked when all differences are close to zero. For generic density and finite $q$ it is not possible to do that for 3 arbitrary $q$, therefore generically coefficients may diverge at worst as $\delta \theta \delta p \Omega^{-2}$, where $\delta \theta  \delta p$ is the region where 2 out of 3 $\Delta \epsilon_i\approx \Omega$. Generically, the measure of this region is  $\Omega^2$ which makes the coefficient finite. For  $q_i=-q_j$ one of the energy differences vanishes, and then $U_i(q_i,0,q_k)\propto \Omega^{-3}\delta \theta d\delta p$, where $\delta \theta \delta p\propto \Omega^2$ is the measure of the area where two differences vanish, thus making the coefficient divergent at most as $\Omega^{-1}$.

The most divergent configuration is $q_i=q_k=-q_j$, since $\delta \theta \delta p\propto \Omega$ and therefore at worst the integral is divergent as $\Omega^{-2}$. Since terms in the sum have opposite uncompensated frequency sum is less divergent than individual terms (see Appendix C for details of the computation) and for the choice of $\Omega_{i}=-\Omega_j=-\Omega_k$ $U(q_i,0,q_i)\propto \Omega^{-3/2}$. Similarly, divergence for the total sum $U$ $g_i\neq g_k$ is less and goes as $\delta \theta \delta p \Omega^{-2}$, which in the worst case scenario results in $\Omega^{-1/2}$ divergence. For $\Omega\propto T_c$ we then can ignore all other terms and use only the most divergent term in the free energy expansion.\\\\
The most divergent term can be simplified to:
\begin{multline}\label{U4:most-divergent}
	U(g_i,0,g_i)=-4\int\frac{ n(\xi_p)(\xi_{p+g_i}-\xi_{p})^3}{(\Omega_i^2+(\xi_{p+g_i}-\xi_{p})^2)^3}
	\\=-2\frac{\d^2 }{\d \Omega^2}\Re \int \frac{n(\xi_p)}{\xi_{p+g_i}-\xi_{p}+i\Omega_i}(1
	+O(\Omega^1))
\end{multline}
An equilibrium state should correspond to the combination of $\Omega_i$-s such that the term is the least divergent: $\Omega_i\approx -\Omega_k$, $\Omega_i\approx -\Omega_j$. Just like with the response, the integral can be divided into the homo- and hetero-parts. The latter \eqref{U4:most-divergent} can be rewritten in the form:
\begin{multline*}
	U(g_i,0,g_i)=2\text{Re}\frac{\d }{i \d \Omega}\int\frac{\delta(\xi_p)}{i\Omega_i+\xi_{p+g_i}-\xi_{p}}
	\\\approx
	-2\sum_{s,s'}\int\frac{d\theta p_s}{\Omega_i^2+m_s^2({p}_s^2+q^2+2qp_s\cos(\theta)-p_{s'}^2)},
\end{multline*}
where $s,s'=i/o$. For the homo-part ($s=s'$) the most divergent part:
\begin{equation}\label{U4:homo}
	U_{4,hom}=-\(\frac{m_e}{\Omega}\)^{1/2}\frac{2^{1/2}\pi}{(m_e q)^3}
	\end{equation}
 The divergence becomes $\Omega^{-3/2}$ at a special point $q=2 p_a$, and $\Omega^{-1/2}$ otherwise. Since the Landau-Ginsburg coefficients can be discontinuous, we do not worry about that feature at $q_c=2p_i$ and take $\Omega^{-1/2}$ behavior to be valid everywhere. \\
Then in the leading order  contribution to $V_4$ is logarithmically divergent:
\begin{equation}\label{U4:het}
	V_{4,het,i}=\frac{1}{2m_e}
	\frac{p_i}{p_i+p_o}
	\frac{(v/(2r_i q_c))^2}
	{(\theta_{io}^2-\frac{iv\Omega}{r_i q_c})^{3/2}}
	\log(\alpha_i\Omega),
\end{equation} 
where $v$ is an inverse average Fermi velocity, $v=1/m_e(p_o+p_i)$ and  $r_{i,1/2}=\frac{p_{i/o}}{2(p_o+p_i)}$ are coefficients of unclear utility.
 If $\delta q_c=0$, the divergence, in full analogy to the homo-part, becomes $\Omega^{-3/2}\log(\Omega)$.\\\\
  It follows then from the minimum energy argument that $\delta q$  should be nonzero\footnote{In fact, a stricter inequality connecting $\sin(\theta_c)$ and the cutoff should hold: $\sin(\theta_c)<$. It is possible that $\delta q$ that corresponds to the minima of energy is, in fact larger, than the one that corresponds to the minimum of the intervalley reponse.}. The presence of homo-part of scattering amplitude contrasts the presented phase from Larkin-Ovchinnikov\cite{larkin1965inhomogeneous}, where, naturally, only hetero-part is present. \\\\
In the condensed phase, excitations have a gap $\propto x_q^2$ that fixes the  infrared divergence. To compensate for its absence in the perturbation theory, we choose a regularization through a sensible cutoff.  For simplicity, we pick $\Omega=T_L$. Then, ignoring difference of $\theta_{c,i}$ from $\theta_{c,o}$, we have for the order parameter far away from the critical point:
\begin{equation}
	M_i^2(q)\propto q^3 \(\frac{T_L}{m_e}\)^{1/2}m_e^2
\end{equation}
For $\delta q$ weakly dependent on $q_c$ (large $q_c$) we conclude that 
	$M_i(q)\propto q^{25/16}$ which vanishes non-analytically at $q=0$. In small $q_c$ limit $\delta q\propto q_c^{1/2}$ and as a result the power changes to $53/32$.\\\\
Since $U(g_i,0,g_i)$ is dominant, and the condensation energy is $\propto -\sum _i\epsilon_q(\epsilon_q/U(g_i,0,g_i))$, the most energetically beneficial is a phase with the largest number of g-vectors, therefore it is the $C_6$ phase that is established (Fig.\ref{fig:crit-temp:TcMq}).

\section{Resistance change}\label{sec:resistance}
When external perturbation does not change topology of the Fermi surface, number of charge species stays the same, correction to the conductivity may only originate through the change of band velocity or quasiparticle wave functions. Naturally, given $SU(2)$ symmetry, presence of homogenous potential (such as pseudo-magnetization $x(q=0)$) in 2D should not change neither mass for nearly quadratic bands nor the matrix elements of $\tau_D$, thus leaving the conductance unchanged. \\\\
On the other hand, spatially varying intervalley potential $x(q)$ obstructs movement of charges, which can be expressed through the change of their velocities and wavefunctions. We should then expect the change of conductance in the perturbative regime to be of the form $\delta \sigma \propto x_q^2\tau_D/\epsilon_{j}$, where  $\epsilon_j$ is the smallest finite characteristic energy for one of the carrier species and summation assumed over $j$. \\\\
 In a regime with week external potential $x_q/\Sigma\ll 1$, correction from the hetero-processes has a form of $x_q^2\tau^2_D$ -- similarly to superconductivity \cite{stepanov2018superconducting}, it appears because of the coupling of electron- and hole-like bands at the verge of the gap opening, and the smallest energy scale of the unperturbed system is disorder self-energy $\Sigma$:
\begin{equation}
	\delta \sigma\propto -x_q^2\tau_D^2.
\end{equation}
Because here electron- and hole-like bands have different Fermi velocities, the total correction, unlike its counterpart in BCS \cite{galitski2001superconducting}, contributions from two different bands do not cancel each other.   We see that because of different nature of Fermi surfaces present and formation of the charge density wave we have non-metallic behavior of conductance. While it seems to be rare, such cases are known \cite{lee1993localized}. \\\\
Once the potential increases, for $\delta q\neq 0$ coupling of different Fermi-surfaces leads to the formation of Fermi arcs in the hetero-channel, thus increasing the number of species. It happens in the regime $x_q/(2\kappa_{io/oi}^{1/2}m_e)<2p_{i/o} q \theta_{c,i/o}^2$, where $\theta_{c,i/o}$ is the central angle for the inner(outer) Fermi-surface. Along the radial direction quasiparticle composing Fermi-arcs are light, while in orbital bands are nearly flat, with mass $\propto (m_e \sin(\theta_{c,i/o}))^{-1}$. Contribution of heavy and effectively 1D linearly dispersive quasiparticles to the conductance is
\begin{equation}
	\delta \sigma_{1,1}\propto -x_q \tau_D(\theta_{c,i}^{-1}+\theta_{c,o}^{-1}),
\end{equation}
Interestingly, it can be larger by its absolute value than one present in the purely partially insulating phase: e.g. $\propto x_q^2 \tau_D /m_e$. . 

One way to interpret this non-negligible correction to the conduction is through formation of the countercurrent: e.g. negative cross terms between electrons and holes in the expression for current.\\
\\In a regime with $x_q\approx \Sigma$, previously linearly dispersive bands, when smeared by the energy scale $\Sigma$, better approximated by quadratically dispersive touching bands, thus leading to 
\begin{equation}
	\delta \sigma_{1,2}\propto -\frac{x_q}{m_e^{1/2}} \tau_D^{3/2}(\theta_{c,i}^{-1}+\theta_{c,o}^{-1}),
\end{equation}
where the power of $3/2$ comes from the divergent density of states. Naturally, as one may guess, we will see below that the power $3/2$ is, in fact, approximate (see Fig. \ref{fig:appendix:conductivity:higher:total-conductance}).

\begin{figure}
\includegraphics[width=1\columnwidth]{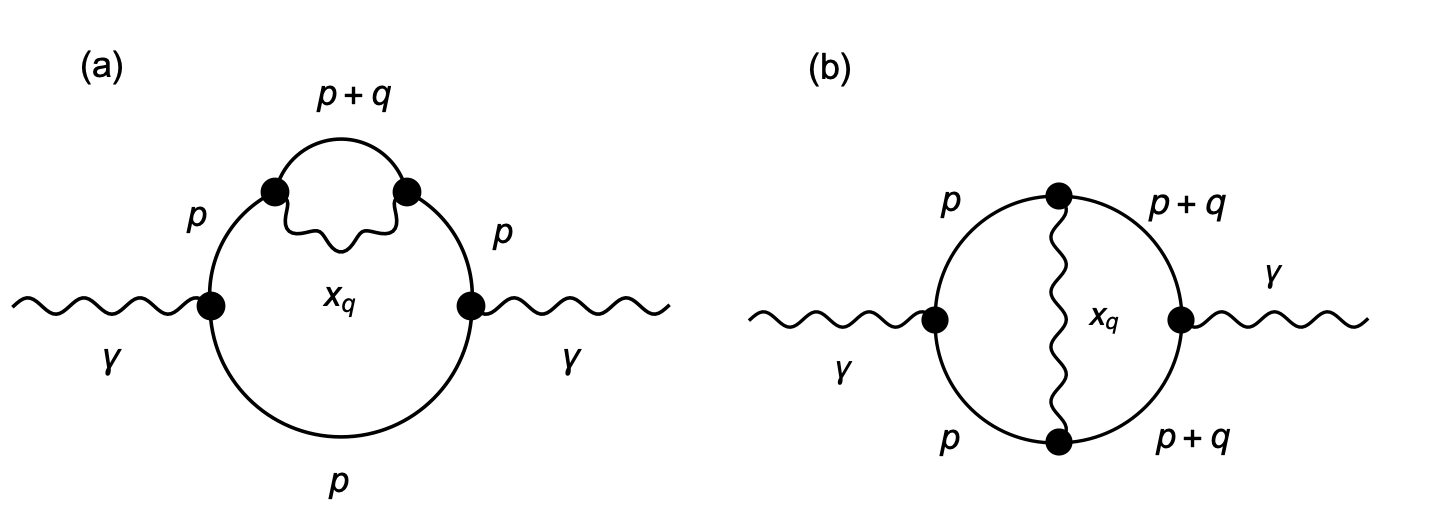}
\caption{Corrections in the leading order in intervalley pseudo-magnetization $x_q$ to the electron part of the conductivity of 2D system with double-well dispersion. Wavy line within the fermion loop denotes propagator $\<x_q x_{-q}\>$. On the symmetry broken side, we take it simply the order parameter squared $x_{q_i}^2$ for $q=q_{i}$ and vanishingly small frequency. Corrections can be divided into a correction to the density of states (a) and a correction to the vertex (b). }
\label{fig:electromagnetic:corrections}
\end{figure}
Finally, once potential gaps out a Fermi arc, all electron-hole pairs in that region of $k$-space become condensed. For $\tau_D m_e\gg 1$, major contribution to $\delta \sigma$ comes from  homo-process, negative, and has normal form $\propto -x_q\tau_D$, as well as correction due to the change of the electron density $\propto x_q^2\tau_D/m_e$.  Having briefly discussed conductance change semi-qualitatively, we now are to find more rigorous arguments through direct computation. As such, we will see that the correction due to the Fermi arcs is, in fact, unimportant and another term $\propto x_q^2\tau_D^2(\theta_{c,i}^{-1}+\theta_{c,o}^{-1})$ dominates.

\subsection{Second order}
For small values of ratio $x_q/\Sigma$ and $q=q_c+\delta q$ it can be shown perturbatively that the power is 2: $\delta \sigma \propto x_q^2\tau_D^{2}$.
The correction to the fermion part of the conductance is given by the sum of two diagrams (see Fig. \ref{fig:electromagnetic:corrections}) can be simplified down to: 
\begin{equation}
  \sigma^{(\Sigma)}_{xx}
  =\frac{3 x_qx_{-q}\Sigma^2}{2\pi}\int \frac{\(v_{p+q}^x+v_{p}^x\)v_{p+q}^x+(\xi_p+\xi_{p+q})\frac{\d v^x_{p+q}}{\d p_x}}{(\xi_p^2+\Sigma^2)(\xi_{p+q}^2+\Sigma^2)^2}
\end{equation}
Situation is similar to scattering of excitons that enter $U_4$ coefficients, except since the $\sigma\propto G_p^2G_{p+q}^2$, characteristic time $\tau_D\propto 1/\Sigma$ enters in a doubled power. Indeed, counting powers of $\Sigma$, we should have $\Sigma^{-4}$ and the area $dp d\theta\propto\Sigma^2$, hence giving $\tau_D^2$-dependence. Because of the smallness of $\delta q$ (or, corollary, because of the fact that the area $dp d\theta\propto (\sin(\theta_{c,i/o}))^{-1}$ is large), hetero-process dominate and enter with a different sign. In the limit of large scattering length ($\tau_D\gg m_e^{-1}$), the leading correction is
\begin{multline}\label{EM:response:second:total:approximate}
	\sigma^{c}_{xx}\propto-\frac{|x_q|^2}{\Sigma^2}  \sum_s \cot(\theta_{c,s})+o(\Sigma/m_e)\\\propto -q^3\frac{m_e^2}{\Sigma^2}\(\frac{T_L}{m_e}\)^{1/2}\sum_s\cot(\theta_{c,s}),
\end{multline}
where $c_1=O(1)$ and $\theta_c\approx 0$ and we used \eqref{U4:homo} (see Appendix D for details of the calculation). Interestingly, terms $O(\Sigma^{-1})$ are effectively absent, since $\Sigma/m_e$ is vanishingly small, while contributions from DOS and vertex correction that go as $\tau_D^2$ do not cancel each other. Because of this term correction to the conductance may not be vanishingly small even in case $x_q/\Sigma<1$, $x_q/m_e\ll 1$.

\subsection{Series resummation}\label{EM:response:infinite}
Deeper inside the partial condensate phase, higher order terms in the expansion in $x_q\tau_D$ become relevant and formula \eqref{EM:response:second:total:approximate} is no longer valid. Note that in the 4-th order, the correction to the conductance are proportional to the coefficients of the Landau-Ginsburg theory $U(q_1,q_2,q_3)$, and hence we assume that the main contribution comes from terms without momentum transfer.  This subset of diagrams is possible to sum up to an infinite order.

Perturbative expansion, for example, can be obtained through a functional integral. Since  $\log(1+G^{-1}(A+x_q))$ is the functional to expand,  all terms come with the same combinatorial coefficient, and hence the correction should have a form of a simple ratio of the form  $x_q^2 \int v_p v_{p/p+q} 1/(x^2_q+g^{-1}_pg^{-1}_{p+q})$, where $g_p=G_p(i\omega=i0+)$. Indeed, one can show \footnote{See Appendix E for proof.} that total correction to conductance $\delta \sigma$ can be divided into three physically distinct terms
\begin{equation}\label{EM:response:infinite:general-formula:vertex}
	\delta \sigma_{odd}=
	\sum_q\frac{|x_q|^2}{2\pi}\int v_p v_{p+q} \Im\(D_q(p)\)^2
\end{equation}
for processes where an odd number of excitons is absorbed/emitted between current vertices, and
\begin{multline}\label{EM:response:infinite:general-formula:DOS}
	\delta \sigma_{even,2}=
	\sum_q\frac{|x_q|^2}{\pi}\int v_p^2 \Im\(g_{p+q}^{-1}D_q(p)\)\Im\(g_{p}D_q(p)\)
\end{multline}
\begin{equation}\label{EM:response:infinite:general-formula:x4}
	\delta \sigma_{even,4}=
	-\sum_q\frac{|x_q|^4}{2\pi}\int v_p^2 \Im\(g_{p}D_q(p)\)^2
\end{equation}
for processes with even number of exciton emission/absorption between current vertices, and $D_q(p)$ stands for $D_q(p)=(g_p^{-1}g_{p+q}^{-1}-x_q^2)^{-1}$ and summation is performed over all reciprocal lattice vectors $q$ that form a crystal lattice \footnote{Note that the origin of asymmetry in $\delta \sigma_{DOS}$ between $p$ and $p+q$ is a convention of how we label current vertex.}. Formula \eqref{EM:response:infinite:general-formula:DOS} simply takes into  account change of the electron Green function $\delta G_{11}$ \cite{altshuler1980interaction} with corrected vertex.

We distinguish three dimensionless parameters:  $\epsilon_{x}=x_q/m_e$, dimensionless energy of an excitation across the Fermi arc $\epsilon_{Q}=p_{i/o} q_c \sin(\theta_{c,i/o})/\kappa_{io/oi}^{1/2}$ as well as $\epsilon_{dis}=\Sigma/m_e$ for the problem. We now briefly describe limiting cases classified by the values of these parameters (there is a plot in appendix -- Fig. \ref{fig:appendix:conductivity:higher:total-conductance}).\\\\
Ratio $\epsilon_{x}/\epsilon_{Q}$ determines whether scaterring of carriers off excitons is energetically allowed. For $\epsilon_{x}/\epsilon_{Q}\ll 1$ correction to conductance has a form $\delta \sigma_{1,1, scat}+\delta \sigma_{1,1, cond}$, where 
\begin{equation} \label{EM:response:infinite:regime-1:scatterring}
	\delta \sigma_{1,1, scat} \propto  -\(\frac{x_q v_i}{\Sigma m_e}\)^2\frac{1}{p_i q_c \theta_{c,i}}-\(\frac{x_q v_o}{\Sigma m_e}\)^2\frac{1}{p_o q_c \theta_{c,o}}
\end{equation}
\begin{equation} \label{EM:response:infinite:regime-1:density}
	\delta \sigma_{1,1, cond} \propto  -\frac{v_i^2\theta_{c,i}}{\Sigma m_e} -\frac{v_o^2\theta_{c,o}}{\Sigma m_e},
\end{equation}
 and the first term corresponds to scattering off excitons, while the second, order parameter independent term simply subtracts condensed fraction from electron density, and we also assumed $\epsilon_{x}\gg \epsilon_{dis}$ for simplicity here.\\\\
For large values of the ratio $\epsilon_{x}/\epsilon_{Q}>1$, scattering off excitons is forbidden, and hence the whole correction comes from the fact that part of the system is in the condensate: 
\begin{equation}
	\delta \sigma_{3,1, cond} \propto  -\frac{v_i^2}{\Sigma m_e}.
\end{equation}
  So that we see that in agreement to naive picture\cite{blinov2024partial} the conductance is reduced by fraction of fermions that form excitons. 
\begin{figure}
\includegraphics[width=1\columnwidth]{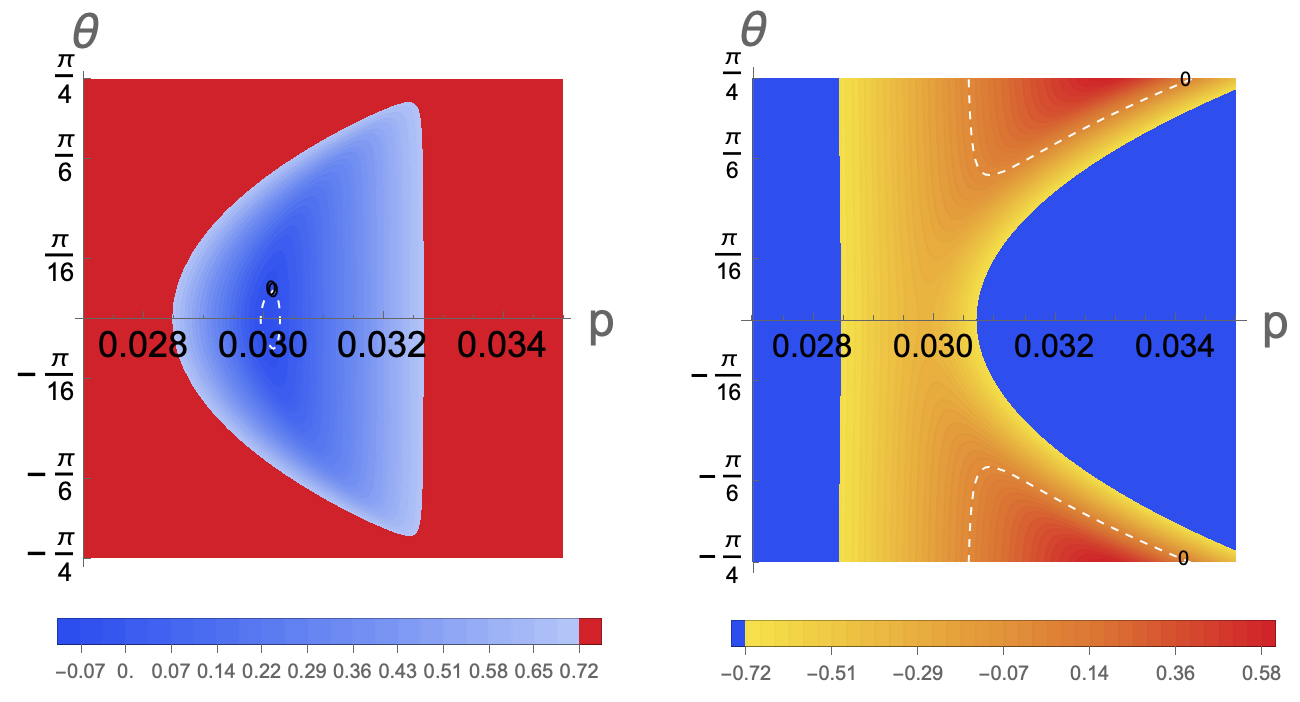}
\caption{Fermi arcs (dashed, white) in the regime with small exciton gap $x_q/(m_e\kappa_{io}^{1/2})<p_{i/o} q_c \theta_{c,i/o}^2$ within the conduction band (left) and valence bands (right)}
\label{fig:electromagnetic:Fermi-arcs}
\end{figure}
For $\epsilon_x/\epsilon_Q<1$ there is also a conductance correction (it appears both in series $\delta \sigma_{odd}$ and $\delta \sigma_{even,2}$) linear in $x_q \Sigma^{-1}$ associated with conversion of electrons into holes and countercurrent generation:
\begin{equation}
	\delta \sigma_{2,1}\propto  -\frac{x_q}{\Sigma m_e^2}\(\frac{v_i(v_i-v_o)}{p_i q \theta_{c,i}}+\frac{v_o(v_o-v_i)}{p_o q_c \theta_{c,o}}\),
\end{equation}
for $x_q\tau_D\gg 1$.\\\\
Once $\epsilon_{dis}$ approaches $\epsilon_{x}$ as a generic rule we would need to multiply each term linear in $x_q$ by $x_q/\sqrt{x_q^2+\Sigma^2}$ or a square of it (see Appendix V for details).

\section{Conclusion}
A phase with partial intervalley exciton condensate is similar to Larkin-Ovchinnikov (LO) phase of superconductivity. It also appears because of the presence of two Fermi surfaces (electron- and hole-like) with different radius ($p_i$ and $p_o$) of a band model, and leads to oscillations of density with period $\approx(p_o-p_i)^{-1}$ much larger than the lattice constant.\\
We showed that unlike the LO-phase in 3D, minimum of energy at least within the weak-coupling theory, corresponds to a phase with $C_6$ symmetric order parameter and not to $C_2$. In contrast to LO-phase, reciprocal lattice vectors of such intervalley crystal are not exactly equal to $p_o-p_i$, but with a small correction $\delta q$ because of the anharmonicity of bands. This fact modifies both order parameter behavior and transport properties qualitatively.\\
As such, coefficients $U_4$, corresponding to scattering of 2 intervalley excitons, diverge as $\Omega^{-1/2}$ and not  $\Omega^{-1}$, which leads to a larger value of the order parameter $x_q$. Interestingly, critical temperature of the phase $\propto x_q$ has two regimes: when $T_c\gg T_1$, the behavior is of MacMillan type $T_c\propto T_L e^{-\frac{m_e}{\theta_c\lambda}}$, while at weakly interacting regime it has a square root dependence on interaction strength $T_{c,L}\propto T_L(-V\Pi_0 -1)^{1/2}$. \\
Order has two components, metallic (coming from the contribution between alike Fermi surfaces) and insulating (coming from the contribution between the different Fermi surfaces). \\
 Because for small $x_q$ the phase has Fermi arcs, Fourier transform of quantum oscillations at very low frequencies should have a maximum, which is consistent to the experiment. However, the latter is also consistent with partially isospin polarized \cite{huang2022spin} order.
What seems to be inconsistent, is that resistance by a factor of $1.5-2$ larger than the resistance of the paramagnetic phase. It is, however, straightforward to explain with partial condensation of excitons, since some of electron-hole pairs condense and do not contribute to the transport on the mean-field level. \\
In the regime where emission of an exciton is permitted by energy conservation $\epsilon_o(p+q)-\epsilon_{i}(p)=\epsilon_x(q)$, charge carriers acquire new mechanism to relax energy with a characteristic time scale $\propto \tau_X$. Then ignoring other changes induced by the presence of the order parameter, by Matthiessen's rule, the correction to the conductivity 
\begin{equation}
	\delta \sigma\propto\frac{\tau_D}{\tau_D/ \tau_X+1}- \tau_D\propto -\tau_D^2/\tau_X
\end{equation}  
for $\tau_X\gg \tau_D$. This is exactly what we obtained in small $x_q$ limit. Remnants of this $\tau_D^2$ behavior can be observed at any doping before Fermi arcs disappear.\\
In this region, change of the resistance, dependent on relative strength of the order parameter $x_q$, inverse scattering time $\tau^{-1}_D$, and characteristic transversal kinetic energy of excitons $m_e q p_i\theta_{c,i}^2/m_e q p_o\theta_{c,o}^2$ have several power regimes $\delta \sigma\propto (\tau_D x_q)^{n} $ with $n$ effectively ranging from $2$ to $1$. It is possible, however, that some of this region is masked by the superconducting phase. Once a system doped below this regime, scattering of charges to neutral collective waves is no longer permitted, and most of the change in the conductance is due to change of carrier concentration, and ratio $\delta \sigma/\sigma_0 \propto -x_q^2/n_o$ is disorder independent. We described this regime before \cite{blinov2024partial}. \\

Another feature of the phase with $C_6$ symmetric order parameter is that in the second order of the order parameter density should have a correction of the form $\delta \rho (r)\propto x_q^2\int e^{i(K-K'-q)r}e^{-i(K-K'+q)r}\rho(q)\rho^*(-q)$ which generates large-period oscillations with periods $10-100 nm$ visible in STM microscopy for sufficiently large systems or Bragg spectroscopy. Third, because electron- and hole-like Fermi surfaces for large values of the displacement field $D>\gamma_1$, where $\gamma_1$ is the energy scale of the hybridization between $A$ and $B$ sublattices of adjacent layers, enhancement of tunneling conductivity in $z$-direction  can be seen at the point of the phase transition.\\ 
As for quantum correction to the conductance, they decay with increase of the scattering length, and thus are not important in clean $p_f l_s\gg 1$ limit. We also note that since excitons have a weak coupling to electrical field $\propto q_c$ and their response is going to be proportional to their density at momentum other than $q=q_c+\delta q$, we expect that some descendants of Aslamasov-Larkin (AL) corrections \cite{aslamasov1968influence} should be less important at low temperatures in the regime of interest for hetero-part of $\delta \sigma$. We do not exclude a possibility, however, that AL-type of corrections to total conductance will cancel term  $(x_q\tau_D)^2$ in \eqref{EM:response:infinite:regime-1:scatterring} -- we leave this question to future studies.\\\\
Finally, it is necessary to remark that classical fluctuations of the order parameter may change the phase transition character \cite{brazovskii1975phase} from second to the first, however since jump of the order parameter is of the order $T_c/m_e\sqrt{\epsilon_\Delta /m_e}\ll x_q/m_e\propto T_c/m_e$, with $\epsilon_\Delta$ being the rotonic minima, most of conclusions of this paper should not be refuted by this fact.

\section{Acknowledgements}
I.V.B. thanks Daniil Antonenko, Allan MacDonald, Andrey Semenov, Nemin Wei and other collegues for valuable suggestions and comments, as well as hospitality and friendly environment of Yale University, Tsinghua University and A. Alikhanyan National Laboratory (Yerevan Physics Institute) where some parts of this work have been done.

\appendix
  \renewcommand{\thefigure}{A.\arabic{figure}}

	\title{Appendix: Weak-coupling theory for partial condensation of mobile excitons in a system with a double-well dispersion}
	
	\author{Igor V. Blinov}
	\email{blinov@utexas.edu}
	\affiliation{
		Department of Physics, The University of Texas at Austin, Austin, Texas 78712, USA}

	\date{July 24, 2025}
	

	\maketitle
	\newpage
	
\onecolumngrid

\tableofcontents
\section{Band structure}
In the following 3 sections, we embark on a journey to evaluate all the coefficients of the Landau-Ginsburg theory for the intervalley charge density wave at an incommensurate wavevector $q=p_o-p_i+\delta q$ using microscopic fermionic theory. First, we are to simplify quartic band-structure to quadratic band-structure with two species of fermions.

 Let me first derive some relations for the band structure. The dispersion is 
 \begin{equation}
 	\epsilon(p)=mp^2+\lambda p^4-\mu,
 \end{equation}
 hence Fermi momentums are
 \begin{equation}
 	p^2_{i,o}=-\frac{m}{2\lambda}\pm\sqrt{\(\frac{m}{2\lambda}\)^2+\frac{\mu}{\lambda}}
 \end{equation}
 So that once 
 \begin{equation}
 	\frac{m^2}{2^2\lambda}>\mu,
 \end{equation}
 we have opening of the inner (electron-like) Fermi-surface.

 \subsection{Bounds}\label{appendix:response:bounds}
Most of the integrals acquire their value very close to one of the Fermi circles. Hence we can divide the integral into 4 parts: when $p,|p+q|\approx p_i$,  $p,|p+q|\approx p_o$ (homo-part) and $p\approx p_i,|p+q|\approx p_o$,   $p\approx p_o,|p+q|\approx p_i$ (hetero-part). More specifically, most of the contribution comes from the region:
\begin{equation}
	-\frac{\delta}{m_e}<p^2-p_{s}^2<\frac{\delta}{m_e},
\end{equation}
\begin{equation}
	-\frac{\delta}{m_e}<(p+q)^2-p_{s'}^2<\frac{\delta}{m_e}.
\end{equation}
Hence integration over angle happens in the range
\begin{equation}
	-\frac{\delta}{m_e}\frac{1}{ p_s q}+\frac{p_{s'}^2-p_s^2-q^2}{2 p_s q}<\cos(\theta)<\frac{p_{s'}^2-p_s^2-q^2}{2 p_s q}+\frac{\delta}{m_e}\frac{1}{ p_s q}.
\end{equation}
Here, we will be interested in the value of the transferred momentum close to $q_c$: $q=q_c+\delta q$. To establish a connection between $\delta q$, $\delta$ we expand in powers of $\delta q$. For concreteness, start with $s'=o$, $s=i$. Expand in vicinity of $\theta_{c,i}$ such that $\cos(\theta_{c,i})=(p_o^2-p_i^2-(q_c+\delta q)^2)/(2p_i (q_c+\delta q))$: 
\begin{equation}
	-\frac{\delta_i}{m_e}\frac{1}{ p_i q}<-\sin(\theta_{c,i})(\theta-\theta_{c,i})<+\frac{\delta_i}{m_e}\frac{1}{ p_i q}.
\end{equation}
which gives: 
\begin{equation}
	\theta_{c,i}^2=\frac{\delta_i}{m_e}\frac{1}{ p_i q}.
\end{equation}
\\\\On the other hand, using the definition above for the central angle $\theta_{c,i}$:
\begin{equation}
	1-\frac{\theta_{c,i}^2}{2}=\(1-\frac{\delta q}{q_c}+\(\frac{\delta q}{q_c}\)^2\)\(1-\delta q\frac{2 q_c+\delta q}{2p_i q_c}\)=1-\frac{\delta q p_o}{q_c p_i}+\frac{\delta q^2}{q_c^2}\(\frac{p_o+p_i}{2 p_i}\),
	\end{equation}
	meaning that 
	\begin{equation}
		\frac{\delta_i}{2m_e}\frac{1}{ p_i q}=\frac{\delta q p_o}{q_c p_i}-\frac{\delta q^2}{q_c^2}\(\frac{p_o+p_i}{2^2 p_i}\).
	\end{equation}
	Similarly, we get for the outer Fermi surface: 
	\begin{equation}
		\theta_{c,o}^2=\frac{\delta_o}{m_e}\frac{1}{ p_o q}.
	\end{equation}
	and from an analogous definition of the outer Fermi surface we can obtain: 
\begin{equation}
		\frac{\delta_o}{2m_e}\frac{1}{ p_o q}=\frac{\delta q p_i}{q_c p_o}-\frac{\delta q^2}{q_c^2}\(\frac{p_o+p_i}{2^2 p_o}\).
	\end{equation}
Note that from the requirement $\delta q_o=\delta q_i$ follows 
\begin{equation}
	\delta_i/p_o^2\approx\delta_o/p_i^2\equiv \delta '.
\end{equation}
	Introducing their arithmetic average $\delta$ we get $\delta=(\delta_i+\delta_o)/2=\delta'(p_o^2+p_i^2)/2=\delta ' |m/\lambda|$. As a result: 
	\begin{equation}
		\theta_{c,i/o}^2=\frac{\delta'}{m_e}\frac{p_{o/i}^2}{q_cp_{i/o}}=\frac{2\delta q p_{o/i}}{q_c p_{i/o}}-\frac{\delta q^2 \<p\>}{q_c^2 p_{i/o}},
	\end{equation}
	so that $\delta q\propto \<p\>\delta '/m_e$. Then we choose $\delta$ such that it minimizes the mistake of the quadratic approximation:
\begin{equation}
	\Delta I=\sum_p \frac{n(\xi(p+q))-n(\xi(p))}{\xi(p+q)-\xi(p)+i\delta}-\sum_{p,a,a'} \frac{n(\xi_a(p+q))-n(\xi_{a'}(p))}{\xi_a(p+q)-\xi_{a'}(p)+i\delta},
\end{equation}
where $\xi(p)=\xi_a(p)+\lambda(p^2-p_a^2)^2$. Since the energy range $\delta$, $\epsilon(p+q)-\epsilon(p)\approx \delta$, then mistake from the range should scale, by the order of magnitude, as $-(\delta-\delta_{1/2})/m_e\delta $ ($\delta_{1/2}$ is the division of momentum range exactly in the middle between $p_i$ and $p_o$) and the mistake from the approximation comes in the second order expansion in $\lambda$, hence should be around $\propto  C \int n(m_e(p^2-p_a^2))\lambda(p^2-p_a^2)^2/m_e^2(p^2-p_a^2)^2\propto  C \lambda\log(\delta)/m_e$, where we ignored the mistake coming from the restriction over the angle range, since it should take care of double-counting. Then the best value should be given by, approximately, 
\begin{equation}
-\frac{\delta_{1/2}}{m_e\delta}+\frac{\lambda}{m_e}C=0,
\end{equation}
where $C$ is of order 1. 
It should be then that $\delta\approx (m_e/\lambda)\delta_{1/2}$. With $\delta_{1/2}\approx(p_o^2-p_i^2)/2=m_e/\lambda$, so that $\delta\approx (m_e/\lambda)^2$.

\section{Response}\label{sec:response}
The generic expression for response is 
\begin{equation}
	\Pi(q,i\Omega)=\sum_n\int d^2p G(p,i\omega_n)G(p+q,i(\omega_n+\Omega)),
\end{equation}
or, after summation over Matsubara frequencies, is:
\begin{equation}
	\Pi(q,i\Omega)=\int d^2p \frac{n(\xi_{p+q})-n(\xi_{p})}{i\Omega+\xi_{p+q}-\xi_{p}}.
\end{equation}

\subsection{Homo-part} 
Homo-part has the form normal for metallic response at finite $q$:  
\begin{equation}
	\Pi(q,i\Omega)=\int d\theta d p p \frac{n(\xi_{p+q})-n(\xi_{p})}{i\Omega+m_e(2p q\cos(\theta)+q^2)}.
\end{equation} 
\paragraph{Zero temperature value}
For small frequencies, response is peaked at $\xi_{p+q}\approx \xi_p$, which for homo-part at temperatures far below Fermi is equivalent to $\xi_p\approx 0$ and hence corrections due to constrain of the integration range over angle discussed in section \ref{appendix:response:bounds} should be $O(\Omega)$. We then extend integration range to $-\pi$ and $\pi$ and represent the response on the inner Fermi surface in the form:
\begin{equation}
	\Pi(q,i\Omega)=-\int  \frac{d\theta d p p n(\xi_{p})}{-i\Omega+m_e(2pq\cos(\theta)+q^2)}-\int \frac{d\theta d p p  n(\xi_{p})}{i\Omega+m_e(2p q\cos(\theta)+q^2)}=-2\text{Re}\(\int \frac{d\theta d p p  n(\xi_{p})}{i\Omega+m_eq(2p \cos(\theta)+q)}\)
\end{equation} 
Which gets us 
\begin{multline}
	\Pi_{ii}(q<2p_i,i\Omega)=
	-\frac{\pi}{m_e q}\(\(\frac{i2^2\Omega }{m_e}-\(\frac{2\Omega }{m_eq}\)^2\)^{1/2}+i\(-\frac{i2^2\Omega }{m_e}+\(\frac{2\Omega }{m_e q}\)^2\)^{1/2}\)
	\\+\frac{\pi}{m_e q}\(\(q^2+\frac{i2^2\Omega}{m_e}-\(\frac{2\Omega}{m_e q}\)^2\)^{1/2}-i\((2p_i)^2-q^2-\frac{i2^2\Omega}{m_e}+\(\frac{2\Omega}{m_e2}\)^2\)^{1/2}\)
	\end{multline}
and
\begin{equation}
	\Pi_{ii}(q>2p_i,i\Omega)=-\frac{\pi}{m_e q}\(\(q^2-(2p_i)^2+\frac{2^2i\Omega}{m_e}-\(\frac{2\Omega}{m_e q}\)^2\)^{1/2}-\(q^2+\frac{i2^2\Omega}{m_e}-\(\frac{2\Omega}{m_e q}\)^2\)^{1/2}\).
	\end{equation}
And identical expressions for the outer-part, except for $p_i\to p_o$. For finite temperatures, we get quadratic dependence on temperature with a characteristic scale of $\epsilon_{F,i}$.

\subsection{Hetero-part}
\paragraph{Zero temperature value}
The hetero-response can be written in the form 
\begin{multline}
	\Pi(q,i\Omega)=-\int d^2p \frac{n(\xi_{p,i})}{-i\Omega-\xi_{p,i}+\xi_{p+q,o}}+\int d^2p \frac{n(\xi_{p,o})}{-i\Omega-\xi_{p+q,i}+\xi_{p,o}}
	\\+\int d^2p \frac{n(\xi_{p,o})}{i\Omega+\xi_{p,o}-\xi_{p+q,i}}-\int d^2p \frac{n(\xi_{p,i})}{i\Omega+\xi_{p+q,o}-\xi_{p,i}}\\
	\equiv -2\text{Re}(I_i(q,i\Omega))+2\text{Re}(I_o(q,i\Omega))
\end{multline}
where the first term responsible for electrons on the inner Fermi-surface, while the second -- on the outer Fermi-surface. The integral over the inner part is 
\begin{equation}
	I_i(q,\Omega)=\int \frac{d^2 p}{(2\pi)^2}\frac{n(\xi_{p,i})}{i\Omega+\xi_{p+q,o}-\xi_{p,i}}=\frac{1}{m_e}\int \frac{d^2 p}{(2\pi)^2}\frac{n(\xi_{p,i})}{\frac{i\Omega}{m_e}-((p+q)^2-p_o^2)-(p^2-p_i^2)}
\end{equation}
The integral over p can be taken either through a substitution $\delta p=p^2-p_i^2$ and linearization in $\delta p$, or directly as an integral over $p$. In zero temperature limit, 
\begin{enumerate}
	\item \textit{Exact integral over $p$}
	\begin{equation}
	I_i(q,\Omega)=\frac{1}{m_e}\int_{0}^{p_i} \frac{d\theta d p}{(2\pi)^2}\frac{p}{\frac{i\Omega}{m_e}-2p^2-2pq\cos(\theta)+p_o^2+p_i^2-q^2}
\end{equation} 
which has roots at $p_{i,1/2}=-\frac{q}{2}\cos(\theta)\pm\frac{1}{2}\sqrt{(q\cos(\theta))^2+2(p_o^2+p_i^2-q^2+i\Omega/m_e)}$.
Note here that since the angle is close to $0$, we can expand in the vicinity of $\theta=0$: 
\begin{multline}
 	p_{i,1/2}=-\frac{q}{2}(1-\theta^2/2)\pm\frac{1}{2}\sqrt{(q(1-\theta^2/2))^2+2(p_o^2+p_i^2-q^2+i\Omega/m_e)}\\=-\frac{q}{2}(1-\theta^2/2)\pm\frac{1}{2}\sqrt{q^2(1-\theta^2)+2(p_o^2+p_i^2-q^2+i\Omega/m_e)}\\=-\frac{q}{2}(1-\theta^2/2)\pm\frac{1}{2}\sqrt{2(p_o^2+p_i^2)-q^2}\\\pm\frac{i\Omega/2m_e}{(2p_o^2+2p_i^2-q^2)^{3/2}}\mp\frac{q^2\theta^2}{2^2\sqrt{2p_o^2+2p_i^2-q^2}}=\kappa_{i,1/2}+q\theta^2r_{i,1/2} \pm i \alpha \Omega ,
 \end{multline}
 which defines coefficient $r_{i,1/2}$ as well as $\alpha$:
 \begin{equation}
 	r_{i,1/2}=\frac{p_{i/o}}{2(p_o+p_i)}
 \end{equation}
  The angle at which the real part of $p_i-p_{i,1/2}$ vanishes we call the central angle
 \begin{equation}
 	\theta_{c,i}^2\equiv \frac{p_i-\kappa_{i,1}}{qr_{i,1}}\approx\(\frac{2\delta q p_{o}}{p_{i} q_c}+\frac{\delta q^2(p_o^2+p_i^2)}{q_c p_{i}(p_o+p_i)^2}\),
 \end{equation}
The integral now can be represented in the form:
	\begin{equation}
	I_i(q,\Omega)=-\frac{1}{2m_e}\int_{0}^{p_i} \frac{d\theta d p}{(2\pi)^2}\frac{p}{(p-p_{i,1})(p-p_{i,2})}=
	-\frac{1}{2m_e}\int_{0}^{p_i} \frac{d\theta d p}{(2\pi)^2}
	\(\frac{p_{i,1}}{p-p_{i,1}}-\frac{p_{i,2}}{p-p_{i,2}}\)\frac{1}{p_{i,1}-p_{i,2}}
\end{equation} 
Integration over $p$ gives:
\begin{equation}
	I_{i}(q,\Omega)\approx\frac{1}{2^2m_e}\int_0^{\theta_{c,i}}
	\frac{d\theta}{(2\pi)^2} \frac{1}{p_{i,1}-p_{i,2}}
	\Bigg(
		p_{i,1}\log\(\frac{(p_i-\bar{p}_{i,1})^2+\alpha^2\Omega^2}{\bar{p}_{i,1}^2+\alpha^2\Omega^2}\)
	-p_{i,2}\log\(\frac{(p_i-\bar{p}_{i,2})^2+\alpha^2\Omega^2}{\bar{p}_{i,2}^2+\alpha^2\Omega^2}
	\)
	+i\pi p_{i,1}
	\Bigg),
\end{equation}
where $p_{i,1/2}=\text{Re}(\bar{p}_{i,1/2})$. Introducing a critical angle: 
\begin{equation}\label{appendix:response:hetero:0T:theta}
	\theta_{c,i/o}^2\equiv \frac{p_i-\kappa_{i,1}}{qr_{i,1}}\approx\(\frac{2\delta q p_{o/i}}{p_{i/o} q_c}\pm\frac{\delta q^2(p_o^2+p_i^2)}{q_c p_{o/i}(p_o+p_i)^2}\)
\end{equation}
or, for small $\delta q$, 
\begin{equation}\label{appendix:response:hetero:0T:theta:approx}
	\theta_{c,i/o}^2=\frac{2\delta q p_{o/i}}{p_{i/o} q_c}.
\end{equation}
we can rewrite the response in the form: 
\begin{equation}
	I_{i}(q,\Omega)=\frac{1}{2^2m_e}\int_0^{\theta_{c,i}}
	\frac{d\theta}{(2\pi)^2} \frac{1}{p_{i,1}-p_{i,2}}
	\Bigg(
		p_{i,1}\log\(\frac{(\theta_{c,i}^2-\theta^2)^2+(\alpha/q r_i)^2\Omega^2}{(\kappa_i/qr_i+\theta^2)^2+(\alpha/q r_i)^2\Omega^2}\)
	-p_{i,2}\log\(\frac{(p_i-p_{i,2})^2+\alpha^2\Omega^2}{p_{i,2}^2+\alpha^2\Omega^2}
	\)
	+i\pi p_{i,1}
	\Bigg).
\end{equation}
which reaches its maximal value for $\theta_{c,i}=\theta_{f}$ and after integration, is equal to
\begin{equation}\label{appendix:response:hetero:0T:inner:finite-Omega}
		I_{i}(q=q_{c}+\delta q,\Omega=0)\approx\frac{\theta_{c,i}}{2^2m_e}\frac{\kappa_{1}}{\kappa_1-\kappa_2}\log\(\frac{\epsilon_c}{\epsilon_{F,i} e^2}\),
\end{equation}
where $\epsilon_c=(\theta_c q_c)^2 m_e$ and $\epsilon_{F} =m_e (p_i+p_o)^2/2^2$, where $\epsilon_c$ plays the role of infra-red cutoff for the relative degree of freedom of an exciton, has a form similar to the density-density response of the 1D atom chain at $q$ and is related to the Peierls distortion\cite{peierls1996quantum}.\\
When frequency kept finite, 
\begin{multline}\label{appendix:response:hetero:0T:inner:finite-Omega}
	I_{het,i}=\frac{\kappa_1 \theta_f}{\kappa_1-\kappa_2}\(\log((\theta_f^2 -\theta_c^2)^2+(\alpha \Omega)^2)-2^2\)\\-\frac{\kappa_1\theta_c'}{\kappa_1-\kappa_2}\(\log((\theta_f-\theta_c)^2+(\alpha\Omega)^2)-\log(\theta_c^2+(\alpha\Omega)^2)-\log((\theta_c+\theta_f)^2+(\alpha\Omega)^2)+\log(\theta_c^2+(\alpha\Omega)^2)\).
\end{multline}
\end{enumerate}

\paragraph{Finite temperature correction: \textbf{inner FS contribution}}
Inner part of the correction due to the finiteness of the temperature can be written as
\begin{multline}
	\Delta I_i(q,\Omega,T)
	=
	\frac{1}{m_e}
	\int^{p_i}_0 
	dp p d\theta\frac{n(m_e(p^2-p_i^2))-1}{2p^2+q^2+2p q\cos(\theta)-\frac{\Delta}{m_e}+\frac{i\Omega}{m_e}}
	\\+
		\frac{1}{m_e}
	\int_{p_i}^{p_\Lambda}dp p d\theta
	\frac{n(m_e(p^2-p_i^2))}{2p^2+q^2+2p q\cos(\theta)-\frac{\Delta}{m_e}+\frac{i\Omega}{m_e}},
\end{multline}
where $\Delta=m_e(p_o^2+p_i^2)$. Changing integration variable to $y=\beta m_e(p^2-p_i^2)$, I get for the correction: 
\begin{multline}
	\Delta I_i(q,\Omega,T)
	=
	\frac{1}{2m_e^2\beta}
	\int^{0}_{-\beta m_ep_i^2} 
	dy d\theta\frac{\bar{n}(y)-1}
	{\frac{2y}{\beta m_e}+2p_i^2+q^2+2(\frac{y}{\beta m_e}+p_i^2)^{1/2} q\cos(\theta)-\frac{\Delta}{m_e}+\frac{i\Omega}{m_e}}
	\\+
		\frac{1}{2m_e^2 \beta }
	\int_{0}^{\beta m_e p_\Lambda^2}dy  d\theta
	\frac{\bar{n}(y)}{\frac{2y}{\beta m_e}+2p_i^2+q^2+2(\frac{y}{\beta m_e}+p_i^2)^{1/2} q\cos(\theta)-\frac{\Delta}{m_e}+\frac{i\Omega}{m_e}},
\end{multline}
where $\bar{n}(y)=(1+e^{y})^{-1}$ which can be rewritten as 
\begin{multline}
	\Delta I_i(q,\Omega,T)
	=
	-\frac{1}{2m_e^2\beta}
	\int^{\beta m_ep_i^2}_{0} 
	dy d\theta\frac{\bar{n}(y)}
	{-\frac{2y}{\beta m_e}+2p_i^2+q^2+2(-\frac{y}{\beta m_e}+p_i^2)^{1/2} q\cos(\theta)-\frac{\Delta}{m_e}+\frac{i\Omega}{m_e}}
	\\+
		\frac{1}{2m_e^2 \beta }
	\int_{0}^{\beta m_e p_\Lambda^2}dy  d\theta
	\frac{\bar{n}(y)}{\frac{2y}{\beta m_e}+2p_i^2+q^2+2(\frac{y}{\beta m_e}+p_i^2)^{1/2} q\cos(\theta)-\frac{\Delta}{m_e}+\frac{i\Omega}{m_e}},
\end{multline}
At a fixed $y$, the denominator, as a function of $\theta$, has two regimes: when the integral has a root at some $\theta$, and the regime of analyticity. Let us define $\theta_{i}$ to be an angle where the denominator vanishes at $y=0$:
	\begin{equation}
		\cos(\theta_i)=\frac{p_o^2-p_i^2-q^2}{2p_i q}.
	\end{equation}	
	Expanding in the vicinity of $\theta=\theta_{c,i}$, I can rewrite the integrals in the form: 
	\begin{multline}
	\Delta I_i(q,\Omega,T)
	=
	-\frac{1}{2m_e^2\beta}
	\int^{\beta m_ep_i^2}_{0} 
	dy d\theta\frac{\bar{n}(y)}
	{-\frac{2y}{\beta m_e}(1+\frac{q}{2p_i}\cos(\theta_i))-2(1-\frac{y}{2\beta m_e p_i^2})p_i q\sin(\theta_i)(\theta-\theta_i)+\frac{i\Omega}{m_e}}
	\\+
		\frac{1}{2m_e^2 \beta }
	\int_{0}^{\beta m_e p_\Lambda^2}dy  d\theta
	\frac{\bar{n}(y)}{\frac{2y}{\beta m_e}(1+\frac{q}{2p_i}\cos(\theta_i))-2(1+\frac{y}{2\beta m_e p_i^2})p_i q\sin(\theta_i)(\theta-\theta_i)+\frac{i\Omega}{m_e}},
\end{multline}
	which after integration gives
		\begin{multline}
	\Delta I_i(q,\Omega,T)
	=
	\frac{1}{2m_e^2\beta}
	\int^{\Lambda}_{0} 
	\frac{dy \bar{n}(y)}{2(1-\frac{y}{2\beta m_e p_i^2})p_iq\sin(\theta_i)}
	\Bigg(
	\\\log\(\(\frac{2y}{\beta m_e}(1+\frac{q}{2p_i}\cos(\theta_i))\)^2+\(\frac{\Omega}{m_e}\)^2\)
	-
	\log\(\(\frac{2y}{\beta m_e}(1+\frac{q}{2p_i}\cos(\theta_i))-2(1-\frac{y}{2\beta m_e p_i^2})p_i q\sin(\theta_i)\theta_i\)^2+\(\frac{\Omega}{m_e}\)^2\)\Bigg)
	\\-
		\frac{1}{2m_e^2 \beta }
	\int_{0}^{\Lambda}dy  \frac{\bar{n}(y)}{2(1+\frac{y}{2\beta m_e p_i^2})p_i q\sin(\theta_i)}
	\Bigg(
	\\\log\(\(\frac{2y}{\beta m_e}(1+\frac{q}{2p_i}\cos(\theta_i))\)^2+\(\frac{\Omega}{m_e}\)^2\)
	-
	\log\(\(\frac{2y}{\beta m_e}(1+\frac{q}{2p_i}\cos(\theta_i))+2(1+\frac{y}{2\beta m_e p_i^2})p_i q\sin(\theta_i)\theta_i\)^2+\(\frac{\Omega}{m_e}\)^2\)\Bigg)
	\\+\frac{\pi i}{2m_e^2\beta p_i q\sin(\theta_i)}\int \frac{dy n(y)}{1-\frac{y}{2\beta m_e p_i^2}}-\frac{\pi i}{2m_e^2\beta p_i q\sin(\theta_i)}\int \frac{dy n(y)}{1+\frac{y}{2\beta m_e p_i^2}},
\end{multline}
The imaginary part  $\propto T^2/T_{F,i}^2$ here will not contribute to the action, and therefore we omit it in what follows. Clearly, there are 2 characteristic temperatures present here. One corresponds to the Fermi scale: $T_{F,i}=m_e p_i^2$, while the other corresponds to mixing of electron-hole excitations: 
		\begin{equation}
		T_{L,i}=\frac{m_e p_i q\theta_{c,i}^2}{1+\frac{q}{2p_i}\cos(\theta_i)+\frac{q}{2p_i}\sin(\theta_{c,i})\theta_{c,i}}\approx \frac{m_e p_o\delta q}{1+\frac{q}{2p_i}}.
	\end{equation}

	\begin{enumerate}[label=\alph*)]
	\item For low temperatures, $T\ll T_{L,i}$, we neglect terms $O((T/T_{F,i})^2)$,		\begin{multline}
	\Delta I_i(q,\Omega,T)
	=
	-\frac{1}{2^2m_e^2\beta}
	\int^{\Lambda}_{0} 
	\frac{dy \bar{n}(y)}{p_iq\sin(\theta_i)}
	\log\(\(-\frac{y}{\beta  p_i q\theta_i^2}(1+\frac{q}{2p_i}\cos(\theta_i)+\frac{ q}{2p_i} \theta_{c,i}^2)+1\)^2+\(\frac{\Omega}{p_i q\theta_i^2}\)^2\)
	\\+
		\frac{1}{2^2m_e^2 \beta }
	\int_{0}^{\Lambda}dy  \frac{\bar{n}(y)}{p_i q\sin(\theta_i)}
	\log\(\(\frac{y}{\beta p_i q\theta_i^2}(1+\frac{q}{2p_i}\cos(\theta_i)+\frac{q}{2p_i}\theta_{c,i}^2)+1\)^2+\(\frac{\Omega}{ p_i q\theta_i^2}\)^2\)
	\\+\frac{\pi i}{2m_e^2\beta p_i q\sin(\theta_i)}\int \frac{dy n(y)}{1-\frac{y}{2\beta m_e p_i^2}}-\frac{\pi i}{2m_e^2\beta p_i q\sin(\theta_i)}\int \frac{dy n(y)}{1+\frac{y}{2\beta m_e p_i^2}},
\end{multline}
 and expand the logarithms in powers of $y T/T_{L,i}$:
 \begin{equation}
	\Delta I_i(q,\Omega,T)
	=
	\frac{1}{12 m_e^2}
		\frac{\pi^2}{ p_iq\sin(\theta_i)}
	\(\frac{T^2}{T_{L,i}}\)\approx\frac{\pi^2 \theta_i }{12m_e(1+\frac{q}{2p_i})}
	\(\frac{T^2}{T_{L,i}^2}\).
\end{equation}

	\item At higher temperatures, $T_{L,i}<T\ll T_{F,i}$ the integrand is peaked at a value of $y$ where the argument of the logarithm vanishes,  which is
	\begin{align}
		y_{1,<}=\frac{\beta p_i q \theta_i^2}{1+\frac{q}{2p_i}\cos(\theta_i)+\frac{q}{2p_i}\theta_{c,i}^2}=\frac{T_{L,i}}{T},\\
		y_{2,<}=-\frac{\beta p_i q \theta_i^2}{1+\frac{q}{2p_i}\cos(\theta_i)+\frac{q}{2p_i}\theta_{c,i}^2}=-\frac{T_{L,i}}{T}.
	\end{align} 
	for the corresponding terms. We then write the integral in the form: 
	\begin{multline}
	\Delta I_i(q,\Omega,T)
	=
	-\frac{1}{2^2m_e^2\beta}
	\int^{\Lambda}_{0} 
	\frac{dy \bar{n}(y)}{p_iq\sin(\theta_i)}
	\Bigg(
	\log\(\(y-\frac{T_{L,i}}{T}\)^2+\(\frac{ \Omega T_\kappa}{2T m_e}\)^2\)\Bigg)
	\\+
		\frac{1}{2^2m_e^2 \beta }
	\int_{0}^{\Lambda}dy  \frac{\bar{n}(y)}{p_i q\sin(\theta_i)}
	\Bigg(
	\log\(\(y+\frac{T_{L,i}}{T}\)^2+\(\frac{ \Omega T_\kappa}{2 T m_e}\)^2\)\Bigg),
\end{multline}
where I additionally introduced $T_\kappa=m_e/(1+\frac{q}{2p_i}\cos(\theta_i))$. We rescale the integration variable $y\to y T_{L,i}/T$ and extend the range to $\infty$, after the division of the integration range we get: 
	\begin{multline}
	\Delta I_i(q,\Omega,T)
	=
	-\frac{T_{L,i}}{2^2m_e^2}
	\int^{\Lambda}_{0} 
	\frac{dy \bar{n}(y)}{p_iq\sin(\theta_i)}
	\Bigg(
	\log\(\(y-1\)^2+\(\frac{ \Omega T_\kappa}{2T_1 m_e}\)^2\)-\log\(\(y+1\)^2+\(\frac{ \Omega T_\kappa}{2T_1 m_e}\)^2\)\Bigg)
	\\=\frac{T_{L,i}}{m_e^2  p_i q \sin(\theta_c)}\int_0^1 \frac{y}{1+e^{y T_{L,i}/T}}+\frac{T_{L,i}}{ m_e^2  p_i q \sin(\theta_c)}\int_1^\infty  \frac{1}{y(1+e^{y T_{L,i}/T})}
\end{multline}
which after division of the latter integral into 2 over ranges $(1,T/T_{L,i})$ and  $(T/T_{L,i},\infty)$:
	\begin{equation}
	\Delta I_i(q,\Omega,T)
\approx\frac{T_{L,i}}{2m_e^2  p_i q \sin(\theta_{c,i})}\(1+\Gamma_0(1)\)+\frac{T_{L,i}}{2m_e^2  p_i q \sin(\theta_{c,i})}\(\log(T/T_{L,i})-\frac{1}{2}\(1-\frac{T_{L,i}}{T}\)\).
\end{equation}
clearly, the linear behavior here may only arise as an approximation to $\log(T/T_{L,i})$: for example while doing an expansion close to 1.

	\end{enumerate}
	
	\paragraph{Finite temperature correction: \textbf{outer FS contribution}}
	The expression for the outer contribution is	\begin{equation}
	I_o(q,\Omega)=\frac{1}{m_e}\int \frac{d^2 p}{(2\pi)^2}\frac{n(-m_e(p^2-p_o^2))}{\frac{i\Omega}{m_e}+2p^2+q^2-p_i^2-p_o^2+2pq\cos(\theta)}=\frac{1}{m_e}\int \frac{d^2 p}{(2\pi)^2}\frac{n(-m_e(p^2-p_o^2))}{\frac{i\Omega}{m_e}+2p^2+q^2+2pq\cos(\theta)-\frac{\Delta}{m_e}}.
\end{equation}
	And therefore calculation goes over the same lines for the outer contribution, hence we simply change $p_i\to p_o$ in the expressions from the previous section. Define 
	\begin{equation}
	\frac{\beta p_o q \theta_o^2}{1+\frac{q}{2p_o}\cos(\theta_{c,o})+\frac{q}{2p_o}\theta_{c,o}^2}\equiv \frac{T_{L,o}}{T}
	\end{equation}
		\begin{equation}
		\frac{\beta p_o q \theta_o^2}{1+\frac{q}{2p_o}\cos(\theta_{c,o})+\frac{q}{2p_o}\theta_{c,o}^2}\equiv -\frac{T_{L,o}}{T}
	\end{equation}
\begin{enumerate}[label=\alph*)]
\item $T\ll T_{L,o}$:
 \begin{equation}
	\Delta I_o(q,\Omega,T)
	=
	\frac{1}{12 m_e^2}
		\frac{\pi^2}{ p_oq\sin(\theta_o)}
	\(\frac{T^2}{T_{L,o}}\)\approx\frac{\pi^2 \theta_o }{12m_e(1+\frac{q}{2p_o})}
	\(\frac{T^2}{T_{L,o}^2}\).
\end{equation}
\item $T_{L,o}<T<T_{F,o}$:
	\begin{equation}
	\Delta I_o(q,\Omega,T)
\approx\frac{T_{L,o}}{2m_e^2  p_o q \sin(\theta_{c,o})}\(1+\Gamma_0(1)\)+\frac{T_{L,o}}{2m_e^2  p_o q \sin(\theta_{c,o})}\(\log(T/T_{L,o})-\frac{1}{2}\(1-\frac{T_{L,o}}{T}\)\).
\end{equation}
\end{enumerate}

\subsection{Critical temperature}\label{sec:response:critical-temperature}
On the mean-field level, the transition temperature given by the solution of the equation
\begin{equation}
	\Pi_0+\Delta \Pi(T_c)+V^{-1}=0.
\end{equation}
The first characteristic temperature, $T_{L,i}\approx m_e p_o \delta q/(1+q/2p_i)$
is of order $10^{-5} m\approx 10^{-1} K$, which, judging by the appearance of the peak in resistance (Fig S6 \cite{zhou2021superconductivity}) is of order of  critical temperature for PIP phase. For critical temperatures below $T_{L,i/o}$, we get: 
\begin{equation}\label{appendix:response:critical-temperature:Tc-low-temperatures}
	T_{c,low}=\sqrt{\frac{\frac{12 m_e}{\pi^2}(-\Pi_0-V^{-1})}{\frac{\theta_i/T_{L,i}^2}{1+\frac{q}{2p_i}}+\frac{\theta_o/T_{L,o}^2}{1+\frac{q}{2p_o}}}}.
\end{equation}
which in the range of interest gives $10^{-4}-10^{-5} m$. For higher temperatures, the formula instead resembles MacMillan:
\begin{equation}
	T_c=T_{L,i}^{\frac{\alpha_i}{\alpha_i+\alpha_o}}T_{L,o}^{\frac{\alpha_o}{\alpha_i+\alpha_o}}e^{-\Pi_0-\tilde{\Pi}-V^{-1}}
\end{equation}
where
\begin{equation}
	\alpha_i=\frac{T_{L,i}}{2m_e^2p_i q\sin(\theta_{c,i})},
\end{equation}
\begin{equation}
	\alpha_o=\frac{T_{L,o}}{2m_e^2p_o q\sin(\theta_{c,o})},
\end{equation}
and 
\begin{equation}
	\tilde{\Pi}=\frac{T_{L,o}}{2m_e^2  p_o q \sin(\theta_{c,o})}\(1+\Gamma_0(1)\)+\frac{T_{L,i}}{2m_e^2  p_i q \sin(\theta_{c,i})}\(1+\Gamma_0(1)\).
\end{equation}
Due to the fact that we can separate two components: electronic-coherent and bosonic condensates, for CDW at $q=q_c+\delta q$, we may talk about two temperature-induced phase transitions on the mean-field level. Indeed, for the hetero-part the critical temperature is the same as \eqref{appendix:response:critical-temperature:Tc-low-temperatures} except with a change $\Pi_0\to \Pi_{het}$
\begin{equation}
	T_{c,het,low}=\sqrt{\frac{\frac{12 m_e}{\pi^2}(-\Pi_{het}-V^{-1})}{\frac{\theta_i/T_{L,i}^2}{1+\frac{q}{2p_i}}+\frac{\theta_o/T_{L,o}^2}{1+\frac{q}{2p_o}}}},
\end{equation}
while for the homo-part the temperature dependence correction to the response function is $(2/3)(1-(q/p_{i/o})^2)^{-3/2}\pi^2(T/T_{F,i/o})^2(1/m_e)$. 
\begin{equation}
	T_{c,hom,low}=T_{hom}\sqrt{\frac{12m_e}{\pi^2}(-\Pi_{hom}-V^{-1})}.
\end{equation}
where $T_{hom}^2=(1-(q/p_i)^2)^{3/2}T_{F,i}^2+(1-(q/p_o)^2)^{3/2}T_{F,o}^2$. Clearly, $T_{c,hom,low}\gg T_{c,het,low}$.

\section{4-th order terms}\label{sec:4th-order}
Coefficients of the Landau-Ginsburg theory of the 4-th order are, generically, divergent. 
\begin{figure}
\includegraphics[width=0.5\columnwidth]{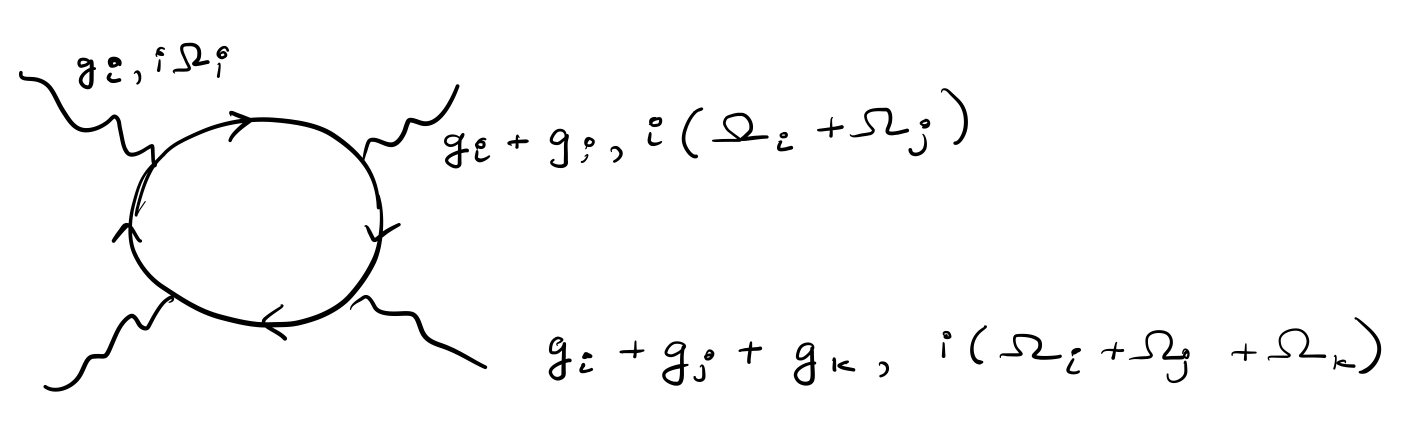}
\caption{Graphic representation of the 4-th order coefficient. }
\label{fig:4-th-order:U4-diagram}
\end{figure}
They are given by the diagrams on Fig.\ref{fig:4-th-order:U4-diagram} and can be written as:
\begin{equation*}
	U(g_i,g_i+g_j,g_i+g_j+g_k)=
	\int d^2 pG(p,i\omega_n) G(p+g_i,i\omega_n+i\Omega_i)G(p+g_i+g_j,i\omega_n+i\Omega_i+i\Omega_j)G(p+g_i+g_j+g_k,i\omega+i\Omega_i+i\Omega_j+i\Omega_k)
\end{equation*}
Using an identity
\begin{equation}
	\frac{1}{i\omega_n-\epsilon_p}\frac{1}{i(\omega_{n}+\Omega)-\epsilon_{p'}}
	=
	\frac{1}{i\Omega-(\epsilon_{p'}-\epsilon_p)}
	\(
	\frac{1}{i\omega_n-\epsilon_p}
	-
	\frac{1}{i(\omega_n+\Omega)-\epsilon_{p'}}
	\),
\end{equation}
we can rewrite the expression above as 
\begin{multline}
	\frac{1}{i\Omega_i-(\epsilon_{p+g_i}-\epsilon_p)}
	\(
	\frac{1}{i\omega_n-\epsilon_p}
	-
	\frac{1}{i(\omega_n+\Omega)-\epsilon_{p+g_i}}
	\)
		\\\frac{1}{i\Omega_k-(\epsilon_{p+g_{i,i,j}}-\epsilon_{p+g_{i,j}})}
	\(
	\frac{1}{i(\omega_n+\Omega_{i,j})-\epsilon_{p+g_{i,j}}}
	-
	\frac{1}{i(\omega_n+\Omega_{i,j,k})-\epsilon_{p+g_{i,j,k}}}
	\),
\end{multline}
where I denoted $g_{i,j}=g_i+g_j$,  $g_{i,j,k}=g_i+g_j+g_k$ and similarly for bosonic frequencies. The latter consists of $4$ terms:
\begin{multline}
	\frac{1}{i\Omega_i-(\epsilon_{p+g_i}-\epsilon_p)}\frac{1}{i\Omega_k-(\epsilon_{p+g_{i,i,j}}-\epsilon_{p+g_{i,j}})}
	\Bigg(
	\frac{1}{i\omega_n-\epsilon_p}\frac{1}{i(\omega_n+\Omega_{i,j})-\epsilon_{p+g_{i,j}}}
		-
	\frac{1}{i\omega_n-\epsilon_p}
	\frac{1}{i(\omega_n+\Omega_{i,j,k})-\epsilon_{p+g_{i,j,k}}}
		\\-
	\frac{1}{i(\omega_n+\Omega_i)-\epsilon_{p+g_i}}\frac{1}{i(\omega_n+\Omega_{i,j})-\epsilon_{p+g_{i,j}}}
	+
	\frac{1}{i(\omega_n+\Omega_i)-\epsilon_{p+g_i}}
	\frac{1}{i(\omega_n+\Omega_{i,j,k})-\epsilon_{p+g_{i,j,k}}}
	\Bigg),
\end{multline}
which can be finally reexpressed as a sum over 4 single-fermion Green's functions with multipliers 
\begin{multline}
	\frac{1}{i\Omega_i-(\epsilon_{p+g_i}-\epsilon_p)}\frac{1}{i\Omega_k-(\epsilon_{p+g_{i,j,k}}-\epsilon_{p+g_{i,j}})}
	\\\Bigg(
	\frac{1}{i\Omega_{i,j}-(\epsilon_{p+g_{i,j}}-\epsilon_p)}
	\(\frac{1}{i\omega_n-\epsilon_p}
	-\frac{1}{i(\omega_n+\Omega_{i,j})-\epsilon_{p+g_{i,j}}}\)
		\\-
	\frac{1}{i\Omega_{i,j,k}-(\epsilon_{p+g_{i,j,k}}-\epsilon_p)}
	\(\frac{1}{i\omega_n-\epsilon_p}-
	\frac{1}{i(\omega_n+\Omega_{i,j,k})-\epsilon_{p+g_{i,j,k}}}\)
		\\-
	\frac{1}{i\Omega_j-(\epsilon_{p+g_{i,j}}-\epsilon_{p+g_i})}
	\(\frac{1}{i(\omega_n+\Omega_i)-\epsilon_{p+g_i}}-\frac{1}{i(\omega_n+\Omega_{i,j})-\epsilon_{p+g_{i,j}}}\)
	\\+
	\frac{1}{i\Omega_{j,k}-(\epsilon_{p+g_{i,j,k}}-\epsilon_{p+g_{i})}}
	\(
	\frac{1}{i(\omega_n+\Omega_i)-\epsilon_{p+g_i}}-
	\frac{1}{i(\omega_n+\Omega_{i,j,k})-\epsilon_{p+g_{i,j,k}}}
	\)
	\Bigg),
\end{multline}
or, explicitly, 
\begin{multline}
	\frac{1}{i\Omega_i-(\epsilon_{p+g_i}-\epsilon_p)}\frac{1}{i\Omega_k-(\epsilon_{p+g_{i,j,k}}-\epsilon_{p+g_{i,j}})}\Bigg(
	\frac{1}{i\omega_n-\epsilon_p}
	\Big(
	\frac{1}{i\Omega_{i,j}-(\epsilon_{p+g_{i,j}}-\epsilon_p)}
	-
	\frac{1}{i\Omega_{i,j,k}-(\epsilon_{p+g_{i,j,k}}-\epsilon_p)}
	\Big)
	\\+
	\frac{1}{i(\omega_n+\Omega_{i,j})-\epsilon_{p+g_{i,j}}}
	\(\frac{1}{i\Omega_j-(\epsilon_{p+g_{i,j}}-\epsilon_{p+g_i})}-	\frac{1}{i\Omega_{i,j}-(\epsilon_{p+g_{i,j}}-\epsilon_p)}\)
	\\
	+
	\frac{1}{i(\omega_n+\Omega_{i,j,k})-\epsilon_{p+g_{i,j,k}}}\(\frac{1}{i\Omega_{i,j,k}-(\epsilon_{p+g_{i,j,k}}-\epsilon_p)}-\frac{1}{i\Omega_{j,k}-(\epsilon_{p+g_{i,j,k}}-\epsilon_{p+g_{i})}}\)
	\\+
	\frac{1}{i(\omega_n+\Omega_i)-\epsilon_{p+g_i}}
	\(
	\frac{1}{i\Omega_{j,k}-(\epsilon_{p+g_{i,j,k}}-\epsilon_{p+g_{i})}}-\frac{1}{i\Omega_j-(\epsilon_{p+g_{i,j}}-\epsilon_{p+g_{i}})}
	\)
	\Bigg)
\end{multline}
Each term here corresponds to an excitation, that started at one of the $p+g_n$ points and hopping around over. Minus would correspond to an opposite direction of hop on the lattice in k-space. After shifting in k-space by one of the g-vectors and performing Matsubara summation, we have 4 integrals in k-space:
\begin{multline}
	U_1=U(g_{i,j},g_k,-g_{i,j,k},g_{i},i\Omega_{i,j},i\Omega_{k,-i,-j},i\Omega_{-2k,-i,-j},i\Omega_{2k,2i,j})\\=
	\int \frac{n(\epsilon_p)}{i\Omega_i-(\epsilon_{p+g_i}-\epsilon_p)}\frac{1}{i\Omega_k-(\epsilon_{p+g_{i,j,k}}-\epsilon_{p+g_{i,j}})}
	\Big(
	\frac{1}{i\Omega_{i,j}-(\epsilon_{p+g_{i,j}}-\epsilon_p)}
	-
	\frac{1}{i\Omega_{i,j,k}-(\epsilon_{p+g_{i,j,k}}-\epsilon_p)}
	\Big)
	\\=
	-\int \frac{n(\epsilon_p)}{i\Omega_i-(\epsilon_{p+g_i}-\epsilon_p)}\frac{1}{i\Omega_k-(\epsilon_{p+g_{i,j,k}}-\epsilon_{p+g_{i,j}})}
	\frac{i\Omega_k-(\epsilon_{p+g_{i,j,k}}-\epsilon_{p+g_{i,j}})}
	{(i\Omega_{i,j}-(\epsilon_{p+g_{i,j}}-\epsilon_p))
	(-i\Omega_{i,j,k}-(\epsilon_p-\epsilon_{p+g_{i,j,k}}))}
	\\
	=V_4(g_{i},g_{ij},g_{i,j,k}|i\Omega_{i},i\Omega_{i,j},i\Omega_{i,j,k})
	,
\end{multline}
in the first line,  the first term apparently corresponds to hoppings $p\to p+g_i\to p\to 
p+g_{i}+g_j\to p+g_{i}+g_{j}+g_{k}\to p$. It seems that for this process to have non-zero contribution one would need to demand that both $g_i+g_j$ and $g_i+g_j+g_k$ belong to the first shell of g-vectors. Note also that the combination $g_i+g_j+g_k$ should not sum to zero, since that would require to have $x(0)$ as one of intervalley fields. Also note that the third argument in $V_4$ is redundant and the reciprocal momentum $g_i$ and corresponding frequency are uncompensated.\\
\\\\ Finally, the second integral is
\begin{multline}
	U_2(g_i,g_j,g_k)=\int 	\frac{n(\epsilon_{p})}{i\Omega_i-(\epsilon_{p}-\epsilon_{p-g_i})}\frac{1}{i\Omega_k-(\epsilon_{p+g_{j,k}}-\epsilon_{p+g_{j}})}
	\(
	\frac{1}{i\Omega_{j,k}-(\epsilon_{p+g_{j,k}}-\epsilon_{p})}-\frac{1}{i\Omega_j-(\epsilon_{p+g_j}-\epsilon_{p})}
	\)
	\\=V_4(-g_i,g_{j,k},g_{j},-i\Omega_i,i\Omega_{j,k},i\Omega_j),
\end{multline}
So that $|g_{j,k}|\leq g$ to make physical both the first and the second terms
\\\\The second terms corresponds to  $p\to p+g_i\to p \to p+g_i+g_j+g_k\to 
p+g_{i}+g_j\to p$, which implies that $g_j+g_k$ and $g_i+g_j$ also should belong to the first shell for it to be nonzero. Alternatively, we can view it as a 5-fermion loop with an additional vertex that goes as $i\Omega_k-(\epsilon_{p+g_{i,j,k}}-\epsilon_{p+g_{i,j}})\to_{g\to 0} i\Omega_k-\bar{v}_p\cdot  \bar{g}_k $. To preserve the change of the valley property, such boson should correspond do either $\sigma_z$ or $\sigma_0$ components. Also note that we omitted frequencies in notation for $V_4$ integrals and the third component, here it is $-g_i$ that is uncompensated, thus cancelling it from the 1st term.\\\\
For the third integral,
\begin{multline}
	U_3(g_i,g_j,g_k)=
	\int 	\frac{n(\epsilon_{p})}{i\Omega_i-(\epsilon_{p-g_j}-\epsilon_{p-g_i-g_j})}\frac{1}{i\Omega_k-(\epsilon_{p+g_{k}}-\epsilon_{p})}
	\(\frac{1}{i\Omega_j-(\epsilon_{p}-\epsilon_{p-g_j})}-	\frac{1}{i\Omega_{i,j}-(\epsilon_{p}-\epsilon_{p-g_{i,j}})}\)
	\\=
	\int 	\frac{n(\epsilon_{p})}{i\Omega_i-(\epsilon_{p-g_j}-\epsilon_{p-g_{i,j}})}\frac{1}{i\Omega_k-(\epsilon_{p+g_{k}}-\epsilon_{p})}
	\frac{1}{i\Omega_j-(\epsilon_{p}-\epsilon_{p-g_{j}})}
	\frac{i\Omega_{i}-(\epsilon_{p-g_j}-\epsilon_{p-g_{ij}})}{i\Omega_{i,j}-(\epsilon_{p}-\epsilon_{p-g_{i,j}})}\\=V_4(g_k,-g_j,-g_{i,j}|i\Omega_k,-i\Omega_j,-i\Omega_{ij})
\end{multline}
so that for the first it is neccessary that $|g_i+g_j|\leq g$ and same for the second, and $g_k$ is uncompensated. \\\\
The fourth integral, in turn: 
\begin{multline}
	U_4(g_i,g_j,g_k)=\int 	\frac{n(\epsilon_{p})}{i\Omega_i-(\epsilon_{p-g_{j,k}}-\epsilon_{p-g_{i,j,k}})}\frac{1}{i\Omega_k-(\epsilon_{p}-\epsilon_{p-g_{k}})}\(\frac{1}{i\Omega_{i,j,k}-(\epsilon_{p}-\epsilon_{p-g_{i,j,k}})}-\frac{1}{i\Omega_{j,k}-(\epsilon_{p}-\epsilon_{p-g_{j,k})}}\)\\=V_4(-g_{k},-g_{i,j,k},-g_{j,k}|-i\Omega_k,-i\Omega_{i,j,k},-i\Omega_{jk}),
\end{multline}
1st term of which would require that $|g_{i,j,k}|\leq g$, while the second necessitates $|g_{j,k}|\leq g$. So that it also compensates extra $g_k$ coming from the third term.
 It is clear that up to the restrictions and bosonic frequency change all 4 terms are equivalent.  Let us now evaluate all $V_4$ terms. \\
\subsection{Estimation of divergences in  $V_4$-term}\label{4-th-order:U4-diagram:divergence}
First, we analyze divergencies of $V_4$ for different combinations of  arguments:  
\begin{equation}
	V_4(g_i,g_{i,j},g_{i,j,k})
	=
	\int \frac{n(\epsilon_p)}{i\Omega_i-(\epsilon_{p+g_i}-\epsilon_p)}
	\frac{1}
	{(i\Omega_{i,j}-(\epsilon_{p+g_{i,j}}-\epsilon_p))
	(i\Omega_{i,j,k}-(\epsilon_{p+g_{i,j,k}}-\epsilon_p))}.
\end{equation}
For low temperatures, the integral is taken over two disjoint regions:
\begin{multline}
	V_4(g_i,g_{i,j},g_{i,j,k})
	=
	\int_0^{p_i} \frac{d^2p}{i\Omega_i-(\epsilon_{p+g_i}-\epsilon_p)}
	\frac{1}
	{(i\Omega_{i,j}-(\epsilon_{p+g_{i,j}}-\epsilon_p))
	(i\Omega_{i,j,k}-(\epsilon_{p+g_{i,j,k}}-\epsilon_p))}
	\\+
		\int_{p_o}^{p_\Lambda} \frac{d^2p}{i\Omega_i-(\epsilon_{p+g_i}-\epsilon_p)}
	\frac{1}
	{(i\Omega_{i,j}-(\epsilon_{p+g_{i,j}}-\epsilon_p))
	(i\Omega_{i,j,k}-(\epsilon_{p+g_{i,j,k}}-\epsilon_p))},
\end{multline}
In such an integral, there can be $\propto \Omega^2$ area where the integrand diverges as $\Omega^{-3}$: for that one would need to require:
\begin{align}\label{4-th-order:U4-diagram:divergence:divergence-condition} 
	\epsilon_p=\epsilon_{p+g_{i,j}},\\
	\epsilon_{p}=\epsilon_{p+g_{i,j,k}},\\
	\epsilon_{p}=\epsilon_{p+g_{i}},
\end{align} 
which is not possible to satisfy without one of $g$-s being zero, and one possible choice is $|g_{ij}|$, while the other, for $C_6$ symmetric case, is $g_{i,j,k}=0$. First consider $|g_{ij}|=0$ alternative. 

Without loss of generality, pick $x$ axis parallel to $g_i$. Based on the roots of the denominator, there could be 3 separate situations:
\begin{enumerate}
	\item $0=\theta_k$, so that $g_k$ is parallel to $g_i$ when $V_4$ is the largest ($V_{4,\parallel}$),
	\item  $0\neq\theta_k$, but there is a region in momentum space of width $\delta p\approx \Omega^{\alpha}$, when $y_{i,l}(p)=y_{k,m}(p)$ ($V_{4,c}$) \footnote{It seems that $\alpha=1$, thus making the difference between $V_{4,c}$ and $V_{4,seo}$ only quantative: two divergences of $\Omega^{-1}$ instead of one.},
	\item  $0\neq\theta_k$ and the two roots of denominators are always separable ($V_{4,sep}$). 
\end{enumerate}
Simple power counting shows that 
\begin{equation}
	V_{4,\parallel}\propto \Omega^{-3}\delta x \delta p\approx \Omega^{-2},
\end{equation}
since areas for the most divergent part grows $\delta x \delta p\propto \Omega$
\begin{equation}
	V_{4,c}\propto \Omega^{-3}\delta x \delta p\approx  \Omega^{-2+\alpha},
\end{equation}
and 
\begin{equation}
	V_{4,sep}\propto \Omega^{-2}\delta x \approx  \Omega^{-1}.
\end{equation}
However, because integrals with uncompensated frequencies come in pairs, there could be mutual cancellation making the divergence slower. In the equilibrium, by minimum energy argument $V_4$ we should  pick the combination of frequencies that gives the coefficient the smallest possible value. 
Since the leading term comes from the choice $g_i=-g_j=g_k$, we choose frequencies to minimize this term.
\begin{equation}
	V_4(g_{i},0,g_i)\propto -\frac{\Omega^2}{i\Omega_i\Omega_{i,j}\Omega_{i,j,k}}v(g_i,0,g_i),
\end{equation}
A similar term from the 2nd integral will read:
\begin{equation}
	V_4(-g_{i},0,-g_i)\propto \frac{\Omega^2}{i\Omega_i \Omega_j\Omega_{j,k}}v(-g_i,0,-g_i),
\end{equation}
for $\Omega\to 0$. If reversal symmetry is not broken, the  $v(-g_i,0,-g_i)=v(g_i,0,g_i)$, and so by choice $\Omega_{ij}\Omega_{i,j,k}=\Omega_j\Omega_{j,k}$ we can cancel the leading divergence. It corresponds to 
\begin{align}\label{4-th-order:U4-diagram:divergence:omegas}
	\Omega_j=-\Omega_i+\delta \Omega'\\
	\Omega_k=\Omega_i-\delta \Omega'+\delta \Omega,
\end{align}
where $\delta \Omega$, $\delta \Omega'\ll \Omega$.  For the third and the fourth terms, we have: 
\begin{equation}
	V_4(g_{i},g_i,0)\propto \frac{\Omega^2}{i\Omega_k\Omega_{j}\Omega_{i,j}}v(g_i,g_i,0),
\end{equation}
\begin{equation}
	V_4(-g_{i},0,-g_i)\propto  -\frac{\Omega^2 v(-g_i,-g,0)}{i\Omega_k\Omega_{ijk}\Omega_{jk}},
\end{equation}
For which choice \eqref{4-th-order:U4-diagram:divergence:omegas} also works.  Let us find, however, combinations that belongs to the second and the third groups. 
Equation $\epsilon(p)=\epsilon$ has two roots as a function of $p$ (absolute value of momentum). So that there will be two combinations of roots that define the second group.\\\\
 The first is given by $(p,g_i)=(p,g_k)$ and hence, for the choice $g_i= g\hat{x} $ 
 \begin{equation}
  \cos(\theta_p)= \cos(\theta_p)\cos(\theta_k)+\sin(\theta_p)\sin(\theta_k),
 \end{equation}
which has, as solutions
\begin{equation}
	\cos(\theta_p)^2=\frac{\sin(\theta_k)^2}{2(1-\cos(\theta_k))}\leq 1
\end{equation}
and the requirment that $\cos(\theta_p)$ has the same sign as $\cos(\theta_k-\theta_p)$. Available values are $\theta_k=\pi n/3$ (with $n=0$ excluded) satisfy the inequality. However, it is clear the $g_k$ should be in the the same half plane as $g_i$, and hence only $\pi/3$ and $5\pi/3$ will satisfy the equality. \\\\
Another solution corresponds to the second root of the dispersion relation $(p+g_i)^2=(p+g_k)^2+\Delta^2_p$, where $\Delta_p^2=2\sqrt{(m/2\lambda)^2+(\mu+\epsilon)/\lambda}$. So that
\begin{equation}
	\cos(\theta_p)=\cos(\theta_p-\theta_k)+\Delta_p^2/(2p g),
\end{equation}
hence for $\theta_k=\pi$ can be easily satisfied on the shell $p=p_o$
\begin{equation}
	\cos(\theta_p)=\Delta^2_p(4p_o g)=\<p_F\>/(2p_o)
\end{equation}
similarly, for $\theta_k=2\pi/3$ we get
\begin{equation}
	\frac{3}{2}\cos(\theta_p)-\frac{\sqrt{3}}{2}\sin(\theta_p)=\<p_F\>/(2p_o),
\end{equation}
the l.h.s. is bounded by $9/4+3/4=3$, which is straightforward to satisfy, since r.h.s is less than 1. It is also clear that once we can satisfy for $2\pi/3$ it is also satisfied for $-2\pi/3$.\\\\
Thus, finally, all 5 vectors belongs to the second group, with 2 corresponding to homo-processes, and 3 -- to hetero. \\\\ 
Finally, it is clear that is impossible to have $\Omega^2$ divergence for $g_{ijk}=0$, since simultaneous satisfaction of \eqref{4-th-order:U4-diagram:divergence:divergence-condition} implies that $\epsilon_{p-g_k}=\epsilon_{p+g_i}$ and since $g_k\neq g_i$, impossible to satisfy for all $p$, rather we have divergence of the second class.

\subsection{Calculation of $V_{4,\parallel}$}
\begin{figure}
\includegraphics[width=0.5\columnwidth]{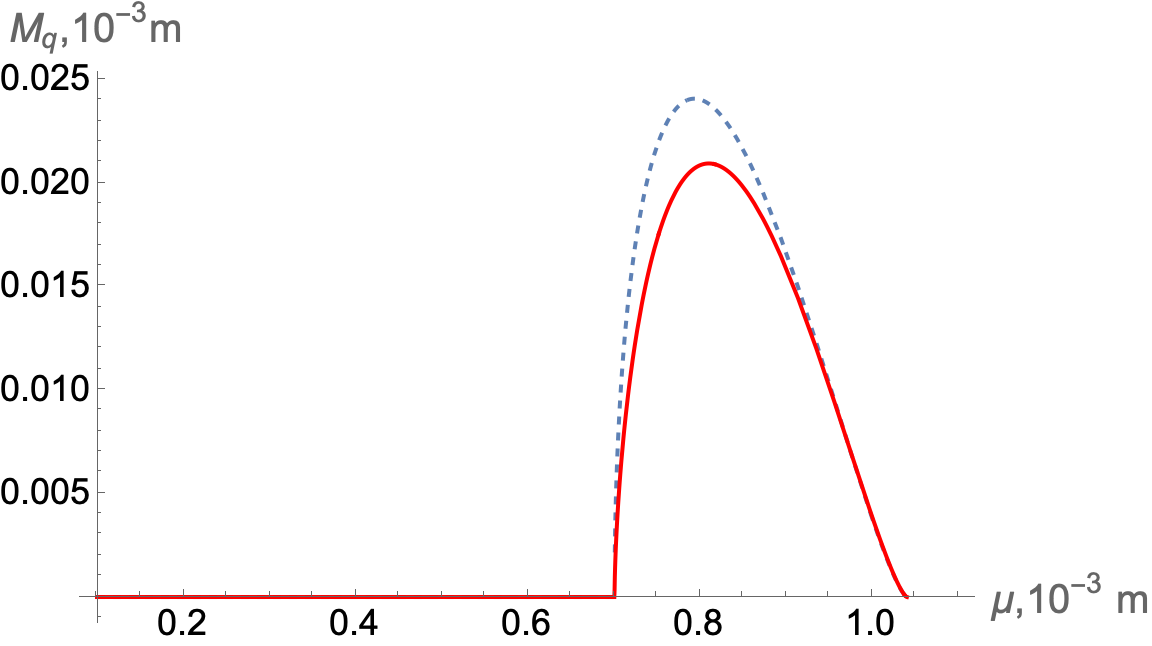}
\caption{Order parameter dependence on the Fermi energy for $\Omega_{IR}=T_L$ (solid line) and $\Omega_{IR}=T_c$ for interaction constant $V=0.45$. Order parameter behavior is nearly the same for both choices.}
\label{fig:4-th-order:U4-diagram:divergence:cutoff-sensitivity}
\end{figure}

Once we established the 4-th order diagram is maximized at $g_i=g_k=-g_j$, we are now to evaluate it. For the choice \eqref{4-th-order:U4-diagram:divergence:omegas} we write the sum of the 4 in the form:
\begin{multline}
	V_{4,\Sigma}=\\-\frac{1}{i\delta \Omega'}\int n(\xi_p)\(\frac{(\xi_{p+g_i}-\xi_p)2i\delta \Omega'+(\delta \Omega'-\Omega)(\Omega+\delta\Omega-\delta \Omega')+\Omega(\Omega+\delta \Omega) }{(i\Omega-(\xi_{p+g_i}-\xi_p))(i(\Omega+\delta\Omega-\delta \Omega')-(\xi_{p+g_i}-\xi_p))(i(\Omega+\delta\Omega)-(\xi_{p+g_i}-\xi_p))(i(-\Omega+\delta\Omega')+(\xi_{p+g_i}-\xi_p))}\)
	\\-\frac{1}{i\delta \Omega}\int n(\xi_p)\(\frac{
	(\xi_{p+g_i}-\xi_p)2i\delta \Omega-(\delta \Omega+\Omega)(\Omega+\delta\Omega-\delta \Omega')+\Omega(\Omega-\delta \Omega') 
	}
	{
	(i\Omega+(\xi_{p+g_i}-\xi_p))(i(\Omega+\delta\Omega-\delta \Omega')+(\xi_{p+g_i}-\xi_p))(i(\Omega+\delta\Omega)+(\xi_{p+g_i}-\xi_p))(-i(-\Omega+\delta\Omega')-(\xi_{p+g_i}-\xi_p))}\)
\end{multline}
Once $O(\delta\Omega^2)$ terms neglected, the 4-th order coefficient can be represented as
\begin{equation}
	V_{4,\Sigma}=2\int\frac{n(\epsilon_p)(\epsilon_{p+g}-\epsilon_p+i\Omega)}{(i\Omega-(\epsilon_{p+g}-\epsilon_p))^4} +2\int\frac{n(\epsilon_p)(\epsilon_{p+g}-\epsilon_p-i\Omega)}{(i\Omega+(\epsilon_{p+g}-\epsilon_p))^4}=-2\frac{\d^2}{\d \Omega^2}\text{Re}\(I\)+\frac{4\Omega }{3}\text{Im}\(\frac{\d ^3}{i\d \Omega^3}I\),  
\end{equation}
where $I$ is the part of the particle-hole response 
\begin{equation}
	I=\int \frac{n(\xi_p)}{i\Omega+\epsilon_{p+g}-\epsilon_p}
\end{equation}
calculated in Section \ref{sec:response}. For the homo-part of the response :
\begin{equation}
	I_{ii}(q<2p_i)=-\frac{\pi}{m_e q}\(\(\frac{i\Omega 2^2}{m_e}-\frac{2^2\Omega^2}{m_e^2q^2}\)^{1/2}-i\(-\frac{i\Omega 2^2}{m_e}+\frac{2^2\Omega^2}{m_e^2q^2}\)^{1/2}\)+f_{ii}\propto -\frac{\pi}{m_e q}(1-i)(a_1'-a_2')+f_{ii},
\end{equation} 
where the last part is weakly dependent on frequency $\Omega$ and $a'=2e^{i\pi/4}(\Omega/m_e+i(\Omega/m_eq)^2)^{1/2}=a'_1+ia'_2$, so that in the leading order $V_{4,hom}=-2^{1/2}\pi(m_e/\Omega)^{1/2}(m_e q)^{-3}$. For the hetero-part, the response is given by \eqref{appendix:response:hetero:0T:inner:finite-Omega}:
\begin{equation}
	I_{io}(q<2p_i)=\frac{1}{2m_e}\frac{\kappa_1 \theta_f}{\kappa_1-\kappa_2}\(\log((\theta_f^2 -\theta_c^2)^2+(\alpha \Omega)^2)\)-\frac{1}{2m_e}\frac{\kappa_1\theta_c'}{\kappa_1-\kappa_2}\(\log((\theta_f-\theta_c)^2+(\alpha\Omega)^2)\)+f_{io},
\end{equation}
where $f_{io}$ is part that depends on $\Omega$ weakly, and $\theta_c'=\sqrt{\theta_c^2-\frac{iv\Omega}{r_i q_c}}$. Then in the leading order  contribution to $V_4$ is logarithmically divergent:
\begin{equation}
	V_{4,het}=
	\frac{1}{2m_e}
	\frac{p_i}{p_i+p_o}
	\frac{(v/r_i q_c)^2}
	{
	(\theta_{io}^2-\frac{iv\Omega}{r_i q_c})^{3/2}
	}\log(\alpha_i\Omega)
	+
	\frac{1}{2m_e}
	\frac{p_o}{p_i+p_o}
	\frac{(v/r_o q_c)^2}
	{
	(
	\theta_{oi}^2-\frac{iv\Omega}{r_o q_c}
	)^{3/2}
	}
	\log(\alpha_o\Omega)
\end{equation} 
Choosing cutoff to be $\Omega\propto T_c$ we should be able to neglect the contribution of the hetero-part whenever
$
	\theta^{3}_{c}m_eq_c^{-2}\gg T_c\log(T_c)^2,
$
which gives a characteristic temperature of order $0.1m_e$ thus by far exceeding actual $T_c\propto m_e 10^{-4}$ in regime of interest. \\\\
Notice again that homo-contribution to the 4-th order scattering amplitude dominates over the hetero-, thus contrasting this case to superconductivity at finite $q$ \cite{larkin1965inhomogeneous}.
As expected for bosonic theories \cite{migdal2018qualitative} 4-th order coefficient diverges at $\Omega\to 0$. We try now to choose a sensible frequency cutoff $\Omega_{IR}$. In the symmetry broken phase, this problem is fixed through opening of the gap $\Delta\propto x_q$ in the excitation spectrum. In a system without a global gap, $\Omega_{IR}\propto T_c$. Because $x_q\propto \Omega_{IR}^{-1/4}$, physical results are not sensitive to the error in the choice of $\Omega_{IR}$ (see Fig. \ref{fig:4-th-order:U4-diagram:divergence:cutoff-sensitivity}). We then pick $\Omega_{IR}=T_L=\sqrt{\frac{12^{1/2}}{\pi}}\sqrt{\(\frac{\theta_{c,i}}{T_{L,i}^2}\)+\(\frac{\theta_{c,o}}{T_{L,o}^2}\)}^{-1}$ for its convenience.

\section{Conductance up to $O(x_q^2)$}\label{sec:2nd-order}
\subsection{Vertex correction}
In this variant, I follow discussion in Mahan \cite{mahan2013many}.  It leads to the absence of $\Omega^0$ term in the imaginary part.
Vertex correction to the response has a form:
\begin{equation}
    \chi^{c}_v(q,i\Omega)=
    x_q(\Omega_1)x_{-q}(-\Omega_1)
     \int v_p v_{p+q}
    G_p(i\omega_n)
    G_{p}(i(\omega_n+\Omega))
    G_{p+q}(i(\omega_n+\Omega+\Omega_1))
    G_{p+q}(i(\omega_n+\Omega_1))
\end{equation}
When the excitonic fields $x_q$ are taken to be classical $(\Omega_1=0)$, the summation over the frequencies can be made into the integration over the three contours. Then 
\begin{multline}
    \chi^{c}_v(q,i\Omega)=
    x_qx_{-q}
    \int_{C_{b1}}\frac{dz}{2\pi i}  v_p v_{p+q}n(z)
    G_p(z)G_{p+q}(z)
    G_{p}(z+i\Omega)
    G_{p+q}(z+i\Omega)
    \\+
     x_qx_{-q}
    \int_{C_{b2}}\frac{dz}{2\pi i}  v_p v_{p+q} n(z)
    G_p(z)G_{p+q}(z)
    G_{p}(z+i\Omega)
    G_{p+q}(z+i\Omega)
\end{multline}
With contour $C_{b1}$ going from $-\infty+i\delta$ to  $\infty+i\delta$, and on the lower part of the cut from  $-\infty-i\delta$ to  $\infty-i\delta$, while $C_{b2}$ going from $-\infty-i\Omega+i\delta$ to  $\infty-i\Omega+i\delta$, and on the lower part of the cut from  $-\infty-i\Omega-i\delta$ to  $\infty-i\Omega-i\delta$. After changing a variable in the second part, I get 
\begin{multline}
    \chi^{c}_v(q,i\Omega)=
    x_qx_{-q}
    \int_{-\infty}^{\infty}\frac{d\epsilon}{2\pi i}  v_p v_{p+q} n(z)
    G_{p}(z+i\Omega)
    G_{p+q}(z+i\Omega)
    \\(
    G_p(z+i\delta)
    G_{p+q}(z+i\delta)
    -
    G_p(z-i\delta)
    G_{p+q}(z-i\delta)
    )
    \\+
     x_qx_{-q}
    \int_{-\infty}^{\infty}\frac{d\epsilon}{2\pi i}
    v_p v_{p+q}  n(z)
    G_p(z-i\Omega)
    G_{p+q}(z-i\Omega)
    \(
    G_{p}(z+i\delta)
    G_{p+q}(z+i\delta)
    -    
    G_{p}(z-i\delta)
    G_{p+q}(z-i\delta)
\),
\end{multline}
which becomes 
\begin{multline*}
    \chi^{c}_v(q,i\Omega)=
    x_qx_{-q}
    \int_{-\infty}^{\infty}\frac{d\epsilon}{2\pi i}  v_p v_{p+q} n(z)
    (G_{p}(z+i\Omega)
    G_{p+q}(z+i\Omega)+G_p(z-i\Omega)
    G_{p+q}(z-i\Omega))
    (
    G_p(z+i\delta)
    G_{p+q}(z+i\delta)
    -
    G_p(z-i\delta)
    G_{p+q}(z-i\delta)
    ).
\end{multline*}
An expression $(
    G_p(z+i\delta)
    G_{p+q}(z+i\delta)
    -
    G_p(z-i\delta)
    G_{p+q}(z-i\delta)
    )$ is analogous to a spectral function $A^{(2)}_{p,p+q}(z)=
    \text{Im}(G_p(z+i\delta)
    G_{p+q}(z+i\delta))=(A_p(z)r_{p+q}(z)+A_{p+q}(z)r_{p}(z))$, where $G_p(z+i\delta)=r_p(z)+iA_p(z)$.
    We now make an analytical continutation $i\Omega\to\Omega+i\delta$ and since it is an imaginary part that contributes to the conductivity \footnote{There must be a part that cancels against the diamagnetic term: absent here}, 
 \begin{multline}
    \pi_v(q,\Omega)=
   2 x_qx_{-q}
    \int_{-\infty}^{\infty}\frac{d\epsilon}{2\pi }  v_p v_{p+q} n(\epsilon)
    (A^{(2)}_{p,p+q}(\epsilon+\Omega)-A^{(2)}_{p,p+q}(\epsilon-\Omega))
 A^{(2)}_{p,p+q}(\epsilon)
 \\=
    2 x_qx_{-q}
    \int_{-\infty}^{\infty}\frac{d\epsilon}{2\pi }  v_p v_{p+q} (n(\epsilon)-n(\epsilon+\Omega))
    A^{(2)}_{p,p+q}(\epsilon+\Omega)
 A^{(2)}_{p,p+q}(\epsilon)
\end{multline}
Stationary response $\Omega\to 0$ is then equal to
 \begin{multline}\label{appendix:vertex:A:response:stationary}
    \pi_v(q,\Omega)=
    2 x_qx_{-q}\frac{\Omega}{2\pi}\int d^2 p
 v_p v_{p+q}
    A^{(2)}_{p,p+q}(\Omega)
 A^{(2)}_{p,p+q}(0)
 =
 \\2 x_qx_{-q}\frac{\Omega}{2\pi} \lim_{\Omega\to 0}\int d^2 p
 v_p v_{p+q}
(A_p(0)r_{p+q}(0)+A_{p+q}(0)r_{p}(0))
(A_p(\Omega)r_{p+q}(\Omega)+A_{p+q}(\Omega)r_{p}(\Omega))
\\= 2 x_qx_{-q}\frac{\Omega}{2\pi} \int d^2 p
 v_p v_{p+q}
(2A_p(0)r_{p+q}(0)A_{p+q}(0)r_{p}(0)
+
A_p^2(0)r_{p+q}^2
+
A^2_{p+q}(0)r_p^2
)
\end{multline}
which should give a positive or negative contribution to the response dependent on whether the homo- (same kind of Fermi surface) or hetero-part is dominant. Unless $A^{(2)}_{p,p+q}(\Omega)$ has a divergent as $\Omega^3$ derivative, there is no $\Omega^0$ term in the imaginary part. The real part has a non-zero coefficient of $\Omega^{-1}$ term. Let us simplify the expression for the vertex correction to the response now. Note that
\begin{equation}\label{appendix:vertex:A:spectral-function}
A^{(2)}_{p,p+q}
=
\frac{\Sigma(\xi_p+\xi_{p+q})}{(\xi_p^2+\Sigma^2)(\xi_{p+q}^2+\Sigma^2)}
=
\bigg(
\frac{
(\xi_p+i\Sigma)(\xi_{p+q}+i\Sigma)-\xi_p\xi_{p+q}+\Sigma^2}
{2i(\xi_p^2+\Sigma^2)(\xi_{p+q}^2+\Sigma^2)}
\bigg)
=
\bigg(
-\frac{
(\xi_p-i\Sigma)(\xi_{p+q}-i\Sigma)-\xi_p\xi_{p+q}+\Sigma^2}
{2i(\xi_p^2+\Sigma^2)(\xi_{p+q}^2+\Sigma^2)}
\bigg)
\end{equation}
which after substitution in \eqref{appendix:vertex:A:response:stationary} leads to
 \begin{equation}
    \pi_v(q,\Omega)=
     x_qx_{-q}\frac{\Omega}{4\pi}\int
 d^2 pv_p v_{p+q}
\bigg(
\frac{1}{(\xi_p^2+\Sigma^2)(\xi_{p+q}^2+\Sigma^2)}
-
\frac{(\Sigma^2-\xi_p\xi_{p+q})^2}{(\xi_p^2+\Sigma^2)^2(\xi_{p+q}^2+\Sigma^2)^2}
\bigg)
 \end{equation}
Note that it gives either positive or negative correction dependent on whether the homo- (same Fermi surface) or the hetero-part dominates. The vertex correction, alternatively, can be expressed in the form:
\begin{equation}\label{appendix:conductance-2:vertex:Im-representation}
	\pi_{2,2}(q,i\Omega)=|x_q|^2\frac{\Omega}{2\pi}\int d^2 p v_p v_{p+q} \Im(G_pG_{p+q})\Im(G_pG_{p+q}),
\end{equation}
which form suggestive of being positive for $v_pv_{p+q}>0$.

\begin{figure}
\includegraphics[width=0.5\columnwidth]{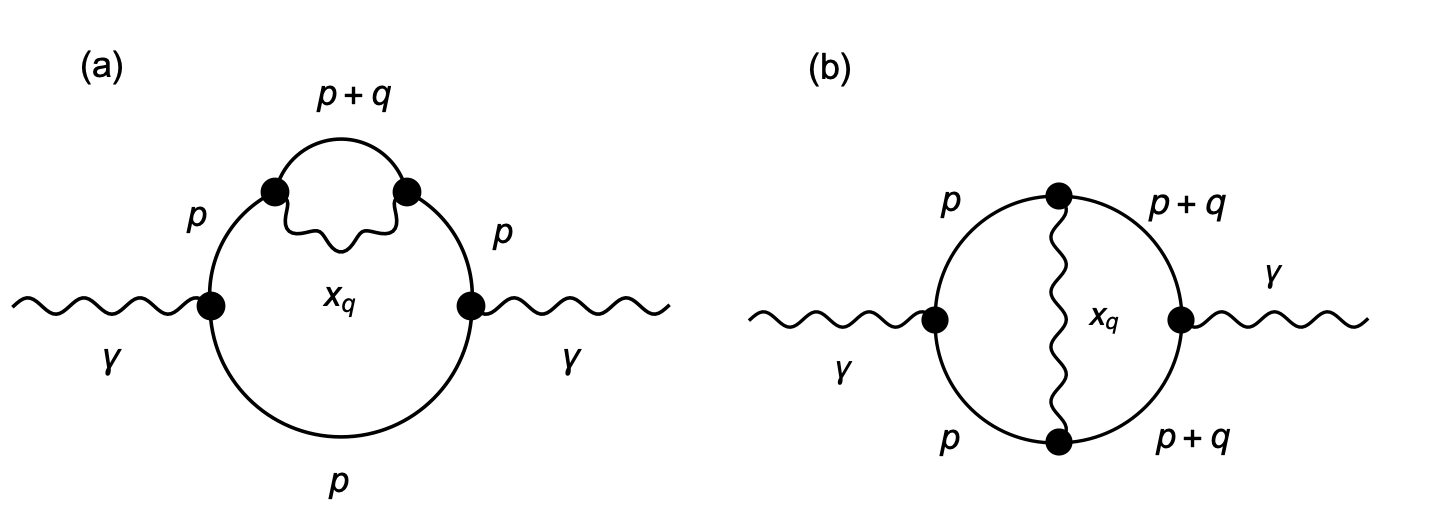}
\caption{Corrections in the leading order in intervalley pseudo-magnetization $x_q$ to the electron part of the conductivity of 2D system with double-well dispersion. Wavy line within the fermion loop denotes propagator $\<x_q x_{-q}\>$. On the symmetry broken side, we take it simply the order parameter squared $x_{q_i}^2$ for $q=q_{i}$ and vanishingly small frequency. Corrections can be divided into a correction to the density of states (a) and a correction to the vertex (b). }
\label{fig:electromagnetic:corrections:vertex}
\end{figure}

\subsection{DOS correction}
The diagram (a) reads
\begin{equation}
    \chi^{c}_{DOS}(q,i\Omega)=
    x_qx_{-q}
     \int v_p v_{p}
    G_p(i\omega_n)
    G_{p+q}(i(\omega_n))
    G_{p}(i(\omega_n))
    G_{p}(i(\omega_n+\Omega)),
\end{equation}
which is equivalent to the expression with integration over z
\begin{multline}
    \chi^{c}_{DOS}(q,i\Omega)=
    x_qx_{-q}
     \bigg(
     \int_{C_{+}}\frac{dz}{2\pi i} v_p v_{p}
    G_p(z)
    G_{p+q}(z)
    G_{p}(z)
    G_{p}(z+i\Omega)
    +
    \int_{C_{+-}} \frac{dz}{2\pi i} v_p v_{p}
    G_p(z)
    G_{p+q}(z)
    G_{p}(z)
    G_{p}(z+i\Omega)
    \\+\int_{C_{-}} \frac{dz}{2\pi i} v_p v_{p}
    G_p(z)
    G_{p+q}(z)
    G_{p}(z)
    G_{p}(z+i\Omega)
    \bigg),
\end{multline}
where the counterclockwise $C_{+/-}$ is the upper (and lower) half-plane bounded by $(-\infty+ i\delta,\infty+ i\delta)$ (and $(\infty-i\delta-i\Omega,-\infty-i\delta-i\Omega)$), while $C_{+-}$ is the contour over the middle rectangualar:

\begin{multline}
    \chi^{c}_{DOS}(q,i\Omega)=
    x_qx_{-q}
     \bigg(
     \int_{-\infty+i\delta}^{\infty+i\delta}
     \frac{d\epsilon}{2\pi i} v_p v_{p}n(z)
    G_p(z)
    G_{p+q}(z)
    G_{p}(z)
    G_{p}(z+i\Omega)
    -
    \int_{-\infty-i\delta}^{\infty-i\delta}
     \frac{d\epsilon}{2\pi i} v_p v_{p} n(z)
    G_p(z)
    G_{p+q}(z)
    G_{p}(z)
    G_{p}(z+i\Omega)
    \\+
    \int_{-\infty-i\Omega}^{\infty-i\Omega}
     \frac{d\epsilon}{2\pi i} v_p v_{p}n(z)
    G_p(z)
    G_{p+q}(z)
    G_{p}(z)
    G_{p}(z+i\Omega)
    -
    \int_{-\infty-i\Omega-i\delta}^{\infty-i\Omega-i\delta}
     \frac{d\epsilon}{2\pi i} v_p v_{p}n(z)
    G_p(z)
    G_{p+q}(z)
    G_{p}(z)
    G_{p}(z+i\Omega)
    \bigg),
\end{multline}
Shifting the contour in the first terms by $\pm i\delta$, in the last terms by $i\Omega\pm\delta$, we obtain: 
\begin{multline}
    \chi^{c}_{DOS}(q,i\Omega)=
    x_qx_{-q}
     \bigg(
     \int_{-\infty}^{\infty}
     \frac{d\epsilon}{2\pi i} v_p v_{p}n(z)
    G_{p}(z+i\Omega)
    (
    G_{p+q}(z+i\delta)G_p(z+i\delta)G_{p}(z+i\delta)-G_{p+q}(z-i\delta)G_p(z-i\delta)G_{p}(z-i\delta))
    \\+
    \int_{-\infty}^{\infty}
     \frac{d\epsilon}{2\pi i} v_p v_{p}n(z)
    G_p(z-i\Omega) G_{p}(z-i\Omega)G_{p+q}(z-i\Omega)
    (
    G_{p}(z+i\delta)
 -
    G_{p}(z-i\delta))
    \bigg),
\end{multline}
In the second term do the analytic continuation, and shift integration variable by $\Omega$
\begin{multline}
    \chi^{c}_{DOS}(q,i\Omega)=
    2x_qx_{-q}
     \bigg(
     \int_{-\infty}^{\infty}
     \frac{d\epsilon}{2\pi} v_p v_{p}n(\epsilon)
    G_{p}(\epsilon+\Omega+i\delta)
\text{Im}\( G_{p+q}(z)G_p(z)G_{p}(z)\)
    \\+
    \int_{-\infty}^{\infty}
     \frac{d\epsilon}{2\pi } v_p v_{p}n(\epsilon+\Omega)
    G_p(z-i\delta) G_{p}(z-i\delta)G_{p+q}(z-i\delta)
\text{Im}\( G_p(z+\Omega)\)
    \bigg),
\end{multline}
Taking the imaginary part of the latter, 
\begin{multline}
    \pi_{DOS}(q,i\Omega)=
    2x_qx_{-q}
     \bigg(
     \int_{-\infty}^{\infty}
     \frac{d\epsilon}{2\pi } v_p v_{p}n(\epsilon)
\text{Im}\( G_{p+q}(z)G_p(z)G_{p}(z)\)
\text{Im}\( G_{p}(z+\Omega)\)
   \\ - 
    \int_{-\infty}^{\infty}
     \frac{d\epsilon}{2\pi } v_p v_{p}n(\epsilon+\Omega)
   \text{Im}\( G_{p+q}(z)G_p(z)G_{p}(z)\)
\text{Im}\( G_{p}(z+\Omega)\)
    \bigg),
\end{multline}
Which becomes in the stationary case $\Omega\to 0$
\begin{equation}\label{appendix:conductance-2:DOS:Im-representation}
    \pi_{DOS}(q,\Omega\to 0)=
    2x_qx_{-q}\frac{\Omega}{2\pi} \int d^2 p
      v_p^2
\text{Im}\( G_{p+q}(0)G_p(0)G_{p}(0)\)
\text{Im}\( G_{p}(\Omega)\)
\end{equation}
whose sign is, generally speaking, undetermined.

\subsection{Explicit expression for $\pi_2(\Omega,q)$}
We also see that the total correction, given by the sum of the DOS- and vertex-corrections, is:
\begin{equation}
    \pi(q,\Omega\to 0)=\pi_{DOS}(q,\Omega)+\pi_{v}(q,\Omega)
\end{equation}
From \eqref{appendix:conductance-2:DOS:Im-representation} it follows that:
\begin{multline}
\sigma_{DOS}(q,\Omega)
=
\frac{2x_q^2 \Sigma^2}{\pi}
\int d^2 p \frac{v_p^2(\xi_p^2-\Sigma^2+2\xi_{p+q}\xi_p)}{(\xi_p^2+\Sigma^2)^3(\xi_{p+q}^2+\Sigma^2)}
=
\frac{2x_q^2 \Omega \Sigma^2}{\pi}
\int d^2 p \frac{v_p^2(2\xi_p^2+2\xi_{p+q}\xi_p-(\xi_p^2+\Sigma^2))}{(\xi_p^2+\Sigma^2)^3(\xi_{p+q}^2+\Sigma^2)}\\=
\frac{2^2x_q^2 \Omega \Sigma^2}{\pi}
\int d^2 p \frac{v_p^2\xi_p(\xi_p+\xi_{p+q})}{(\xi_p^2+\Sigma^2)^3(\xi_{p+q}^2+\Sigma^2)}
-\frac{2x_q^2 \Omega \Sigma^2}{\pi}
\int d^2 p \frac{v_p^2(\xi_p^2+\Sigma^2)}{(\xi_p^2+\Sigma^2)^3(\xi_{p+q}^2+\Sigma^2)}
\end{multline}
while from  \eqref{appendix:conductance-2:vertex:Im-representation}
\begin{equation}
\sigma_{v}(q,\Omega)
=
\frac{x_q^2 \Sigma^2}{\pi}
\int d^2 p \frac{v_p v_{p+q}(\xi_p+\xi_{p+q})^2}{(\xi_p^2+\Sigma^2)^2(\xi_{p+q}^2+\Sigma^2)^2}
\end{equation}
for $\alpha,\gamma=1$. For the $x-x$ response, $v_p=\d \xi/\d p_x=\cos(\theta)\d\xi/\d p$. Integrating by parts, we can rewrite the first part of the DOS-part as:
\begin{equation}
\Sigma^2\int dp_y \int  \frac{dp_x}{(\xi_{p}^2+(\alpha\Sigma)^2)^2} \bigg(
-\frac{v_p(\xi_p+\xi_{p+q})(2\xi_{p+q}v_{p+q})}{(\xi_{p+q}^2+(\gamma\Sigma)^2)^2}+\frac{v_p'(\xi_p+\xi_{p+q})+v_p(v_p+v_{p+q})}{(\xi_{p+q}^2+(\gamma\Sigma)^2)}
\bigg)
\end{equation}
After the resummation of 2 terms we have:
\begin{equation}
	 \sigma(q,\Omega\to 0)
	 =
\frac{\Sigma^2x_q^2}{\pi}\int \frac{d^2p}{(\xi_{p}^2+(\alpha\Sigma)^2)^2} \bigg(\frac{v'_{x,p}(\xi_p+\xi_{p+q})+v_p(-v_p+v_{p+q})}{(\xi_{p+q}^2+(\gamma\Sigma)^2)}\bigg).
	 \end{equation}
The latter can be represented as a derivative over parameter $\alpha$:
\begin{equation}
	 \sigma(q,\Omega\to 0)
	 =
-\frac{x_q ^2 	 }{\pi}\frac{1}{2\alpha}\frac{\d}{\d \alpha}\int \frac{d^2p}{\xi_{p}^2+(\alpha\Sigma)^2} \bigg(\frac{v'_{x,p}(\xi_p+\xi_{p+q})+v_p(-v_p+v_{p+q})}{\xi_{p+q}^2+(\gamma\Sigma)^2}\bigg).
	 \end{equation}

\subsection{Estimate}
For $\Sigma\ll m_e$, most of the contribution comes from the vicinity of the Fermi surface. Hence we can divide the integral into 4 parts: when $p,|p+q|\approx p_i$,  $p,|p+q|\approx p_o$ (homo-part) and $p\approx p_i,|p+q|\approx p_o$,   $p\approx p_o,|p+q|\approx p_i$ (hetero-part). More specifically, the most of the contribution comes from region:
\begin{equation}
	-\frac{\Sigma}{m_e}<p^2-p_{s}^2<\frac{\Sigma}{m_e},
\end{equation}
\begin{equation}
	-\frac{\Sigma}{m_e}<(p+q)^2-p_{s'}^2<\frac{\Sigma}{m_e}.
\end{equation}
Hence the value is acquired in the range
\begin{equation}
	-\frac{\Sigma}{m_e}\frac{1}{ p_s q}+\frac{p_{s'}^2-p_s^2-q^2}{2 p_s q}<\cos(\theta)<\frac{p_{s'}^2-p_s^2-q^2}{2 p_s q}+\frac{\Sigma}{m_e}\frac{1}{ p_s q}.
\end{equation}
\begin{enumerate}[label={\alph*)}]
	\item For the homo-part, $s'=s$, the area over the angle is 
\begin{equation}
	-\frac{\Sigma}{m_e}\frac{1}{ p_s q}-\frac{q}{2 p_s}<\cos(\theta)<-\frac{q}{2 p_s}+\frac{\Sigma}{m_e}\frac{1}{ p_s q}.
\end{equation}
Hence homo-part of the response contributes mostly at smaller hole concentration, when $q_c=p_o-p_i<2p_s$. Define angle $\cos(\theta_s)=-q/2p_s$, then when such angle exist,
\begin{equation}
	-\frac{\Sigma}{m_e}\frac{1}{ p_s q}<-\sin(\theta_s)\delta \theta<\frac{\Sigma}{m_e}\frac{1}{ p_s q}.
\end{equation}
And hence we can estimate the homo-part ($q>2p_f$) of the response to be 
\begin{equation}
	\sigma_{homo}\approx \frac{x_q^2}{2\pi \Sigma^2}(\cot(\theta_{s,i})+\cot(\theta_{s,o})).
\end{equation}
For small $q$ ($q<2p_f$) case, 
\begin{equation}
	\sigma_{homo}\approx \frac{x_q^2}{2\pi Q_{ii}^2 \Sigma m_e}(\cot(\theta_{s,i})+\cot(\theta_{s,o})).
\end{equation}

\item For the hetero-part instead, the angle region is 
\begin{equation}
	-\frac{\Sigma}{m_e}\frac{1}{ p_s q}+\frac{p_{s'}^2-p_s^2-(p_o^2+p_i^2-2p_op_i+2\delta qq_c)}{2 p_s q}<\cos(\theta)<\frac{p_{s'}^2-p_s^2-(p_o^2+p_i^2-2p_op_i+2\delta qq_c)}{2 p_s q}+\frac{\Sigma}{m_e}\frac{1}{ p_s q},
\end{equation}
and hence for the inner-outer ($s'=o$, $s=i$) processes, we have the angle close to $0$ \begin{equation}
	\cos(\theta_{io})=\frac{p_o-p_i-\delta qq_c/p_i}{q_c+\delta q}=
	1-\frac{\delta q}{q_c}-\frac{\delta q}{p_i}
	\end{equation}
	for the outer-inner processes ($s'=i$, $s=o$), instead, angle is close to $\pi$: 
	\begin{equation}
		\cos(\theta_{oi})=\frac{p_op_i-p_o^2-\delta q q_c}{p_o (q_c+\delta q)}=-1-\frac{\delta q}{p_o}+\frac{\delta q}{q_c}
	\end{equation}
	In both cases then, independently on the doping, for any finite $\delta q$,  angles $\theta_{io}$ and $\pi-\theta_{io}$ are small:
	\begin{align}
		\sin(\theta_{io})\approx \sqrt{2\delta q(q_c^{-1}+p_i^{-1})}\\
		\sin(\theta_{oi})\approx \sqrt{2\delta q(q_c^{-1}-p_o^{-1})}
	\end{align}
	
	Hence the "value acquisition" area should be larger than the one of the homo-processes apart from a special point $2p_i=q_c$:
	\begin{align}
		-\frac{\Sigma}{m_e p_i q_c}<\sin(\theta_{io})\delta \theta<\frac{\Sigma}{m_e p_i q_c},\\
		-\frac{\Sigma}{m_e p_o q_c}<\sin(\theta_{oi})\delta \theta<\frac{\Sigma}{m_e p_o q_c}.
			\end{align}
	The corresponding contributions to the conductivity correction is 
	\begin{equation}
			\sigma_{het}\approx -\frac{x_q^2}{\pi q \Sigma^2}\<p_f\>\(\cot(\theta_{c,i})+\cot(\theta_{c,o})\),
\end{equation}
\end{enumerate}
We then expect that the correction to the conductance goes as $\tau_d^2$ and the processes with scattering between different Fermi surfaces (hetero-processes) dominate by factor of $\theta_{c,i}^{-1}\propto(\delta q/q_c)^{-1/2}\approx 10$.

\subsection{Accurate calculation}
To obtain a better approximation for conductance, let us look once again at the total expression:
\begin{equation}
	 \sigma_{xx}(q,\Omega\to 0)
	 =
\frac{\Sigma^2x_q^2}{\pi}\int \frac{d^2p}{(\xi_{p}^2+(\alpha\Sigma)^2)^2} \bigg(\frac{v'_{x,p}(\xi_p+\xi_{p+q})+v_p(-v_p+v_{p+q})}{(\xi_{p+q}^2+(\gamma\Sigma)^2)}\bigg).
\end{equation}
The latter can be represented through the derivative:
\begin{equation}
		 \sigma_{xx}(q,\Omega\to 0)
	 =
-\frac{x_q^2}{2\pi\alpha}\frac{\d }{\d \alpha}\int \frac{d^2p}{\xi_{p}^2+(\alpha\Sigma)^2} \bigg(\frac{v'_{x,p}(\xi_p+\xi_{p+q})+v_p(-v_p+v_{p+q})}{\xi_{p+q}^2+(\gamma\Sigma)^2}\bigg).
\end{equation}
\begin{enumerate}[label=\alph*)]
	\item Hetero-processes
	\\ For particles residing on different Fermi-surfaces, we get, in $\delta p$-approximation:
	\begin{equation}
		\sigma_{io}=-\frac{x_q^2}{\alpha \pi m_e^2}
		\frac{\d}{\d \alpha} \int \frac{d \delta p d\theta}{\delta p^2+(\frac{\alpha \Sigma}{m_e})^2}\frac{\cos(\theta)(\delta p(1-\kappa_{io})-Q)-2(p_i+\frac{\delta p}{2p_i})\cos(\theta)(q+2(p_i+\frac{\delta p}{2p_i})\cos(\theta))}{(\delta p \kappa_{io}+Q)^2+\(\frac{\gamma \Sigma}{m_e}\)^2}
	\end{equation}
	where variable $Q\equiv p_i^2+q^2-p_o^2+2p_i q\cos(\theta)$.
	Integration over $\delta p$ can be performed through the residues, which are situated at $\delta p_{3/4}=\pm i\alpha \Sigma/m_e$,  $\delta p_{1/2}=-Q/\kappa_{io}\pm i\gamma \Sigma/(m_e\kappa_{io})$:
		\begin{multline}
		\sigma_{io}=
		-\frac{x_q^2}{\alpha m_e}\frac{\d }{\d \alpha}\int \frac{d\theta}{\alpha \Sigma}\frac{-Q-2p_iq-2^2p_i^2\cos(\theta)+\frac{i\Sigma \alpha}{m_e}(1-\kappa_{io}-\frac{q}{p_i}-4\cos(\theta))\cos(\theta)}{(Q+\frac{i\Sigma\alpha}{m_e}\kappa_{io})^2+\(\frac{\gamma \Sigma}{m_e}\)^2}
		\\
		-\frac{x_q^2\kappa_{io}}{\alpha m_e}\frac{\d }{\d \alpha}\int \frac{d\theta}{\gamma \Sigma}\frac{-Q-2p_iq-2^2p_i^2\cos(\theta)+\(\frac{i\gamma \Sigma}{m_e\kappa_{io}}-\frac{Q}{\kappa_{io}}\)\(1-\kappa_{io}-\frac{q}{p_i}-4\cos(\theta)\)\cos(\theta)}{(Q-\frac{i\Sigma\alpha}{m_e}\kappa_{io})^2+\(\frac{\gamma \Sigma}{m_e}\)^2}.
	\end{multline}
	 After linearization $Q=-2p_i q\sin(\theta_{c,i})(\theta-\theta_{c,i})$ we can change integration variable to $Q$. We can shift integration contour above to $i\gamma\Sigma/m_e$ in the second term without crossing the poles. In the first term, however, contour shift below to $-i\alpha \kappa_{io}\Sigma/m_e$ leads to crossing the pole at $Q_{1}^{(1)}=-i\Sigma \alpha \kappa_{io}/m_e+i\gamma \Sigma/m_e$. Doing subsequent variable changes in both terms we move integration limits to their initial values. Then the integral has the form:
	\begin{multline}
		\sigma_{io}=-\frac{x_q^2}{2\alpha m_e}\frac{\d }{\d \alpha}\int \frac{dQ'}{\alpha \Sigma p_i q \sin(\theta_{c,i})}\frac{-Q-2p_iq-2^2p_i^2\cos(\theta)+\frac{i\Sigma \alpha}{m_e}(1-\kappa_{io}-\frac{q}{p_i}-4\cos(\theta))\cos(\theta)}{(Q')^2+\(\frac{\gamma \Sigma}{m_e}\)^2}
		\\
		-\frac{x_q^2\kappa_{io}}{2\alpha m_e}\frac{\d }{\d \alpha}\int \frac{dQ''}{\gamma \Sigma p_i q\sin(\theta_{c,i})}\frac{-Q-2p_iq-2^2p_i^2\cos(\theta)+\(\frac{i\gamma \Sigma}{m_e\kappa_{io}}-\frac{Q}{\kappa_{io}}\)\(1-\kappa_{io}-\frac{q}{p_i}-4\cos(\theta)\)\cos(\theta)}{(Q'')^2+\(\frac{\gamma \Sigma}{m_e}\)^2}.
	\end{multline}
	where $Q_1^{(1)}=-i\alpha \Sigma\kappa_{io}/m_e+i\gamma\Sigma/m_e$. Which implies that the correction to the conductivity in the leading order goes as 
	\begin{equation}
		\sigma_{io}=-\frac{x_q^2\pi\<p_f\> \cot\(\theta_{c,i}\)}{q\Sigma^2}+O(x_q^2/(m_e\Sigma)),
	\end{equation}
	and an analogous contribution from outer-inner processes.
			\item Homo-processes
			\\\\For particles residing on the same Fermi surface,
			\begin{equation}
\sigma_{ii}=-\frac{x_q^2}{\alpha \pi m_e^2}
		\frac{\d}{\d \alpha} \int \frac{d \delta p d\theta}{\delta p^2+(\frac{\alpha \Sigma}{m_e})^2}\frac{(\delta p(1+\kappa_{ii})+Q_{ii})+2(p_i+\frac{\delta p}{2p_i})\cos(\theta)q}{(\delta p \kappa_{ii}+Q_{ii})^2+\(\frac{\gamma \Sigma}{m_e}\)^2},
\end{equation}
where $Q_{ii}=2p_i q\cos(\theta)+q^2$. Integration over residues 
\begin{align}
	\delta p_{1,2}=\pm \frac{i\alpha \Sigma}{m_e},\\
	\delta p_{3,4}=-\frac{Q_{ii}}{\kappa_{ii}}\pm\frac{i\gamma \Sigma}{m_e}.
\end{align}
gives:
			\begin{multline}
\sigma_{ii}=-\frac{x_q^2}{2\alpha  m_e \Sigma}
		\frac{\d}{\d \alpha} \int \frac{ d\theta}{\alpha }\frac{(\frac{i\alpha \Sigma}{m_e}(1+\kappa_{ii})+Q_{ii})+2(p_i+\frac{1}{2p_i}\frac{i\alpha \Sigma}{m_e})\cos(\theta)q}{(\frac{i\alpha \Sigma}{m_e} \kappa_{ii}+Q_{ii})^2+\(\frac{\gamma \Sigma}{m_e}\)^2}\\
		-\frac{x_q^2}{2\alpha  m_e \Sigma}
		\frac{\d}{\d \alpha} \int  \frac{ d\theta}{\gamma}  \kappa_{ii}\frac{(\delta p_{3}(1+\kappa_{ii})+Q_{ii})+2(p_i+\frac{1}{2p_i}\delta p_{3})\cos(\theta)q}{(Q_{ii}-\frac{i\gamma \Sigma\kappa_{ii}}{m_e})^2+\(\frac{\alpha \Sigma \kappa_{ii}}{m_e}\)^2}.
\end{multline}
\\ In the regime $q>2p_i$ no pole present near the real axis (up to terms of order $\Sigma/m_e$), thus making the contribution $\propto x_q^2(m_e \Sigma)^{-1}$.  Poles are at:
\begin{equation}
	Q^{(1)}_{\pm}=\frac{i\Sigma}{m_e}\frac{-\alpha(1-\frac{Q_0}{2p_i^2})\pm \gamma}{1+\frac{i\alpha \Sigma}{2m_e p_i^2}},
\end{equation}
and 
\begin{equation}
	Q^{(2)}_{\pm}=\frac{i\Sigma}{m_e}\(1-\frac{Q_0}{2p_i^2}\)\frac{\(\gamma\pm \alpha\)}{1-\frac{i\Sigma}{2m_ep_i^2}(\gamma\pm \alpha)},\end{equation}
	where $Q_0=p_i^2-p_o^2+q^2=-2p_iq_c+2q_c\delta q$. For 
				\begin{multline}
\sigma_{ii}=-\frac{x_q^2}{2\alpha  m_e \Sigma}
		\frac{\d}{\d \alpha} \int \frac{ d\theta}{\alpha }\frac{(\frac{i\alpha \Sigma}{m_e}(1+\kappa_{ii})+Q)+2(p_i+\frac{1}{2p_i}\frac{i\alpha \Sigma}{m_e})\cos(\theta)q}{(1+\frac{i\alpha\Sigma}{2m_ep_i^2})^2(Q-Q_+^{(1)})(Q-Q_-^{(1)})}\\
		-\frac{x_q^2}{2\alpha  m_e \Sigma}
		\frac{\d}{\d \alpha} \int  \frac{ d\theta}{\gamma}  \kappa_{ii}\frac{(\delta p_{3}(1+\kappa_{ii})+Q_{ii})+2(p_i+\frac{1}{2p_i}\delta p_{3})\cos(\theta)q}{(1-\frac{i(\gamma-\alpha)\Sigma}{2m_ep_i^2})(1-\frac{i(\gamma+\alpha)\Sigma}{2m_ep_i^2})(Q-Q_+^{(2)})(Q-Q_-^{(2)})}.
\end{multline}
In the leading in $\Sigma$ approximation then we get from the pole integration:
				\begin{equation}
\sigma_{ii}\approx
		\frac{\pi x_q^2}{2\Sigma^2}
		\cot(\theta_{ii}) (1+\frac{q}{p_i}\cos(\theta_{ii})),
\end{equation}
which is smaller than the hetero-contribution since $\cos(\theta_{ii})=q/(2p_i)<0.5$ in the regime of interest (See Fig. \ref{fig:electromagnetic:corrections:vertex:kappa-io}) by a factor of 5-10. 
	\end{enumerate}

\section{Conductance up to $O(x_q^n)$}
In this section, we obtain an expression for conductance up to an infinite order in $x_q$ through series summation.  

	\subsection{4-th order}\label{appendix:conductivity:higher:4-th}
	In the 4-th order, there will be 3 types of diagrams. An analog of the vertex correction is:
	\begin{equation}\label{appendix:conductivity:higher:4:gen-exp-2}
		\pi_{2,4}(q,\Omega\to 0)
		=|x_q|^4\int v_p v_{p+q}
		G_p (i\omega_n)G_{p+q}(i\omega_n)
		G_{p+q}(i(\omega_n+\Omega))G_p(i(\omega_n+\Omega)) G_{p+q}(i(\omega_n+\Omega))
		G_p (i(\omega_n+\Omega))
	\end{equation}
	and shows up when only one $x$-s is one of the sides. There are 2 such diagrams. Then there is an analog of the DOS-correction (with all $x$-lines on one side) that reads: 
		\begin{equation}\label{appendix:conductivity:higher:4:gen-exp-1}
		\pi_{1,4}(q,\Omega\to 0)
		=|x_q|^4\int v_p^2
		G_p (i\omega_n)
		G_{p}(i(\omega_n+\Omega))
		G_{p+q}(i(\omega_n+\Omega))
		G_p(i(\omega_n+\Omega))
		G_{p+q}(i(\omega_n+\Omega))
		G_p (i(\omega_n+\Omega))
	\end{equation}
	and an additional diagram
	\begin{equation}\label{appendix:conductivity:higher:4:gen-exp-1}
		\pi_{3,4}(q,\Omega\to 0)
		=|x_q|^4\int v_p^2
		G_{p}(i\omega_n)
		G_{p+q}(i\omega_n)
		G_{p+q}(i(\omega_n+\Omega))
		G_p (i(\omega_n+\Omega))
		G_{p+q}(i(\omega_n+\Omega))
		G_p(i(\omega_n+\Omega)),
	\end{equation}
	that has a two legs on each side, both come with a factor of 3. An explicit expression for the vertex correction is:
		\begin{equation}
		\pi_{2,4}(q,\Omega\to 0)
		=|x_q|^4\sum\int v_p v_{p+q}
		G_p(i\omega_n) G_{p+q}(i\omega_n)
		G_{p+q}(i\omega_n+i\Omega)G_p (i\omega_n+i\Omega) G_{p+q}(i\omega_n+i\Omega)
		G_p (i\omega_n+i\Omega).
	\end{equation}
After doing the same manipulations as for the 2-nd order correction, we get for the vertex correction: 
		\begin{equation}
		\pi_{2,4}(q,\Omega\to 0)
		=|x_q|^4\frac{\Omega}{2\pi}\sum\int v_p v_{p+q}
		\Im((G_p G_{p+q})^2)\Im(G_p G_{p+q}).
	\end{equation}
	The DOS-correction analogously can be written as: 
		\begin{equation}
		\pi_{1,4}(q,\Omega\to 0)
		=|x_q|^4
		\int v_p^2
		G_p 
		G_{p}
		G_{p+q}
		G_p
		G_{p+q}
		G_p =
		\frac{|x_q|^4 \Omega}{2\pi}
		\int v_p^2
		\Im(G_{p,>})
		\Im(G_{p,>}
		G_{p+q,>}
		G_{p,>}
		G_{p+q,>}
		G_{p,>})
	\end{equation}
	and, similarly,
			\begin{equation}
		\pi_{3,4}(q,\Omega\to 0)
		=
		\frac{|x_q|^4\Omega}{2\pi}
		\int v_p^2
		\Im(G_{p,>}G_{p+q,>}G_{p,>})
		\Im(
		G_{p,>}
		G_{p+q,>}
		G_{p,>}
		),
	\end{equation}
	where for shortness $G_{p,>}$ without the frequency argument denotes  $G_{p,>}(i\omega_n\to 0)$.
	
	\subsection{6-th order}\label{appendix:conductivity:higher:6-th}
	
	In the 6-th order, there will be 4 separate diagrams. Among them, there will be 2 vertex-like corrections:
	\begin{multline}\label{appendix:conductivity:higher:generating:2:6}
		\pi_{2,6}(q,\Omega\to 0)
		=|x_q|^6
		\int v_p v_{p+q}
		G_p(i\omega_n) 
		G_{p+q}(i\omega_n) 
		G_{p+q}(i\omega_n+i\Omega) 
		G_{p}(i\omega_n+i\Omega)
		\\G_{p+q}(i\omega_n+i\Omega)
		G_{p}(i\omega_n+i\Omega) 
		G_{p+q}(i\omega_n+i\Omega)
		G_{p}(i\omega_n+i\Omega) 
		=
		\frac{|x_q|^6\Omega}{2\pi}
		\int v_p v_{p+q}
		\Im(G_p 
		G_{p+q} )
		\Im(
		(G_{p+q}
		G_{p})^3)
	\end{multline}
and 
	\begin{multline}
		\pi_{4,6}(q,\Omega\to 0)
		=|x_q|^6
		\int v_p v_{p+q}
		G_p(i\omega_n) 
		G_{p+q}(i\omega_n) 
		G_{p}(i\omega_n) 
		G_{p+q}(i\omega_n)
		\\G_{p+q}(i\omega_n+i\Omega)
		G_{p}(i\omega_n+i\Omega) 
		G_{p+q}(i\omega_n+i\Omega)
		G_{p}(i\omega_n+i\Omega) 
		\\=
		\frac{|x_q|^6\Omega}{2\pi}
		\int v_p v_{p+q}
		\Im((G_p 
		G_{p+q})^2)
		\Im((G_{p+q}
		G_{p})^2)
		),
	\end{multline}
	while DOS correction is: 
	\begin{multline}
		\pi_{1,6}(q,\Omega\to 0)
		=|x_q|^6
		\int v_p^2
		G_p(i\omega_n) 
		G_{p}(i\omega_n+i\Omega) 
		G_{p+q}(i\omega_n+i\Omega) 
		G_{p}(i\omega_n+i\Omega)
		\\G_{p+q}(i\omega_n+i\Omega)
		G_{p}(i\omega_n+i\Omega) 
		G_{p+q}(i\omega_n+i\Omega)
		G_{p}(i\omega_n+i\Omega)
		\\=
		\frac{|x_q|^6\Omega}{2\pi}
		\int v_p^2
		\Im(G_{p,>} )
		\Im((G_{p,>} 
		G_{p+q,>} )^3
		G_{p,>})
		),
	\end{multline}
and
	\begin{multline}
		\pi_{3,6}(q,\Omega\to 0)
		=|x_q|^6
		\int v_p^2
		G_p(i\omega_n) 
		G_{p+q}(i\omega_n)
		G_{p}(i\omega_n) 
		G_{p}(i\omega_n+i\Omega)
		\\G_{p+q}(i\omega_n+i\Omega)
		G_{p}(i\omega_n+i\Omega) 
		G_{p+q}(i\omega_n+i\Omega)
		G_{p}(i\omega_n+i\Omega)
		\\=
		\frac{|x_q|^6\Omega}{2\pi}
		\int v_p^2
		\Im(G_{p,>}G_{p+q,>}
		G_{p,>})
		\Im(
		G_{p,>}
		G_{p+q,>}
		G_{p,>}
		G_{p+q,>}
		G_{p,>}	
		)
		).
	\end{multline}

	\subsection{Combinatorics}\label{appendix:conductivity:higher:combinatorics}
All terms originate from the expansion of the effective action of the form
\begin{equation}
	S_{coup}=\text{Tr}(\log(1+G^{-1}(A+x_q)))=...+\frac{1}{2k}\text{Tr}((G^{-1}(A+x_q))^{2k})+...
\end{equation}
Beside the linear coefficient coming from the Taylor expansion of the logarithm, each type of the correction (e.g. \eqref{appendix:conductivity:higher:generating:2:6}) in a given order comes with combinatorial coefficient.Term that corresponds to the order $2k$ has $x_q$ in power of $2n$ and $A$ in power of 2.  In the second order, resulting terms are
\begin{multline}\label{appendix:conductivity:higher:combinatorics:perturbation}
	\text{Tr}((G^{-1}(A+x_q))^2)_{e.m}=\text{Tr}((Ax_q)^2)+\text{Tr}((x_qA)^2)+\text{Tr}(Ax_q^2 A)+\text{Tr}(A^2 x_q^2)+\text{Tr}(x_q^2A^2)+\text{Tr}(x_q A^2 x_q)\\=2\text{Tr}((Ax_q)^2)+4\text{Tr}(A^2 x_q^2),
\end{multline}
where for convenience $G^{-1}$ are implicit on the r.h.s.. Presence of the trace merges terms distinguished by the cyclic permutations. Another words, we can define an equivalence relation between 2 permutations $a_{i}\sim_1 a_j$ through a cyclic permutation. Moreover, permutations inside the elements of the same kind ($x$ or $A$) are equivalent, which defines another equivalence relation. The quantity invariant under the cyclic permutation is the minimal distance between the vector potentials. The last sum then represents the sum of the elements of the quotient group of a permutation group $P_{2n+2}$ of $2n+2$ elements with respect to the cyclic permutations $(P_C)$ as well as permutations inside the class of elements ($P_{2n}$, $P_{2}$). 
\\
In order $2n$, there are $n+1$ elements of the quotient group . Let us pick a single element $v_i$ of the group $P_{2n+2}$ that belongs to the equivalence class $i$. All elements of the equivalence class make an orbit of an element $v_i$ w/r to $C$.\\\\
Then this class will have $2n+2-r$ elements, where $r$ is a number of solutions of the equations of the type:
\begin{equation}\label{appendix:conductivity:higher:combinatorics:degeneracy}
	C^k v_i=v_i,
\end{equation}
where $C^k$ is the cyclic permutation of the order $k$. The cyclic permutation can be represented as a matrix $2n+2\times 2n+2$ with ones at the upper sub-diagonal. 
\begin{equation}
	C=\begin{pmatrix}
	 0& 1 & ...\\
	 0 & 0 & 1 & ...\\
	 ...\\
	 1 & 0 & 0 & ...
	\end{pmatrix}
\end{equation}
Let us represent each term in \eqref{appendix:conductivity:higher:combinatorics:perturbation} by vector $u_i$: every 1 will denote $A$ and $0$ denotes $x_q$.  Then equation \eqref{appendix:conductivity:higher:combinatorics:degeneracy} will be equivalent to $C^k v_i u_i=v_i u_i=u_j$: $r$ then is equal to the number of eigenvectors of $C^{k=1,...}$ that belong to the set of vectors with $\sum_j u_j =2$. Any vector should return to itself after doing $2n+2$ cyclic permutations, hence all eigenvalues are of the form $\lambda_{j=1...n+1}= \pm e^{i 2 \pi j/(n+1)}$. For $k=1$, there is a single eigenvector of value 1,  the one filled with $1$-s. The only chance to have a vector with 1 ones invariant under some power of $C$ is to have ones at a distance of $n+1$, such that $C^{n+1}v_{inv}=v_{inv}$. Obviously, there will be n+1 such states.  \\Hence there will be a single equivalence class with $r\neq 0$ with $n+1$ elements.\\
We have then $n$ classes with $2n+2$ elements and one with $n+1$. \\\\ 
For even $n$, there are $n/2$ classes with odd distance ("vertex"-correction) and $n/2+1$ with even distance ("DOS"-) between current vertices \footnote{Can be proven by induction, since upon addition of new $x_q^2$, two new distances will be added, one would correspond to odd number of $x_q$ between current vertices, and another to odd.}. Hence vertex correction will come with combinatorial coefficient
\begin{align}
	N_{DOS,even}=(2n+2)n/2+n+1=(n+1)^2,\\
	N_{vertex,even}=(2n+2)n/2=n(n+1).
\end{align}
For "DOS" correction, except for the symmetric term, all other terms can be divided into two (distance $k$ and $2n-k-1$), each then coming with coefficient $n+1$. Or, taking into account $1/2(k+1)$ of logarithm, we get $1/2$ for each term in the sum.\\
Similarly for n odd, there are $(n-1)/2+1$ classes with ("vertex"-) correction, and $(n+1)/2$ with "DOS"-correction. 
\begin{align}
	N_{DOS,odd}=(n+1)^2,\\
	N_{vertex,odd}=n(n+1).
\end{align}
By the same argument, there is a coefficient of $1/2$ in front of each term of "vertex"-correction.
\\\\Overall, there must be $(n+1)^2$- "DOS"-like elements, and $n(n+1)$  "vertex"-like terms. \\\\

\subsection{Total correction to conductance}\label{appendix:conductivity:higher}
It is possible to write an equation for the response function, for vertex-like (vertices at different momentum) and DOS-like (vertices at the same momentum) terms separately. Namely, we can do series resummation for two parts separately without solving Bethe-Solpeter. Since part of the correction comes from the change in the spectrum of the quasiparticles, it is instructive to start with finding corrected Green function through the Dyson equation. As before, we implicitly take $q$ equal to $q_c+\delta q$ by the absolute value. 

\begin{figure}
\includegraphics[width=1\columnwidth]{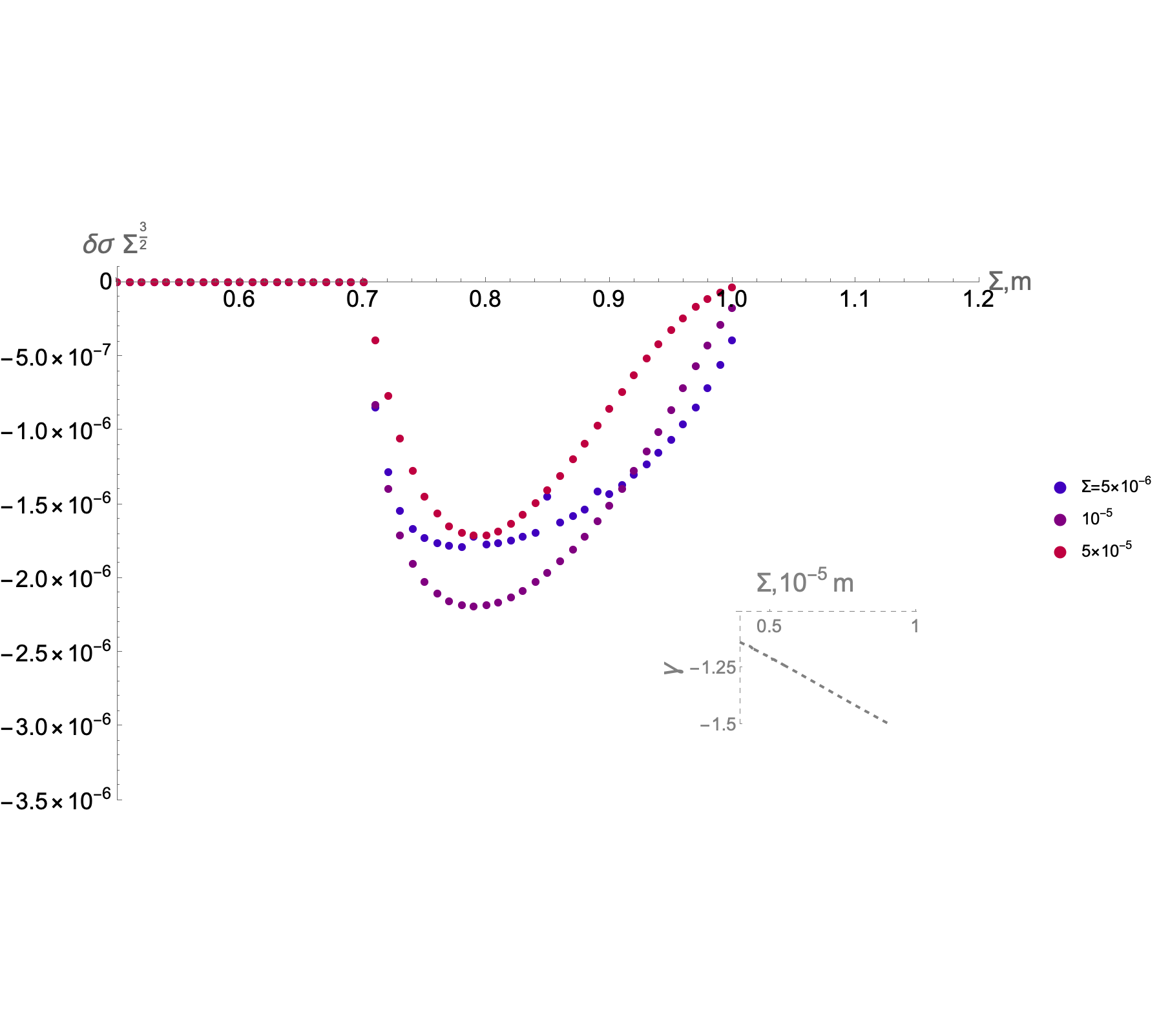}
\caption{Scaled correction (each point is mmltiplied by $\Sigma^{3/2}=\tau_D^{-3/2}$) to conductance as a function of the chemical potential for 3 different self-energies in the experimentally relevant range. Inset: effective power $\zeta\equiv\Sigma\d \log(\delta \sigma)/\d \Sigma $. Note that $Q_{c}<Q_{max}$ region ($0.7 m \lessapprox\mu\lessapprox 0.75 m$) is dominated by $\delta\sigma _{even,4}$ terms, so that countercurrent part that dominates $\delta\sigma _{even,2}$, $\delta\sigma _{odd}$  is effectively absent. }
\label{fig:appendix:conductivity:higher:total-conductance}
\end{figure}

\subsubsection{Self-energy correction}\label{appendix:conductivity:higher:dyson}
In the calculation of higher-order correction to the conductance, we neglected terms such that $q$ brought by the excitonic fields is different, since those, analogously to the 4-th order correction, are negligibly small. Equivalently, we consider correction to the conductance being composed out of $n$ equal terms, where $n$ is the number of reciprocal lattice vectors of the order parameter $x_q$.
Spectrum of the electrons that contribute to the change in the conductance can be modeled by the solution of the Dyson equation with a single $q$ vector. It has a form of a $2\times 2$ matrix equation: 
\begin{align}
	G(p,p')=G_0(p)+G_0(p)\Sigma_q G(p+q,p')\\
	G(p+q,p')=G_0(p+q)+G_0(p+q)\Sigma_{-q} G(p,p')
\end{align}
 whose solution is 
 \begin{equation}
 	G(p,p')=(1-G_0(p)\Sigma_qG_{0}(p+q)\Sigma_{-q})^{-1}\(G_0(p)+G_0(p)\Sigma_pG_0(p+q)\),
 \end{equation}  
 where $\Sigma_{q}$ is a matrix in $k$ and valley spaces. 
For valley symmetric system $[G_0(p),\Sigma_q]=0$ and hence 
 \begin{equation}
 	G(p,p')=((G_{0}(p+q)G_0(p))^{-1}-\Sigma_q^2)^{-1}\(G^{-1}_{0}(p+q)+\Sigma_q\).
 \end{equation}  
 Using explicit form of the intervalley self-energy and Green's function, I obtain: 
 \begin{equation}
 	G(p,p')=
 	\frac{\(G^{-1}_{0}(p+q)+x_q\)}{(G_0(p+q)G_0(p))^{-1}-x_q^2}.
 \end{equation} 
 
Then the correction to the Green's function is equal to
 \begin{equation}\label{appendix:conductivity:higher:dyson:deltaG11}
 	\delta G_{11}(p,p')\equiv G_{11}(p,p')-G_{0}(p)=\frac{x_q^2G^{-1}_{0}(p+q)}{(G_0(p+q)G_0(p))^{-1}-x_q^2},
 \end{equation} 
  \begin{equation}
 	\delta G_{12}(p,p')=\frac{x_q}{(G_0(p+q)G_0(p))^{-1}-x_q^2}.
 \end{equation} 
Thus the quasiparticle spectrum is given by
 \begin{equation}
 	\lambda_{\pm}(p,q)=\frac{\xi_p+\xi_{p+q}}{2}\pm
 	\sqrt{\frac{(\xi_p-\xi_{p+q})^2}{2^2}+x_q^2}\equiv\<\epsilon_p\>\pm \Delta(p)
 \end{equation} 
 where dispersion of the original particles $\xi_p$ is allowed to have a finite imaginary part. 
As before, we divide contributions from hetero-processes (when $G_0(p+q)$ and $G_0(p)$ are on different Fermi surfaces) and homo-processes (same Fermi surface). 
\begin{figure}
\includegraphics[width=1\columnwidth]{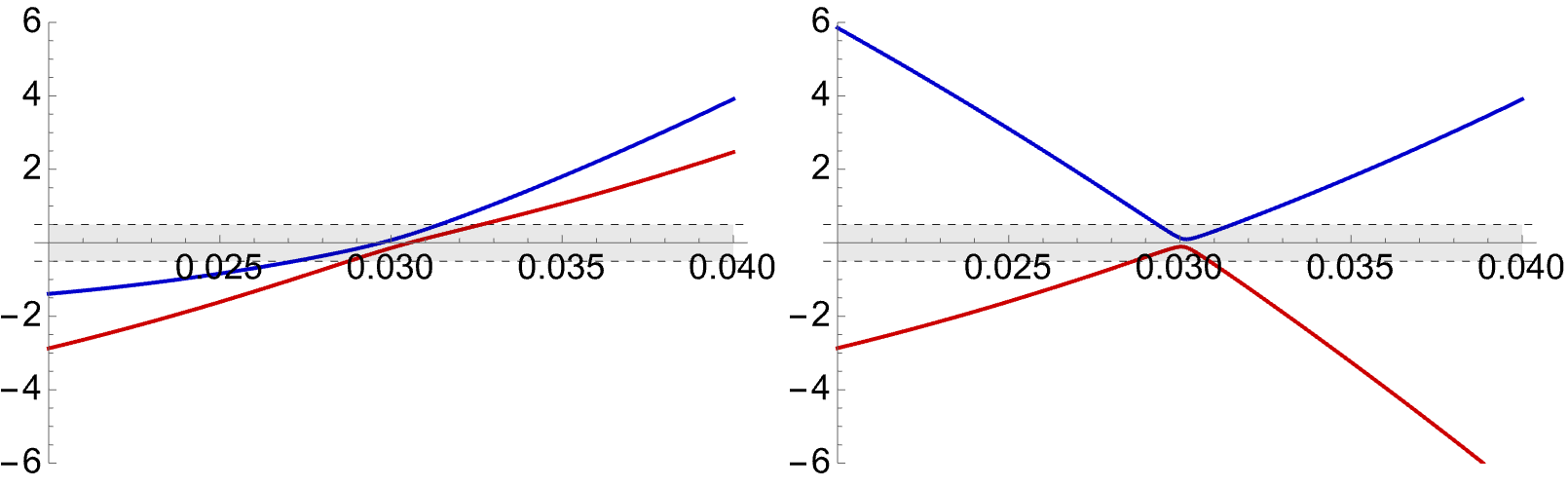}
\caption{Band structure of quasiparticles  originating from the inner-inner processes ($\theta=\cos^{-1}(-q/2p_i)$) and inner-outer processes ($\theta\approx 0$). $\mu=7.1\times 10^{-4} m$ here. We see here that the contribution from the inner-outer processes to the conductance is, in fact, of insulator for large gap $x_q\gg \Sigma$, while for smaller gap $x_q\approx \Sigma$ it should be analogous to the semi-metal. Gray zone close to the Fermi energy $(\epsilon=0)$ is to denote smearing induced by the disorder self-energy ($\Sigma=5\times10^{-5} m$ here).}
\label{fig:electromagnetic:corrections:dyson:quasiparticle-bands}
\end{figure}

\paragraph{Hetero-processes}

For quasiparticles originating from mixing different Fermi-surfaces, there is a gap opening around nesting sites.
Indeed, let us find a equation of the Fermi surface around these. Using $\xi_p=m_e(p^2-p_i^2)$, $\xi_{p+q}=-m_e(p^2-p_i^2+2(p-p_i)q\cos(\theta)+Q)$, which from $\<\epsilon_p\>=\Delta(p)$\footnote{The latter apparently expresses an energy conservation between electron-hole pairs and an exciton formation.} gets us
\begin{equation}\label{appendix:conductivity:higher:dyson:hetero:Fermi-condition}
m_e^2(Q+2\tilde p q\cos(\theta))^2=\((2x_q)^2+m_e^2\(2\tilde p(p+p_i)+2\tilde p q\cos(\theta)+Q\)^2\),
\end{equation}
which after resolving quadratic equation:
\begin{equation}
	p-p_i=\tilde p\approx-\frac{Q}{4p_i(1+\frac{q}{p_i}\cos(\theta))}\pm \sqrt{\(\frac{Q}{4p_i(1+\frac{q}{p_i}\cos(\theta))}\)^2-\frac{x_q^2}{4m_e^2p_i^2(1+\frac{q}{p_i}\cos(\theta))}},
\end{equation}
 defines a closed curve in $(p,\theta)$-space where a constant energy surface intersects 0. 

 \begin{figure}
\includegraphics[width=1\columnwidth]{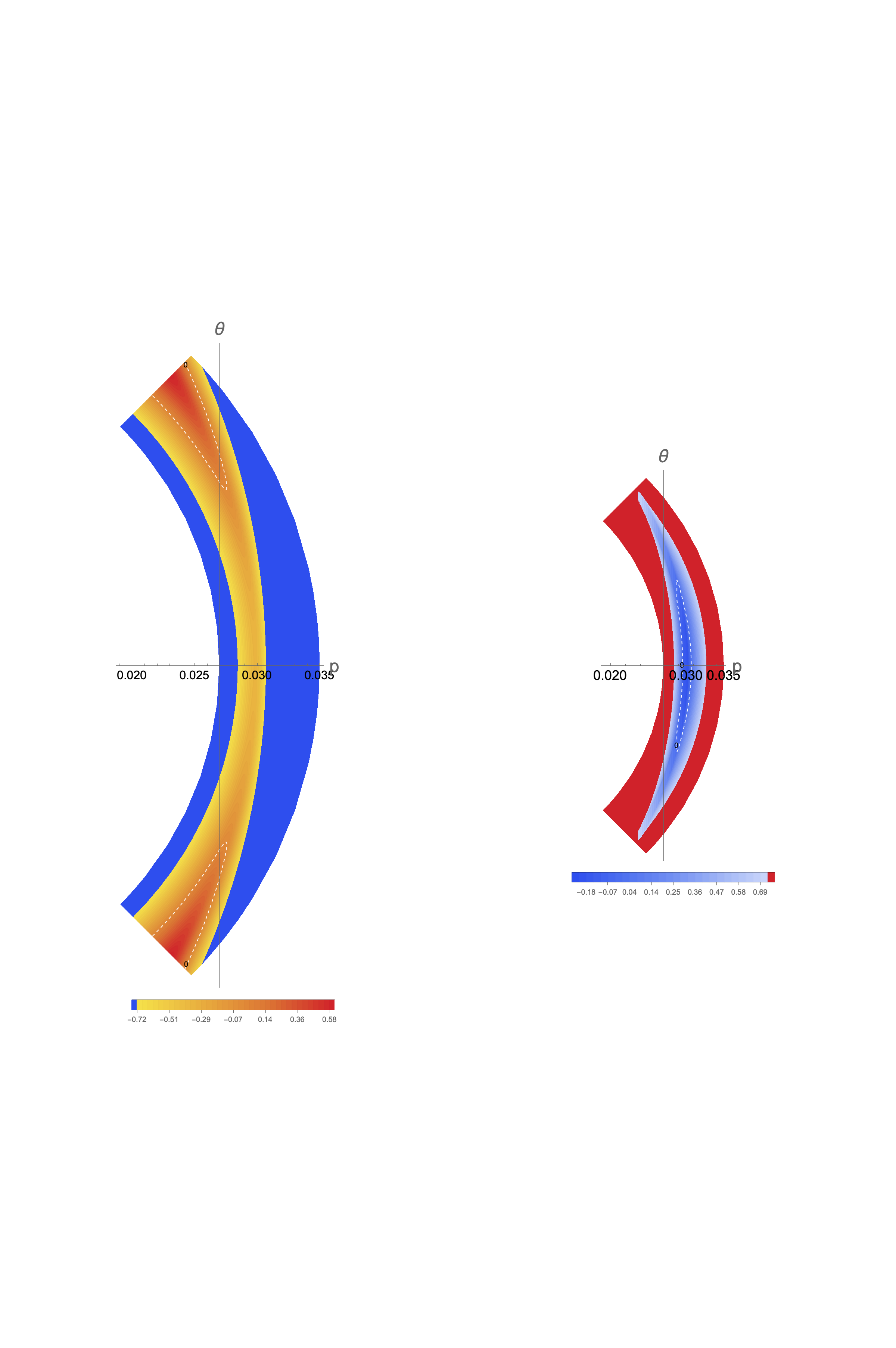}
\caption{Fermi arcs in the valence (left) and the conduction (right) bands for $\mu=7.23\times 10^{-4}m$.}
\label{fig:electromagnetic:corrections:dyson:Fermi-arcs}
\end{figure}

 Clearly, it has solutions only for some $\theta$, namely whenever
 \begin{equation}
 	\frac{Q}{2\sqrt{\kappa_{io}}}>\frac{x_q}{m_e}.
 \end{equation}
 Because $Q\leq Q(0)$, a regime with Fermi arcs exists only for:
  \begin{equation}
 	\frac{p_i q\theta_{c,i}^2}{2}\geqq\frac{x_q}{m_e},
 \end{equation}
 so that the hole has enough transversal energy to absorb an exciton during a one-dimensional jump from $\theta=0 \to \theta_{c,i}$. Since for small $\theta_{c,i}$, Fermi arcs will be concentrated in small region of $k$-space, for $\Sigma\approx m_e Q$ the quasiparticles will behave essentially as one-dimensional.
  \\ Additionally, note that with respect to the momentum perpendicular to the Fermi surface $p_\perp\propto p_i\theta$ the dispersion is linear with the density of states
 \begin{equation}\label{appendix:conductivity:higher:dyson:hetero:nu-theta}
 	\nu_{het}=\nu_\theta\propto \frac{1}{m_e}\frac{1}{2p_iq\sin(\theta_{c,i})},
 \end{equation}
 and therefore has large (by a factor of $\propto \theta_{c,i}^{-1}\approx 10$) contribution to transport processes in comparison to the homo-processes.  
 
\begin{figure}
\includegraphics[width=0.5\columnwidth]{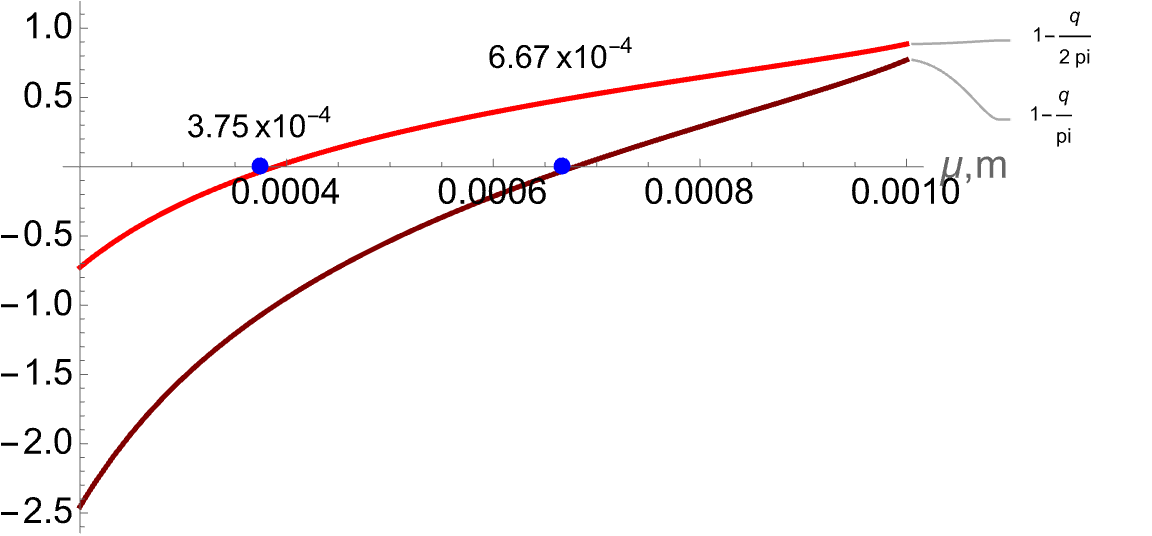}
\caption{Parameters $\kappa_{io,min}=1-(q/p_i)$ as well as $(\kappa_{io,max}+1)/2$ as function of density. Densities at which these parameters can have zeros are irrelevant for the present study. }
\label{fig:electromagnetic:corrections:vertex:kappa-io}
\end{figure}
 
 For large values of 
  $x_q$, contribution from the hetero-processes is gapped for all $\theta$. 

\begin{figure}
\includegraphics[width=0.5\columnwidth]{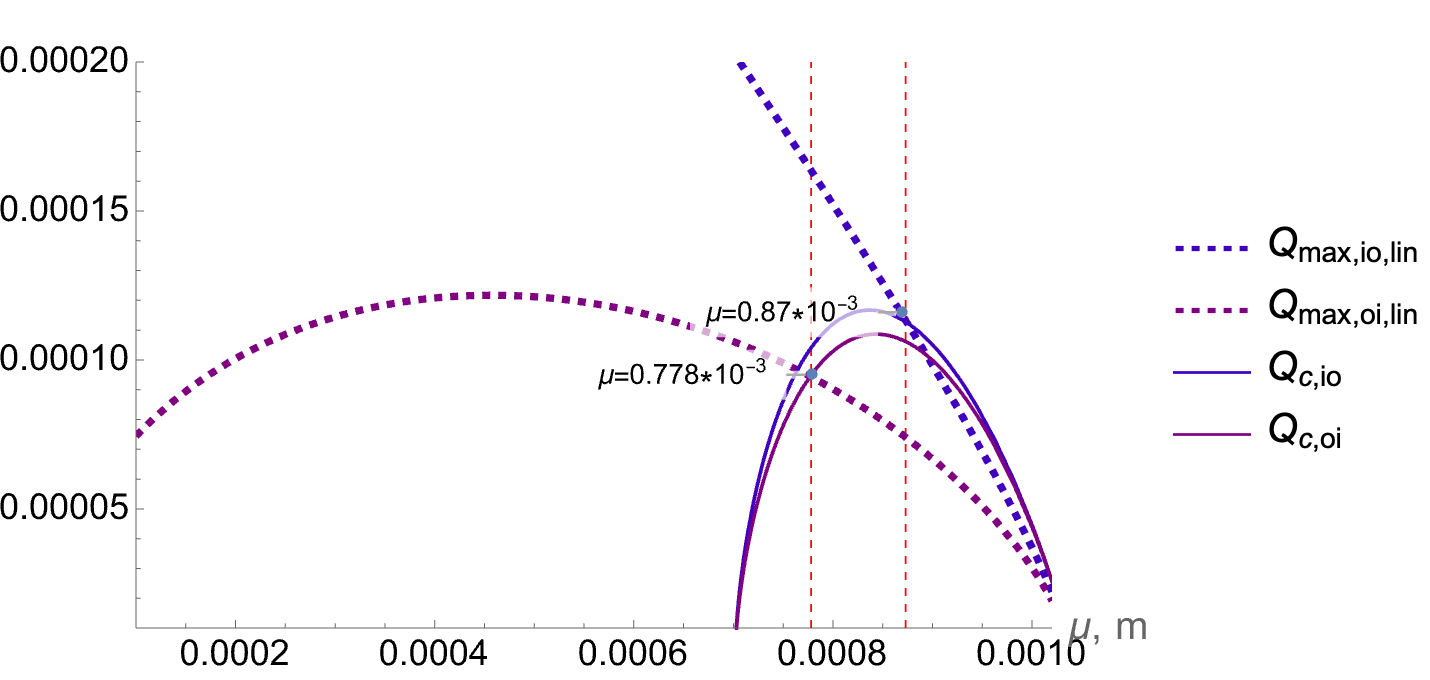}
\caption{Points where change of regime happens: for Fermi energy large than  $\mu_c$ ($\mu_c=0.778\times 10^{-3}$ for outer-inner processes, and $\mu_c=0.87\times 10^{-3}$ for inner-outer processes) Fermi arcs disappear.} 
\label{fig:electromagnetic:corrections:vertex:regimesMs}
\end{figure}

\paragraph{Homo-processes}
For the same quasiparticles originating from alike Fermi-surfaces, the quasiparticles are metallic, since they always intersect the zero-energy level. Using $\xi_p=m_e(p^2-p_i^2)$, $\xi_{p+q}=m_e(p^2-p_i^2+2(p-p_i)q\cos(\theta)+Q_{ii})$, where $Q_{ii}=2p_i q\cos(\theta)+q^2$ so that for $\<\epsilon_p\>=\Delta(p)$ we get:
\begin{equation}
	\frac{m_e^2}{2^2}\(2p^2-2p_i^2+2(p-p_i)q\cos(\theta)+Q_{ii}\)^2=\(\frac{m_e^2}{2^2}(Q_{ii}+2(p-p_i)q\cos(\theta))^2+x_q^2\),
\end{equation}
which is different from \label{appendix:conductivity:higher:dyson:hetero:Fermi-condition} by the sign change $x_q^2\to - x_q^2$, hence there always be two closed Fermi surface given by
\begin{equation}
	p-p_i=\tilde p\approx-\frac{Q_{ii}}{4p_i(1+\frac{q}{p_i}\cos(\theta))}\pm \sqrt{\(\frac{Q_{ii}}{4p_i(1+\frac{q}{p_i}\cos(\theta))}\)^2+\frac{x_q^2}{4m_e^2p_i^2(1+\frac{q}{p_i}\cos(\theta))}}
\end{equation}
which close to nesting $Q_{ii}\approx 0$ simply results into  $p-p_i\approx \pm x_q/(2m_ep_i\kappa_{io}^{1/2})\ll p_i$, so that we conclude that the mass of the quasiparticles composed out of particles residing on the same surfaces remains nearly the same
\begin{equation}
	\nu_{hom}\propto \frac{1}{m_e}.
\end{equation}

\subsubsection{DOS-correction}
Unlike the vertex-correction, DOS-correction has only odd powers of Green function and forms the expansion of the form:
\begin{multline}\label{appendix:conductivity:higher:total:DOS:conductance:general}
	\sigma_{even}=\frac{|x_q|^2}{\pi}\int v_p^2 \Im(G_{p,>})\Im(G_{p,>}G_{p+q,>}G_{p,>})
	\\
	+\frac{|x_q|^4}{\pi}\int v_p^2 \Im(G_{p,>})\Im((G_{p,>}G_{p+q,>})^2G_{p,>})+\frac{|x_q|^4}{2\pi}\int v_p^2 \Im(G_{p,>}G_{p,>}G_{p+q,>})\Im(G_{p,>}G_{p+q,>}G_{p,>})+
	\\
	\frac{|x_q|^6}{\pi}\int v_p^2 \Im(G_{p,>})\Im((G_{p,>}G_{p+q,>})^3G_{p,>})
	+
	\frac{|x_q|^6}{\pi}\int v_p^2 \Im(G_{p,>}G_{p+q,>}G_{p,>})\Im((G_{p,>}G_{p+q,>})^2G_{p,>})
	\\=
	\frac{|x_q|^2}{2\pi}\int v_p^2 \text{Im}(G_p) \text{Im}(G_pG_{p+q}G_p(1-x_q^2(G_pG_{p+q}))^{-1})
	\\
	+
	\frac{|x_q|^2}{2\pi}\int v_p^2 \text{Im}(G_p G_{p+q}G_p) \text{Im}(G_p(1-x_q^2(G_pG_{p+q}))^{-1})
	\\
	+
	\frac{|x_q|^4}{2\pi}\int v_p^2 \text{Im}((G_p G_{p+q})^2G_p) \text{Im}(G_p(1-x_q^2(G_pG_{p+q}))^{-1})+...
	\\=
	\frac{|x_q|^2}{2\pi}\int v_p^2 \text{Im}(G_p(1-x_q^2(G_pG_{p+q}))^{-1}) \text{Im}(G_pG_{p+q}G_p(1-x_q^2(G_pG_{p+q}))^{-1})+\frac{|x_q|^2}{2\pi}\int v_p^2 \text{Im}(G_p) \text{Im}(G_pG_{p+q}G_p(1-x_q^2(G_pG_{p+q}))^{-1})
	\\=\frac{|x_q|^2}{2\pi}\int v_p^2 \(\text{Im}(G_p(1-x_q^2(G_pG_{p+q}))^{-1}) +\text{Im}(G_p)\)\text{Im}(G_pG_{p+q}G_p(1-x_q^2(G_pG_{p+q}))^{-1})
	\\=\frac{|x_q|^2}{2\pi}\int v_p^2 \(\text{Im}((2G_p-G_pG_{p+q}G_px_q^2)(1-x_q^2(G_pG_{p+q}))^{-1}) \)\text{Im}(G_pG_{p+q}G_p(1-x_q^2(G_pG_{p+q}))^{-1})
	\\=
	\frac{|x_q|^2}{\pi}\int v_p^2 \text{Im}(G_p(1-x_q^2(G_pG_{p+q}))^{-1}) \text{Im}(G_pG_{p+q}G_p(1-x_q^2(G_pG_{p+q}))^{-1})
	\\-
	\frac{|x_q|^4}{2\pi}\int v_p^2 \text{Im}(G_pG_{p+q}G_p(1-x_q^2(G_pG_{p+q}))^{-1}) \text{Im}(G_pG_{p+q}G_p(1-x_q^2(G_pG_{p+q}))^{-1})
	\\=\frac{|x_q|^2}{\pi}\int v_p^2 \text{Im}(G_{p+q}^{-1}((G_pG_{p+q})^{-1}-x_q^2)^{-1}) \text{Im}(G_p((G_pG_{p+q})^{-1}-x_q^2)^{-1})
	\\-
	\frac{|x_q|^4}{2\pi}\int v_p^2 \text{Im}(G_p((G_pG_{p+q})^{-1}-x_q^2)^{-1}) \text{Im}(G_p((G_pG_{p+q})^{-1}-x_q^2)^{-1})
\end{multline}
 It is also clear that it has a meaning of the conductance calculated with $\delta G_{1,1}$ \eqref{appendix:conductivity:higher:dyson:deltaG11}  and a corrected single vertex(twice), plus additional term with both vertex corrected and bare Green's functions. Let me call the former $\sigma_{DOS}$ and the latter $\sigma_{v,2}$. We now calculate $\sigma_{DOS}$: \\

Explicit form is 
\begin{equation}
	 	\sigma_{DOS}=-\frac{|x_q|^2\Sigma^2}{\pi}
	 	\int v_p^2
	 	 \(\frac{(\xi_{p+q}\xi_{p}-|x_q|^2-\Sigma^2)-(\xi_{p+q}+\xi_p)\xi_{p+q}}{(\xi_{p+q}\xi_{p}-|x_q|^2-\Sigma^2)^2+\Sigma^2(\xi_{p+q}+\xi_p)^2}\) 
	 	 \(\frac{
	 	 (\xi_{p+q}\xi_{p}-|x_q|^2-\Sigma^2)+(\xi_{p+q}+\xi_p)\xi_p
	 	 }
	 	 {(\xi_{p+q}\xi_{p}-|x_q|^2-\Sigma^2)^2+\Sigma^2(\xi_{p+q}+\xi_p)^2}\frac{1}{\xi_p^2+\Sigma^2}\)
\end{equation}	
or, introducing parameters $\alpha$ and $\gamma$
\begin{multline}
	 	\sigma_{DOS}=-\frac{|x_q|^2\Sigma^2}{\pi}
	 	\int v_p^2
	 	 \(\frac{
	 	 (\xi_{p+q}\xi_{p}-|x_q|^2-\Sigma^2)^2-(\xi_{p+q}+\xi_p)^2\xi_p\xi_{p+q}
	 	 -(\xi_{p+q}\xi_{p}-|x_q|^2-\Sigma^2)(\xi_{p+q}^2-\xi_p^2)}
	 	 {(\gamma^2(\xi_{p+q}\xi_{p}-|x_q|^2-\Sigma^2)^2+\alpha^2\Sigma^2(\xi_{p+q}+\xi_p)^2)^2}\frac{1}{\xi_p^2+\Sigma^2}\)\\=
	 	 \frac{\d }{\gamma \d \gamma}\frac{|x_q|^2\Sigma^2}{2\pi\gamma^2}
	 	\int v_p^2
	 	 \frac{
	 	 1}
	 	 {(\xi_{p+q}\xi_{p}-|x_q|^2-\Sigma^2)^2+\frac{\alpha^2}{\gamma^2}\Sigma^2(\xi_{p+q}+\xi_p)^2}\frac{1}{\xi_p^2+\Sigma^2}
	 	 \\
	 	 -
	 	 \frac{\d }{\alpha \d \alpha}\frac{|x_q|^2}{2\pi \gamma^2}
	 	\int v_p^2
	 	 \frac{
	 	 \xi_p\xi_{p+q}}
	 	 {(\xi_{p+q}\xi_{p}-|x_q|^2-\Sigma^2)^2+\frac{\alpha^2}{\gamma^2}\Sigma^2(\xi_{p+q}+\xi_p)^2}\frac{1}{\xi_p^2+\Sigma^2}
	 	 \\
	 	 +\frac{|x_q|^2}{\pi}\frac{\Sigma^2}{2x_q}\frac{\d }{\d x_q}\frac{1}{2\gamma^2}
	 	\int v_p^2
	 	 \frac{
	 	 (\xi_{p+q}^2-\xi_p^2)\gamma^2}
	 	 {(\xi_{p+q}\xi_{p}-|x_q|^2-\Sigma^2)^2+\frac{\alpha^2}{\gamma^2}\Sigma^2(\xi_{p+q}+\xi_p)^2}\frac{1}{\xi_p^2+\Sigma^2}
\end{multline}	
We then define three generating functions $g_{DOS,i}(\bar{p}=(\gamma,x_q,\alpha))$ such that:
\begin{equation}
	\sigma_{DOS}=(x_q\Sigma)^2\frac{\d }{2p_i\d p_i}g_{DOS,i}(\bar{p}), 
\end{equation}
or
\begin{equation}
	g_{DOS,i}(\bar{p})=\frac{1}{\pi \gamma^2}\int \frac{
	 	 v_p^2  f_i(p,\theta)}
	 	 {(\xi_{p+q}\xi_{p}-|x_q|^2-\Sigma^2)^2+\frac{\alpha^2}{\gamma^2}\Sigma^2(\xi_{p+q}+\xi_p)^2}\frac{1}{\xi_p^2+\Sigma^2}
\end{equation}
where $f_1=1$,   $f_2=(\xi_{p+q}^2-\xi_p^2)/2$,  $f_3=-\xi_p\xi_{p+q}/\Sigma^2$. 

	\paragraph{Hetero-contribution}
	 We start with the i-o process since these are the processes that by a factor of $\sin(\theta_c)^{-1}$ larger. Clearly, the poles are the same as before (See \ref{appendix:conductivity:higher:vertex}) except for $\alpha\ \to \alpha / \gamma$ and  
	 \begin{equation}
	g_{DOS,i}=\frac{1}{\pi \gamma^2}
	 	\int v_p^2 \frac{f_i(\delta p)}{\delta p^2+\frac{\Sigma^2}{m_e^2}}
	 	\frac{1}{(-(\kappa_{io}\delta p+Q)\delta p-\frac{|x_q|^2}{m_e^2}-\frac{\Sigma^2}{m_e^2})^2+\frac{\alpha^2\Sigma^2}{\gamma^2m_e^2}(\delta p-\kappa_{io}(\delta p +Q))^2}
	 \end{equation}
	 the integrand has additional poles at $\delta p_{3/4}=\pm i\Sigma/m_e$, as well as poles common to the vertex correction:
	 \begin{equation}\label{appendix:conductivity:higher:total:DOS:hetero:generating:residues:exact}
	\delta p_{+,1/2}=-\frac{Q+\frac{i \Sigma\alpha q}{\gamma m_e p_i}\cos(\theta)}{2\kappa_{io}}\pm\sqrt{\frac{1}{(2\kappa_{io})^2}\(Q+\frac{i \Sigma\alpha q}{m_e p_i\gamma  }\cos(\theta)\)^2-\frac{\Sigma^2}{m_e^2\kappa_{io}}
	-\frac{|x_q|^2}{m_e^2 \kappa_{io}}-\frac{i \alpha  \Sigma Q}{m_e\gamma \kappa_{io} }}
\end{equation}
Because $x_q/m_e\ll p_i q $, with latter being a scale of $Q$, the integrand is still peaked close to $\theta_{c,i}$, hence we neglect angle dependence of $\kappa_{io}$: $\cos(\theta)\approx \cos(\theta_{c,i})$ (the regime with high density $\kappa_{io}=0$ can be a topic of separate study, but it is irrelevant for experimentally accessed regime, see Fig. \ref{fig:electromagnetic:corrections:vertex:kappa-io}). We still can expand $Q$ in vicinity of it. \\\\
Then there will be an anoother critical angle $\theta_{crit}$, at which sign of the phase changes:
\begin{equation}
	Q_{c,io}=\frac{2x_q\kappa_{io}^{1/2}}{m_e},
\end{equation}
expansion of $Q_{io}$ in the vicinity of the (first) critical angle gets us $Q_{io}(\theta)=-2p_i q \sin(\theta_{c,i})(\theta-\theta_{c,i})$. Assuming $\Sigma\ll x_{q}$, there will be 2 distinct regimes $Q_{io}(0)<Q_{c,io}$, $Q_{c,io}>Q_{io}$. 

\begin{figure}
\includegraphics[width=0.5\columnwidth]{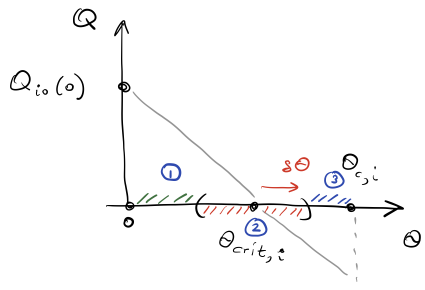}
\caption{Different regions in the small $x_q$ case: $Q_{io}(0)<Q_{c,io}$. }
\label{fig:electromagnetic:corrections:vertex:Q-regimes}
\end{figure}

Then the integral over $\delta p$:
	 \begin{multline}\label{appendix:conductivity:higher:total:DOS:hetero:generating:residues}
	g_{DOS,i}=\frac{1}{\gamma^2 m_e^6}
	 	\int  \frac{v_p^2f_i(\frac{i\Sigma}{m_e})}{\frac{\Sigma}{m_e}}
	 	\frac{1}{(-(\kappa_{io}\frac{i\Sigma}{m_e}+Q)\frac{i\Sigma}{m_e}-\frac{|x_q|^2}{m_e^2}-\frac{\Sigma^2}{m_e^2})^2+\frac{\alpha^2\Sigma^2}{\gamma^2m_e^2}(\frac{i\Sigma}{m_e}-\kappa_{io}(\frac{i\Sigma}{m_e} +Q))^2}
	 	\\+
	 	\frac{2i}{\gamma^2\kappa_{io}^2m_e^6}
	 	\int \frac{v_p^2}{\delta p_{+,2}-\delta p_{-,1}}
	 	 \bigg(\frac{f_i(\delta p_{+,2})}{\delta p_{+,2}^2+\frac{\Sigma^2}{m_e^2}}
	 	\frac{1}{(\delta p_{+,2}-\delta p_{+,1})(\delta p_{+,2}-\delta p_{-,2})}
	 	-
 \frac{f_i(\delta p_{-,1})}{\delta p_{-,1}^2+\frac{\Sigma^2}{m_e^2}}
	 	 	\frac{1}{(\delta p_{-,1}-\delta p_{+,1})(\delta p_{-,1}-\delta p_{-,2})}\bigg)
	 \end{multline}
	 For convenience, define two auxiliary functions $\<f_i\>$ and $\delta f_i$ such that $f_i(\delta p_{+,2})=\<f_i\>+\delta f_i$ and $f_i(\delta p_{-,1})=\<f_i\>-\delta f_i$, then the whole generating function can be written as a sum of 3 qualitatively different terms: 
	 
	 	 \begin{multline*}\label{appendix:conductivity:higher:total:DOS:hetero:generating:residues}
	g_{DOS,i}=\frac{1}{\pi\gamma^2m_e^6}
	 	\int  \frac{v_p^2f_i(\frac{i\Sigma}{m_e})}{\frac{2\Sigma}{m_e}}
	 	\frac{1}{(-(\kappa_{io}\frac{i\Sigma}{m_e}+Q)\frac{i\Sigma}{m_e}-\frac{|x_q|^2}{m_e^2}-\frac{\Sigma^2}{m_e^2})^2+\frac{\alpha^2\Sigma^2}{\gamma^2 m_e^2}(\frac{i\Sigma}{m_e}-\kappa_{io}(\frac{i\Sigma}{m_e} +Q))^2}
	 	\\+
	 	\frac{i}{\pi\gamma^2\kappa_{io}^2m_e^6}
	 	\int \frac{v_p^2 \<f_i\>}{\delta p_{+,2}-\delta p_{-,1}}
	 	 \bigg(\frac{1}{\delta p_{+,2}^2+\frac{\Sigma^2}{m_e^2}}
	 	\frac{1}{(\delta p_{+,2}-\delta p_{+,1})(\delta p_{+,2}-\delta p_{-,2})}
	 	-
 \frac{1}{\delta p_{-,1}^2+\frac{\Sigma^2}{m_e^2}}
	 	 	\frac{1}{(\delta p_{-,1}-\delta p_{+,1})(\delta p_{-,1}-\delta p_{-,2})}\bigg)
	 	 	\\
	 	 	+
	 	\frac{i}{\pi\gamma^2\kappa_{io}^2m_e^6}
	 	\int \frac{v_p^2 \delta f_i}{\delta p_{+,2}-\delta p_{-,1}}
	 	 \bigg(\frac{1}{\delta p_{+,2}^2+\frac{\Sigma^2}{m_e^2}}
	 	\frac{1}{(\delta p_{+,2}-\delta p_{+,1})(\delta p_{+,2}-\delta p_{-,2})}
	 	+
 \frac{1}{\delta p_{-,1}^2+\frac{\Sigma^2}{m_e^2}}
	 	 	\frac{1}{(\delta p_{-,1}-\delta p_{+,1})(\delta p_{-,1}-\delta p_{-,2})}\bigg),
	 \end{multline*}
	 where the most significant difference from the vertex part is the presence of $\delta f_i\propto O(\Sigma/m_e)$ part. For completeness, write new functions in terms of poles:
\begin{align}
	\<f_1\>=1\\
	\delta f_1=0\\
		\<f_2\>=\frac{m_e^2}{2}\((\kappa_{io}^2-1)(\delta p_{+,2}^2+\delta p_{-,1}^2)+2Q\kappa_{io}(\delta p_{+,2}+\delta p_{-,1})+Q^2\)\\
	\delta f_2=\frac{m_e^2}{2}\((\kappa_{io}^2-1)(\delta p_{+,2}+\delta p_{-,1})+2Q\kappa_{io}\)(\delta p_{+,2}-\delta p_{-,1})\\
	\<f_3\>=\frac{m_e^2}{2\Sigma^2}\(\kappa_{io}\(\delta p_{+,2}^2+\delta p_{-,1}^2\)+Q(\delta p_{+,2}+\delta p_{-,1})\)\\
		\delta f_3=\frac{m_e^2}{2\Sigma^2}\(\kappa_{io}\(\delta p_{+,2}+\delta p_{-,1}\)+Q\)(\delta p_{+,2}-\delta p_{-,1})
\end{align}
We now take each of the 5 integrals over angle in three regimes, distinct by the behavior of the poles. To define the boundaries, let us write the expression for the poles in the form that clearly separates phase and the absolute value:
	 \begin{multline}\label{appendix:conductivity:higher:total:DOS:hetero:generating:residues:exact:absolute-phase-form}
	\delta p_{+,1/2}=-\frac{Q+\frac{i \Sigma\alpha q}{m_e p_i \gamma}\cos(\theta)}{2\kappa_{io}}\\\pm
	\(\frac{1}{(2\kappa_{io})^2}\(Q^2-Q_{c,io}^2-\(\frac{\Sigma }{m_e}\frac{q}{p_i}\frac{\alpha}{\gamma}\cos(\theta)\)^2-\(\frac{2\Sigma}{m_e}\)^2\kappa_{io}\)^2
+
\(\frac{\Sigma}{m_e}\frac{\alpha}{\gamma}\frac{Q}{\kappa_{io}}\)^2\(1+\frac{q}{2p_i}\cos(\theta)\)^2
\)^{1/4}e^{i\phi(\theta)/2},\end{multline}
where phase $\phi$:
\begin{equation}
	\phi(\theta)=\tan^{-1}\(\frac{-\frac{2\Sigma}{m_e}\frac{\alpha}{\gamma}Q\(1+\frac{q}{2p_i}\cos(\theta)\)}{Q^2-Q_{c,io}^2-\(\frac{\Sigma }{m_e}\frac{q}{p_i}\frac{\alpha}{\gamma}\cos(\theta)\)^2-\(\frac{2\Sigma}{m_e}\)^2\kappa_{io}}\).
\end{equation}
\begin{figure}
\includegraphics[width=1\columnwidth]{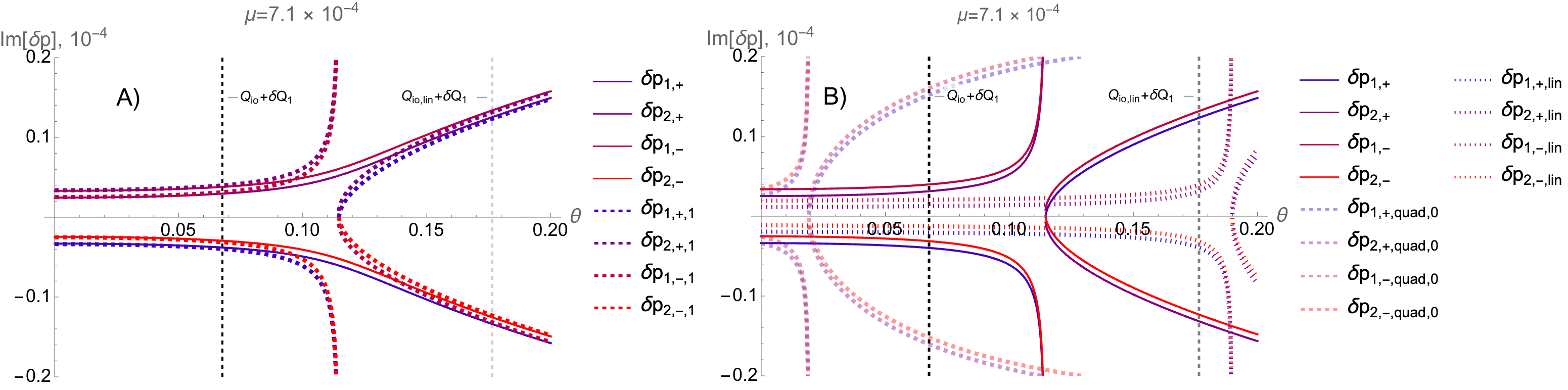}
\caption{Region 1: comparison of the exact poles \eqref{appendix:conductivity:higher:total:DOS:hetero:generating:residues:exact} as a function of $\theta$ with approximate expression \eqref{appendix:conductivity:higher:total:DOS:hetero:generating:residues:regime-1} in the region 1 with linear expansion of Q (A, dotted), quadratic expansion of Q (A, dashed) in the vicinity of $\theta=\theta_{c,i}$ and expansion near $\theta=0$ (B dotted, semi-transparent). We use linear approximation in the following calculation. Vertical line to denote the end of the region 1.}
\label{fig:electromagnetic:corrections:DOS:poles:dp-1-comparison}
\end{figure}
As discussed before, based on the phase behavior, we distinguish 3 different regimes: positive real part, negligible imaginary part, region close to the $Q_{c,io}$, negative real part, negligible imaginary part. Given the disorder is not too strong ($Q_{c,io}\geq \Sigma/m_e$) we set boundaries to:
\begin{equation}\label{appendix:conductivity:higher:total:DOS:hetero:generating:residues:boundaries:1}
	Q^2-Q^2_{c,io}\geq \frac{2\Sigma Q}{m_e}\approx \frac{2\Sigma Q_{c,io}}{m_e}
\end{equation}
for the first region,
\begin{equation}\label{appendix:conductivity:higher:total:DOS:hetero:generating:residues:boundaries:2}
	-\frac{2\Sigma Q_c}{m_e}<Q^2-Q^2_{c,io}<\frac{2\Sigma Q_{c,io}}{m_e}
\end{equation}
for the second, and
\begin{equation}\label{appendix:conductivity:higher:total:DOS:hetero:generating:residues:boundaries:3}
	Q^2-Q^2_{c,io}<-\frac{2\Sigma Q_{c,io}}{m_e}
\end{equation}
for the third. 
\begin{enumerate}[label={\arabic*)}]
	\item  First, for $x_q/m_e<Q(0)/(2\kappa_{io})$ poles exhibit again all 3 possible behaviors (see  \ref{appendix:conductivity:higher:vertex}) \begin{enumerate}
\item As such, in the first region the poles are 
\begin{equation}\label{appendix:conductivity:higher:total:DOS:hetero:generating:residues:regime-1}
	\delta p_{+,1/2}=-\frac{1}{2\kappa_{io}}\(Q+\frac{i \Sigma\alpha q}{m_e p_i \gamma}\cos(\theta)\)
\pm
\frac{1}{2\kappa_{io}}\(Q^2
	-Q_c^2\)^{1/2}\(1-\frac{i(\kappa_{io}+1)\frac{\alpha\Sigma Q}{m_e\gamma}}{Q^2-Q_c^2}\),
\end{equation}
meaning that poles present in the upper half-plane are $\delta p_{+,2}$ and $\delta p_{-,1}$. Additionally, denote 
\begin{align}
	\frac{1}{\delta p_{+,2}^2+\frac{\Sigma^2}{m_e^2}}=\<s\>+\delta s,\\
	\frac{1}{\delta p_{-,1}^2+\frac{\Sigma^2}{m_e^2}}=\<s\>-\delta s.
\end{align}

 Note right away a derivative with respect to $\alpha$ or $\gamma$ of these functions
 \begin{equation}
 	\frac{\d }{\d(\alpha/\gamma)}\(\<s\>/\delta s\)= \frac{\d }{\d \delta p}(\<s\>/\delta s)\Bigg|_{\delta p=\delta p_{+,2}}\frac{\d  p_{+,2}}{\d (\alpha/\gamma)}+\frac{\d }{\d \delta p}(\<s\>/\delta s)\Bigg|_{\delta p=\delta p_{-,1}}\frac{\d  p_{-,1}}{\d (\alpha/\gamma)}
 \end{equation}
 will add at least a power of $\propto \Sigma/(Q^2-Q_c^2)^{1/2}\propto\Sigma^{1/2}$. A derivative w/r to $x_q$ does not change the power of $\Sigma$. By an identical argument, $\d \<f_i\>(\delta f_i)/(\d \alpha(\gamma)) $   also adds at least a power of $\Sigma^{1/2}$. We then conclude that the dominant contribution in the small $\Sigma/m_e$ limit comes from derivative w/r to $\alpha$ of the ratio difference or, if vanishing, derivatives of $f_i/s$ w/r to $\alpha$:
  \begin{multline*}
\label{appendix:conductivity:higher:total:DOS:hetero:generating:residues}
	g_{DOS,i,1}=\frac{1}{\pi\gamma^2m_e^6}
	 	\int  \frac{v_p^2f_i(\frac{i\Sigma}{m_e})}{\frac{2\Sigma}{m_e}}
	 	\frac{1}{(-(\kappa_{io}\frac{i\Sigma}{m_e}+Q)\frac{i\Sigma}{m_e}-\frac{|x_q|^2}{m_e^2}-\frac{\Sigma^2}{m_e^2})^2+\frac{\alpha^2\Sigma^2}{\gamma^2}(\frac{i\Sigma}{m_e}-\kappa_{io}(\frac{i\Sigma}{m_e} +Q))^2}
	 	\\+
	 	\frac{1}{2\pi\gamma^2\kappa_{io}^2m_e^6}
	 	\int \frac{v_p^2 \<f_i\>\<s\>}{\delta p_{+,2}-\delta p_{-,1}}
	 	 \bigg(
	 	\frac{1}{(\delta p_{+,2}-\delta p_{+,1})\text{Im}(\delta p_{+,2})}
	 	-
	 	 	\frac{1}{\text{Im}(\delta p_{-,1})(\delta p_{-,1}-\delta p_{-,2})}\bigg)\\+
	 	 	\frac{1}{2\pi\gamma^2\kappa_{io}^2m_e^6}
	 	\int \frac{v_p^2 \<f_i\>\delta s}{\delta p_{+,2}-\delta p_{-,1}}
	 	 \bigg(
	 	\frac{1}{(\delta p_{+,2}-\delta p_{+,1})\text{Im}(\delta p_{+,2})}
	 	+
	 	 	\frac{1}{(\delta p_{-,1}-\delta p_{+,1})\text{Im}(\delta p_{-,1})}\bigg)
	 	 	\\
	 	 	+
	 	\frac{1}{2\pi\gamma^2\kappa_{io}^2m_e^6}
	 	\int \frac{v_p^2 \delta f_i \<s\>}{\delta p_{+,2}-\delta p_{-,1}}
	 	 \bigg(
	 	\frac{1}{(\delta p_{+,2}-\delta p_{+,1})\text{Im}(\delta p_{+,2})}
	 	+
	 	 	\frac{1}{\text{Im}(\delta p_{-,1})(\delta p_{-,1}-\delta p_{-,2})}\bigg)
	 	 	\\
	 	 		 	 	+
	 	\frac{1}{2\pi\gamma^2\kappa_{io}^2m_e^6}
	 	\int \frac{v_p^2 \delta f_i \delta s}{\delta p_{+,2}-\delta p_{-,1}}
	 	 \bigg(
	 	\frac{1}{(\delta p_{+,2}-\delta p_{+,1})\text{Im}(\delta p_{+,2})}
	 	-
	 	 	\frac{1}{\text{Im}(\delta p_{-,1})(\delta p_{-,1}-\delta p_{-,2})}\bigg).
	 \end{multline*}
	 
\begin{figure}
\includegraphics[width=0.5\columnwidth]{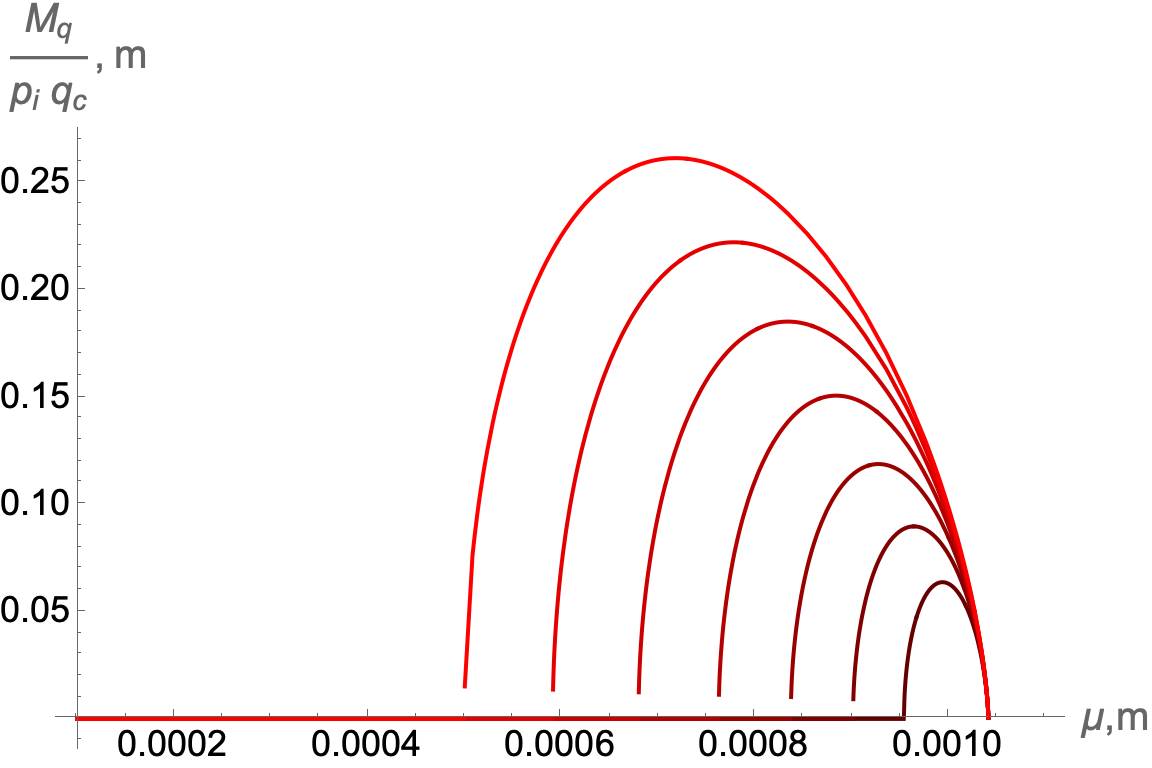}
\caption{Plot of the order parameter as a function of the chemical potential $\mu$ in units of mass, normalized to $p_i q_c$ }
\label{fig:appendix:conductivity:higher:total:DOS:hetero:MqTopiQC}
\end{figure}

Substituting approximate expressions for poles now, we get 
\begin{multline*}
	g_{DOS,i,1}=\frac{2^2 \kappa_{io}^3}{\gamma m_e^5 \Sigma\alpha}\int d\theta \frac{v_p^2(\<f_i\>\<s\>+\delta s \delta f_i)}{\sqrt{Q^2-Q_c^2}+\frac{i\Sigma}{m_e}\frac{\alpha}{\gamma}\frac{q}{p_i}\cos(\theta)}
	\frac{(\kappa_{io}+1)Q+\frac{i\Sigma}{m_e}\frac{\alpha}{\gamma}\frac{(\kappa_{io}+1)Q}{\kappa_{io}\sqrt{Q^2-Q_c^2}}\frac{q}{p_i}\cos(\theta)}{\frac{(\kappa_{io}+1)^2Q^2}{Q^2-Q_c^2}-(\frac{q}{p_i}\cos(\theta))^2}\frac{1}{Q^2-Q_c^2+\(\frac{\alpha \Sigma(\kappa_{io}+1)Q}{\gamma m_e \kappa_{io}\sqrt{Q^2-Q_c^2}}\)^2}\\+
	\frac{2^2 \kappa_{io}^3}{\gamma m_e^5 \Sigma\alpha}\int d\theta \frac{v_p^2(\<f_i\>\delta s+ \<s\> \delta f_i)}{\sqrt{Q^2-Q_c^2}+\frac{i\Sigma}{m_e}\frac{\alpha}{\gamma}\frac{q}{p_i}\cos(\theta)}
	\frac{\sqrt{Q^2-Q_c^2}\frac{q}{p_i}\cos(\theta)+\frac{i\Sigma \alpha}{m_e\gamma}\frac{(\kappa_{io}+1)^2Q^2}{\kappa_{io}(Q^2-Q_c^2)}\frac{q}{p_i}\cos(\theta)}{\frac{(\kappa_{io}+1)^2Q^2}{Q^2-Q_c^2}-(\frac{q}{p_i}\cos(\theta))^2}\frac{1}{Q^2-Q_c^2+\(\frac{\alpha \Sigma(\kappa_{io}+1)Q}{\gamma m_e \kappa_{io}\sqrt{Q^2-Q_c^2}}\)^2}.
\end{multline*}
 where as before $Q=p_i^2-p_o^2+q^2+2p_i q\cos(\theta)$. For small $Q_c\ll 2p_i q_c$ (see Fig. \ref{fig:appendix:conductivity:higher:total:DOS:hetero:MqTopiQC}) it should be still true that the integral is peaked in vicinity of $\theta_{c,i}$, and hence we can linearize $Q=-2p_i q\sin(\theta_{c,i})(\theta-\theta_{c,i})$ (see Fig. \eqref{fig:electromagnetic:corrections:DOS:poles:dp-1-comparison} for comparison of $\theta$-dependence of poles in linear approximation of Q and other approximations). Since the minimum possible value of the difference $Q_c^2-Q^2\propto 2 Q_c \Sigma/m_e$, for $\Sigma\lessapprox Q_c(p_i/q_c)\approx 10^{-4}$, it should be possible to approximate the expression above by 
\begin{multline}
	g_{DOS,i,1}=\frac{2^2 \kappa_{io}^3}{\gamma m_e^5 \Sigma\alpha}\frac{1}{\kappa_{io}+1}\int \frac{dQ}{p_i q_c \sin(\theta_{c,i})} \frac{v_p^2(\<f_i\>\<s\>+\delta s \delta f_i)}{\sqrt{Q^2-Q_c^2}}
	\frac{1}{Q}\\+
	\frac{2^2 \kappa_{io}^3}{\gamma m_e^5 \Sigma\alpha}\frac{1}{(\kappa_{io}+1)^2}\int \frac{dQ}{ p_i q_c \sin(\theta_{c,i})}\frac{v_p^2(\<f_i\>\delta s+ \<s\> \delta f_i)}{Q^2}
\frac{q_c}{p_i}\cos(\theta_{c,i}).
\end{multline}
In the numerator, we neglect $Q$-dependence and simply use $Q\approx Q_c$ and also concentrate on $i=3$ (derivative w/r to $\alpha$). Then  $\delta p_{3,-}\approx \delta p_{4,+}\approx -Q_c/(2\kappa_{io})$, and hence 
\begin{equation}
	\<f_3\>\<s\>(Q=Q_c)=-\frac{m_e^2\kappa_{io}}{\Sigma^2}.
\end{equation}
Then 
\begin{multline}
	g_{DOS,i,1}=\frac{2\kappa_{io}^3}{\gamma m_e^5 \Sigma\alpha}\frac{1}{\kappa_{io}+1}\int \frac{dQ}{p_i q_c \sin(\theta_{c,i})} \frac{v_p^2\<f_i\>\<s\>}{\sqrt{Q^2-Q_c^2}}\approx 
	\frac{2^4\kappa_{io}^4}{\gamma m_e \Sigma^3\alpha}\frac{1}{\kappa_{io}+1}\frac{1}{\sqrt{2Q_c}}\int \frac{dQ}{q_c \sin(\theta_{c,i})} \frac{p_i(\cos(\theta_{c,i}))^2}{\sqrt{Q-Q_c}}\frac{1}{Q}
	\\=
	\frac{2^5 \kappa_{io}^4}{\gamma \alpha m_e \Sigma^3}\frac{1}{\kappa_{io}+1}\frac{1}{\sqrt{2}Q_c}\frac{p_i\cos(\theta_{c,i})^2}{q_c\sin(\theta_{c,i})}\tan^{-1}\(\frac{\sqrt{Q_0-Q_c}}{Q_c^{1/2}}\).
\end{multline}
As a consequence, conductance in this region is:
\begin{equation}
	\sigma_{DOS,het,1}\approx\frac{ 2^3x_q  }{\Sigma m_e^2}\frac{\kappa_{io}^{7/2}}{\kappa_{io}+1}\frac{v_i v_o}{p_i q\sin(\theta_{c,i})}\tan^{-1}\(\frac{\sqrt{Q_0-Q_c}}{Q_c^{1/2}}\),
\end{equation}
which, in fact, is comparable to the quadratic correction in the range $x_q/\Sigma\approx 1$. Note that correction has a positive sign, which is somewhat counterintuitive. It is also clear that in the real part of the correction there are no terms $O(\Sigma/m_e)$.\\
Therefore, the sign change as a function of $\Sigma/m_e$ should come from first term, with poles coming from the original quasiparticles unaffected by the potential. We expand on the sign change at the end of this section.

\item The middle region, just as before, can be divided into two subregions: left-vicinity of $Q=Q_c$ (a) and right vicinity of this point. 
\begin{figure}
\includegraphics[width=1.\columnwidth]{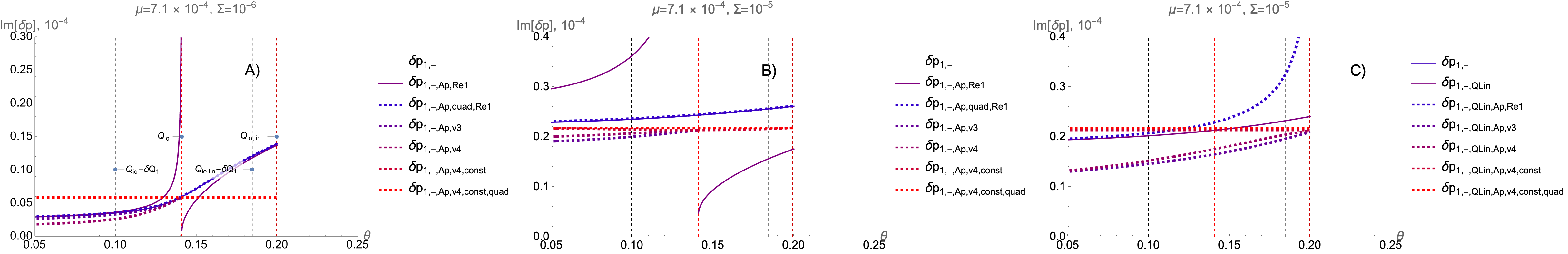}
\caption{Region 2a: comparison of the exact poles \eqref{appendix:conductivity:higher:total:DOS:hetero:generating:residues:exact} for regions 2a  as a function of $\theta$ with approximate expression \eqref{appendix:conductivity:higher:total:DOS:hetero:generating:residues:regime-2a}. Vertical lines to denote boundaries of the region 2a or 2b (gray -- $Q_{io,lin}+\delta Q_1$, with linear approximation of $Q$, black -- $Q_{io,lin}$, linear $Q$, red -- exact $Q_{io}$ on A, and an analogous labeling used for B) and C). Disorder self-energy is $\Sigma=10^{-6}$ on A), and $\Sigma=10^{-5}$ for B) and C). On C), we used linear approximation for $Q$: $Q=2 q p_i cos(\theta_{c,i})(\theta-\theta_{c,i})$. }
\label{fig:electromagnetic:corrections:DOS:poles:dp-2a-comparison}
\end{figure}

\begin{enumerate}[label={\alph*)}]

\item In the left-middle region, the imaginary part of the pole is no longer negligible in comparison to the real part. For concreteness, we pick a region of size $\Sigma$ defined through $0<Q^2-Q^2_c<2A_1Q_{c,io}(\Sigma/m_e)$. In this region, we approximate roots by:
\begin{multline}\label{appendix:conductivity:higher:total:DOS:hetero:generating:residues:regime-2a}
	\delta p_{+,1/2,iia}=-\frac{Q_c+\frac{i\Sigma \alpha q }{p_i\gamma m_e }\cos(\theta)}{2\kappa_{io}}
	\\\pm\frac{1}{\kappa_{io}}\(\frac{\Sigma}{m_e}\)^{1/2}\(\(\frac{\alpha Q_c}{2\gamma}\(\kappa_{io}+1\)\)^2+\(\frac{\Sigma}{m_e}\)^2\(\kappa_{io}+\(\frac{q}{p_i}\frac{\alpha}{2\gamma}\cos(\theta_{c,i})\)^2\)^2\)^{1/4}e^{-i\pi\text{sign}(\Sigma)/4}.
\end{multline}
and take $A_1=1/4$. For convenience, let me denote:
\begin{multline}
	\tau_C=m_e^{-1}\(\(\frac{\alpha Q_c}{2\gamma}\(\kappa_{io}+1\)\)^2+\(\frac{\Sigma}{m_e}\)^2\(\kappa_{io}+\(\frac{q}{p_i}\frac{\alpha}{2\gamma}\cos(\theta_{c,i})\)^2\)^2\)^{-1/2}
	\\=
	\(\(\frac{\alpha}{\gamma\tau_X}\)^2+\(\frac{1}{\tau_D}\)^2\(\kappa_{io}+\(\frac{q}{p_i}\frac{\alpha}{2\gamma}\cos(\theta_{c,i})\)^2\)^2\)^{-1/2}
	,
\end{multline}
so that we obtain an expression
\begin{equation}\label{appendix:conductivity:higher:total:DOS:hetero:generating:residues:regime-2a-tau}
	\delta p_{+,1/2,iia}=-\frac{Q_c+\frac{i\Sigma \alpha q }{p_i\gamma m_e }\cos(\theta)}{2\kappa_{io}}
	\pm\frac{1}{\kappa_{io}m_e}\(\tau_C \tau_D\)^{-1/2}e^{-\frac{i\pi}{4}\text{sign}(\Sigma)}.
\end{equation}

\begin{figure}
\includegraphics[width=1.\columnwidth]{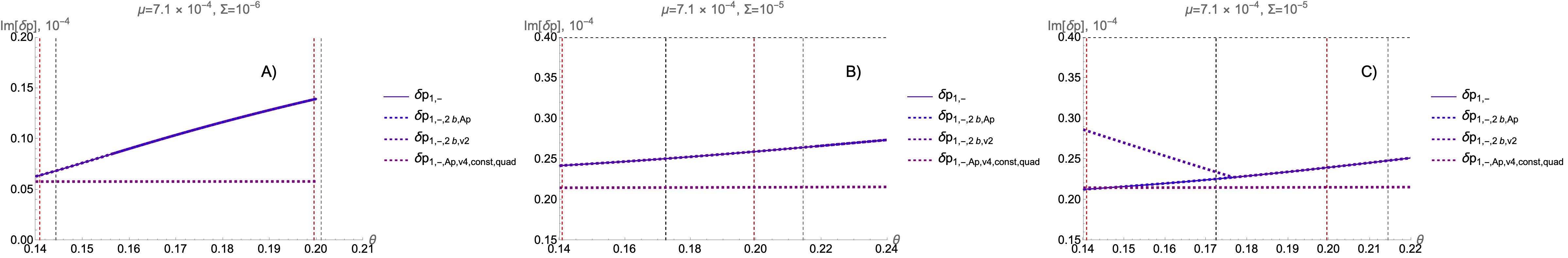}
\caption{Region 2b: comparison of the exact poles \eqref{appendix:conductivity:higher:total:DOS:hetero:generating:residues:exact} for regions 2b  as a function of $\theta$ with approximate expression \eqref{appendix:conductivity:higher:total:DOS:hetero:generating:residues:regime-2a}. Vertical lines to denote boundaries of the region 2a or 2b (red -- $Q_{io}-\delta Q_1$ (dark red with linear approximation of $Q$), black -- $Q_{io}$ (gray for linear approximation) on A, and an analogous labeling used for B) and C). Disorder self-energy is $\Sigma=10^{-6}$ on A), and $\Sigma=10^{-5}$ for B) and C). On C), we used linear approximation for $Q$: $Q=2 q p_i cos(\theta_{c,i})(\theta-\theta_{c,i})$. }
\label{fig:electromagnetic:corrections:DOS:poles:dp-2b-comparison}
\end{figure}

\item   In the right vicinity of $\theta_{crit,i}$ ($Q_c$), defined by $-2A_1Q_c\frac{\Sigma}{m_e}<Q^2-Q^2_c<0$, roots are 
 \begin{equation}\label{appendix:conductivity:higher:total:DOS:hetero:generating:residues:regime-2b}
	\delta p_{+,1/2,iib}\approx-\frac{Q+\frac{i\Sigma \alpha q }{p_i\gamma m_e }\cos(\theta)}{2\kappa_{io}}
	\pm\(\frac{\alpha \Sigma}{\gamma m_e}\frac{\kappa_{io}+1}{2 \kappa_{io}^2}Q_c\)^{1/2}\(\frac{1-\frac{iQ_c}{\delta Q}\frac{\Sigma}{m_e} (\kappa_{io}+1)}{1+\frac{iQ_c}{\delta Q}\frac{\Sigma}{m_e} (\kappa_{io}+1)}\)^{1/4}e^{-i\frac{\pi}{2}\text{sign}(\Sigma)}.
\end{equation}
	
	The problem, however, is that $\delta Q/Q_c$ of order 1 and we cannot expand the bracket. We then use constant approximation (See Fig.\ref{fig:electromagnetic:corrections:DOS:poles:dp-2a-comparison}, \ref{fig:electromagnetic:corrections:DOS:poles:dp-2b-comparison}) in the whole region close to $Q_c$.
	
	 The generating function then is
	\begin{equation}
		g_{DOS,het,2}=
		\frac{2^{7/2}\pi \kappa_{io}^4 p_i}{\gamma^2 q_c}\frac{\tau_D(\tau_D\tau_C)^{3/2}}{ \sin(\theta_{c,i})}
		A_1\(\frac{1}{2-\(\frac{q}{p_i}\frac{\alpha}{\gamma}\cos(\theta_{c,i})\)^2\frac{\tau_C}{\tau_D}}\).
	\end{equation}
	the correction to conductance is 
		\begin{multline}
		\sigma_{DOS,het,2}\approx 
		\frac{2^{5/2}A_1\pi}{\sin(\theta_{c,i})}\frac{p_i}{q}\frac{\kappa_{io}^4}{\gamma^2}\frac{x_q^2}{m_e^2}\frac{(\tau_D\tau_C)^{1/2}}{2-\(\frac{q}{p_i}\frac{\alpha}{\gamma}\cos(\theta_{c,i})\)^2\frac{\tau_C}{\tau_D}}\(\frac{3}{2}\frac{\d \tau_C}{\d \alpha}+\frac{\tau_C\(\frac{2\alpha \tau_C}{\tau_D}+\frac{\d \tau_C}{\d \alpha}\frac{\alpha^2}{\tau_D}\)}{2-\(\frac{q}{p_i}\frac{\alpha}{\gamma}\cos(\theta_{c,i})\)^2\frac{\tau_C}{\tau_D}}\(\frac{q}{p_i\gamma}\cos(\theta_{c,i})^2\)\)
			\end{multline}
where the derivative of $\tau_C$ can be written as:  
	\begin{equation}
	\frac{\d \tau_C}{\d \alpha}=-\tau_C^3\frac{\alpha}{\gamma^2}\(\frac{1}{\tau_X^2}+\frac{1}{\tau_D^2}\(\frac{q}{p_i}\cos(\theta)\)^2\),
	\end{equation}
	where we also recalled that $Q_c=2x_q\kappa_{io}^{1/2}/m_e$. 	
	\end{enumerate}
\begin{figure}
\includegraphics[width=0.5\columnwidth]{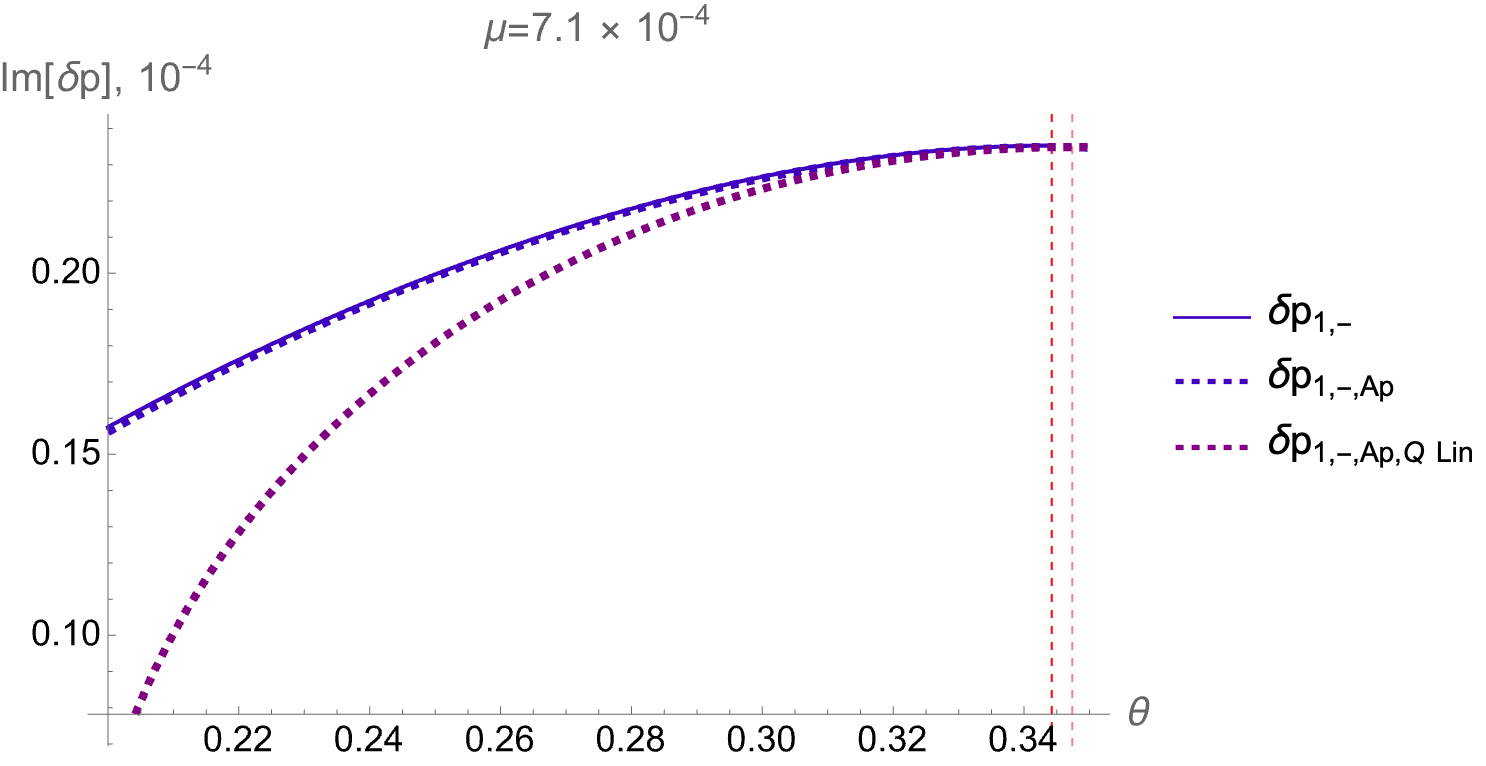}
\caption{Region 3: comparison of the exact poles (\eqref{appendix:conductivity:higher:total:DOS:hetero:generating:residues:exact}, blue, solid) for region 3 as a function of $\theta$ with approximate expression (\eqref{appendix:conductivity:higher:total:DOS:hetero:generating:residues:regime-3}, dashed lines). Vertical lines to denote boundaries of the region 3 (gray (black) -- $Q_{io,lin}-\delta Q_1$ in linear approximation of $Q$ ($Q_{io}-\delta Q_1$), and pink (red) to denote angle corresponding to $Q=0$. We use the linear upper boundary in the calculation. $\Sigma=10^{-6}$ here.}
\label{fig:electromagnetic:corrections:DOS:poles:dp-2b-comparison}
\end{figure}
\item In the region to the far right from $Q_c$, roots are:
\begin{equation}\label{appendix:conductivity:higher:total:DOS:hetero:generating:residues:regime-3}
	\delta p_{+,1/2,iii}=-\frac{Q+\frac{i\Sigma q}{m_e p_i}\frac{\alpha}{\gamma}\cos(\theta)}{2\kappa_{io}}\mp i \sqrt{\(\frac{Q_c}{2\kappa_{io}}\)^2-\(\frac{Q}{2\kappa_{io}}\)^2}\(1-i\frac{(\kappa_{io}+1)\frac{\alpha\Sigma Q}{m_e\gamma\kappa_{io}^2}}{4\(\(\frac{Q}{2\kappa_{io}}\)^2
	-\frac{|x_q|^2}{m_e^2 \kappa_{io}}\)}\)\text{sign}(\Sigma).
\end{equation} 
With properties as in region I:
\begin{align}
	Re(\delta p_{+,1/2,iii})=Re(\delta p_{-,1/2,iii}),\\
	Im(\delta p_{+,1/2,iii})=-Im(\delta p_{-,1/2,iii}).
\end{align}
After some simplifications, generation function in the third region becomes:
\begin{multline}
	g_{i}\approx\frac{2^4 \pi}{\gamma^2 m_e^6}
	\int d\theta \frac{v_p^2\<f_i\>\<s\>\kappa_{io}^3 (Q_c^2-Q^2)}{(\kappa_{io}+1)Q+\frac{i q}{p_i}\cos(\theta)(Q_c^2-Q^2)^{1/2}}\\\frac{\frac{iq}{p_i}\cos(\theta)(Q_c^2-Q^2)+(\kappa_{io}+1)Q(Q_c^2-Q^2)^{1/2}}{\((Q_c^2-Q^2)^2+\((\kappa_{io}+1)\frac{\alpha \Sigma Q_c}{\gamma m_e}\)^2\)\(Q_c^2-Q^2-(\frac{\Sigma}{m_e}\frac{q}{p_i}\frac{\alpha}{\gamma}\cos(\theta))^2\)}.
\end{multline}
real part of which, in negligence of $\((\Sigma/m_e)(q/p_i)\cos(\theta))^2\)$ in comparison to $Q_c \delta Q$ takes the form
\begin{equation}
	g_{i}\approx \frac{2^5 \pi\tau_D^{5/2}}{\gamma^2 m_e^{3/2}}\frac{\kappa_{io}^4(p_i/q)}{\sin(\theta_{c,i})Q_c^{3/2}}\frac{1}{(\kappa_{io}+1)^{1/2}}\(\frac{2\gamma }{\alpha}\)^{1/2},
\end{equation}
where we extended integration range from $0$ to $\infty$, which should be a reasonable approximation as long as $Q_c$ and $\Sigma/m_e$ ($\delta Q$ and $\Sigma/m_e$) are well separated. So that the resulting contribution to the conductance is
\begin{equation}
	\sigma_{3}\approx -2^2 \pi x_q^{1/2}\tau_D^{1/2}\frac{\kappa_{io}^{13/4}(p_i/q)}{\sin(\theta_{c,i})}\frac{1}{(\kappa_{io}+1)^{1/2}}.
\end{equation}
\end{enumerate}

\end{enumerate}

\paragraph{ Homo-contribution}
Close to the Fermi surface, for homo-processes (without loss of generality we look at the inner-inner case) dispersion can be written in the form $\xi_{p+q}=m_e(\kappa\delta p+Q_{ii}(\theta))$ and $\xi_p=m_e\delta p$, where $Q_{ii}(\theta)\equiv q^2+2p_i q \cos(\theta)$ and, as before, $\kappa\equiv 1+(q/p_i)\cos(\theta)$. The generating function is then: 
\begin{equation}
	g_i\equiv \frac{1}{\pi\gamma^2 m_e^6}\int\frac{ v_p^2 f_i(\alpha,\gamma,x_q,\delta p)}{\delta p^2+\frac{\Sigma^2}{m_e^2}}\frac{1}{\(\delta p(\kappa_{ii}\delta p+Q_{ii})-\frac{\Sigma^2+x^2_q}{m_e^2}\)^2+\(\frac{\Sigma \alpha }{m_e \gamma}\)^2\(\delta p(\kappa_{ii}+1)+Q_{ii}\)^2}.
\end{equation}
The poles then are 
\begin{equation}
	\delta p_{+,3/4}=-\frac{Q_{ii}-\frac{i\Sigma\alpha}{m_e \gamma}(\kappa_{ii}+1)}{2\kappa_{ii}}\pm\sqrt{\(\frac{Q_{ii}-\frac{i\Sigma\alpha}{m_e \gamma}(\kappa_{ii}+1)}{2\kappa_{ii}}\)^2+\frac{x_q^2+\Sigma^2}{m_e^2\kappa_{ii}}+\frac{i\Sigma\alpha}{m_e \gamma \kappa_{ii}}Q_{ii}}.
\end{equation}
Note that since here (effectively) angular kinetic energy $\propto Q_{ii}^2/(2\kappa_{ii})$  enters with the same sign as the gap $x_q^2/m_e^2$, there is no point where the real part of the expression under the root changes sign. We then anticipate, based on the discussion in the hetero-contribution, that the homo-contribution will have linear $\tau_D$ behavior. Additionally, in the regime $Q^2+Q_c^2\gg 2\kappa_{ii}\Sigma^2/m_e^2$ we can expand:
\begin{equation}
	\delta p_{+,3/4}\approx -\frac{Q_{ii}-\frac{i\Sigma\alpha}{m_e \gamma}(\kappa_{ii}+1)}{2\kappa_{ii}}\pm\sqrt{\(\frac{Q_{ii}}{2\kappa_{ii}}\)^2+\(\frac{Q}{2\kappa_{ii}}\)^2}\(1+\frac{\frac{i\Sigma}{2^2m_e}\frac{\alpha}{\gamma}\frac{\kappa_{ii}-1}{\kappa_{ii}^2}Q_{ii}}{\(\frac{Q_{ii}}{2\kappa_{ii}}\)^2+\(\frac{Q_c}{2\kappa_{ii}}\)^2}\).
\end{equation}
After integrating over the residues we have, naturally, the same expression:
\begin{multline}
	g_i\equiv \frac{1}{\gamma^2 m_e^5\Sigma}\int\frac{v_p^2 f_i(\alpha,\gamma,x_q,\frac{i\Sigma}{m_e})}{\kappa^2(\frac{i\Sigma}{m_e}-\delta p_{+,3})(\frac{i\Sigma}{m_e}-\delta p_{-,3})(\frac{i\Sigma}{m_e}-\delta p_{+,4})(\frac{i\Sigma}{m_e}-\delta p_{-,4})}
	\\+
	 \frac{2i}{\gamma^2 m_e^6}\int\frac{ v_p^2 (\<f_i\>\<s\>+\delta f_i\delta s)}{\kappa^2(\delta p_{+,3}-\delta p_{+,4})}
	 \(\frac{1}{\delta p_{+,3}-\delta p_{-,3}}\frac{1}{\delta p_{+,3}-\delta p_{-,4}}-\frac{1}{\delta p_{+,4}-\delta p_{-,3}}\frac{1}{\delta p_{+,4}-\delta p_{-,4}}\)
	 \\+
	 \frac{2i}{\gamma^2 m_e^6}\int\frac{ v_p^2 (\<f_i\>\delta s+\delta f_i\<s\>)}{\kappa^2(\delta p_{+,3}-\delta p_{+,4})}
	 \(\frac{1}{\delta p_{+,3}-\delta p_{-,3}}\frac{1}{\delta p_{+,3}-\delta p_{-,4}}+\frac{1}{\delta p_{+,4}-\delta p_{-,3}}\frac{1}{\delta p_{+,4}-\delta p_{-,4}}\),
\end{multline}
where averages $\<f\>, \<s\>$ and differences $\delta f$, $\delta s$ have the same meaning as before, but with new expressions for poles. Since $\propto \<f\> \<s\>$ is by a factor of $\tau_D$ larger than everything else, we approximate the whole expression by it. After doing substitution, it becomes:
\begin{equation}
	g_i\approx
	 \frac{2\pi }{\gamma  \alpha  m_e^5 \Sigma}\int\frac{ v_p^2 \<f_i\>\<s\> \kappa}{\sqrt{Q^2+Q_c^2}}
	 \(\frac{1}{\kappa +1+\frac{Q(\kappa-1)\alpha/\gamma}{\sqrt{Q^2+Q_c^2}}}\frac{1}{\sqrt{Q^2+Q_c^2}+\frac{i\Sigma}{m_e}\frac{\alpha}{\gamma}(\kappa+1)}+\frac{1}{\kappa +1-\frac{Q(\kappa-1)\alpha/\gamma}{\sqrt{Q^2+Q_c^2}}}\frac{1}{\sqrt{Q^2+Q_c^2}-\frac{i\Sigma}{m_e}\frac{\alpha}{\gamma}(\kappa+1)}\),
\end{equation}
which in the limit of small $q/p_i$ can be represented as 
\begin{multline}\label{appendix:conductivity:higher:total:DOS:homo:generating:small-qpi}
	g_i\approx
	\frac{2\pi }{\gamma  \alpha  m_e^5 \Sigma}\int v_p^2 \<f_i\>\<s\> \kappa
	 \bigg(\frac{1}{(\kappa +1)\sqrt{Q^2+Q_c^2}+Q(\kappa-1)\frac{\alpha}{\gamma}}\frac{1}{\sqrt{Q^2+Q_c^2}+\frac{i\Sigma}{m_e}\frac{\alpha}{\gamma}(\kappa+1)}
	 \\+\frac{1}{(\kappa +1)\sqrt{Q^2+Q_c^2}-Q(\kappa-1)\frac{\alpha}{\gamma}}\frac{1}{\sqrt{Q^2+Q_c^2}-\frac{i\Sigma}{m_e}\frac{\alpha}{\gamma}(\kappa+1)}\bigg)
\end{multline}
As before, we expect major contribution to come from $\<f_3\>$. In negligence of $\Sigma/m_e$ w/r to $\sqrt{Q^2+Q_c^2}$ $\<f_3\>$ can be approximated by
\begin{equation}
	\<f_i\>=-\frac{m_e^2}{(2\Sigma)^2}\frac{Q_c^2}{\kappa_{ii}},
\end{equation} 
\begin{equation}
	\<s\>=\frac{(2\kappa_{ii})^2\(2(Q/Q_c)^2+1\)}{Q_c^2}.
\end{equation}
Then correction to conductivity is
\begin{equation}
	\delta\sigma_{DOS,hom}=\frac{2  x_q^2 v_i^2(2\kappa_{ii})^{3/2}}{Q_c\Sigma m_e^3 J_{ii}}\tan^{-1}\(\frac{(2\kappa_{ii})^{1/2}Q_{max}}{(\kappa_{ii}+1)Q_c}\)=\frac{2^{3/2} x_q v_i^2(\kappa_{ii})^{5/2}}{\Sigma m_e^2 J_{ii}}\tan^{-1}\(\frac{(2\kappa_{ii})^{1/2}Q_{max}}{(\kappa_{ii}+1)Q_c}\),
\end{equation}
where $J_{ii}=2p_i q \sin(\theta_{i,i})=2p_i q\sqrt{1-\(\frac{q}{2p_i}\)^2}$, which is also positive, surprisingly \footnote{There could be a factor of $2^{1/2-3/2}$ mistake here}. Note also that for large $\Sigma$ (and small $Q_{max}$) $Q_c=(2^2\kappa_{io}/m_e)(x_q^2+\Sigma^2)^{1/2}$ can be a better approximation, which will give a larger power effectively. To take it more rigorously, we make a substitution $z=e^{i\theta}$:\begin{equation}
	g_3\approx -\frac{2}{\gamma \alpha  m_e \Sigma^3}\frac{\kappa_{ii}^3}{\kappa_{ii}+1}\frac{p_i}{q}\oint \frac{dz}{iz}\(2-\frac{Q_c^2 z^2/(p_iq_c)}{(q/p_i z+(z^2+1))^2+z^2(Q_c/p_i q_c)^2}\).
\end{equation}
The first term has poles at $z_1=0$, the second has non-vanishing poles at 
\begin{align}\label{appendix:conductivity:higher:total:DOS:homo:generating:poles}
	z_{1/2,+}=-\frac{q^2+i Q_c}{2p_i q_c}\pm\sqrt{\(\frac{q^2+i Q_c}{2p_i q_c}\)^2-1}\equiv -\frac{q^2+i Q_c}{2p_i q_c}\pm\frac{1}{2p_i q_c}D,\\
		z_{1/2,-}=-\frac{q^2-i Q_c}{2p_i q_c}\pm\sqrt{\(\frac{q^2-i Q_c}{2p_i q_c}\)^2-1}\equiv -\frac{q^2-i Q_c}{2p_i q_c}\pm\frac{1}{2p_i q_c}D^*,
\end{align}
among which $|z_{1,+}|<1$ and $|z_{1,-}|<1$. Taking the integral over residues, we obtain:
\begin{equation}\label{appendix:conductivity:higher:total:DOS:homo:generating:result:implicit}
	g_3\approx -\frac{(2^2\pi) }{\gamma \alpha  m_e \Sigma^3}\frac{\kappa_{ii}^3}{\kappa_{ii}+1}\frac{p_i}{q}\(2-\frac{Q_c^2}{z_{1,+}-z_{1,-}}\(\frac{z_{1,+}}{(z_{1,+}-z_{2,-})(z_{1,+}-z_{2,+})}-\frac{z_{1,-}}{(z_{1,-}-z_{2,-})(z_{1,-}-z_{2,+})}\)\).
\end{equation}
Or, after using \eqref{appendix:conductivity:higher:total:DOS:homo:generating:poles}: 
\begin{equation}\label{appendix:conductivity:higher:total:DOS:homo:generating:result:explicit}
	g_3\approx -\frac{(2^2\pi) }{\gamma \alpha  m_e \Sigma^3}\frac{\kappa_{ii}^3}{\kappa_{ii}+1}\frac{p_i}{q}\(2-\frac{Q_c^2/2}{-iQ_c+(D-D^*)/2}\(\frac{-q^2-iQ_c+D}{D(-iQ_c+(D+D^*)/2)}-\frac{-q^2+iQ_c+D^*}{D^*(iQ_c+(D+D^*)/2)}\)\).
\end{equation}

\paragraph{Sign-change of the DOS-correction} \label{appendix:conductivity:higher:DOS:sign-change}
As the scattering lifetime decreases and becomes comparable to $x_q$, first term coming from the pole $\delta p=i\Sigma/m_e$ can no longer be neglected. Again, we divide correction into hetero- and homo-contributions. Let us perform an expansion into powers of $Q m_e\tau_2$, where $\tau_2=(x_q^2+\Sigma^2)^{-1/2}$, or, equivalently, $Q m_e\tau_C$
\begin{enumerate}[label=(\alph*)]
\item Hetero-contribution
The generating function in zeroth order is
\begin{equation}
	g^{(0)}_{DOS,i\to o}=\frac{1}{m_e\Sigma}
	\int \frac{v_p^2 f_i^{(0)}}{\gamma^2(x_q^2+\Sigma^2(1-\kappa_{io}))^2-\alpha^2\Sigma^4(1-\kappa_{io})^2},
\end{equation}
where $f^{(0)}_1=1$, $f^{(0)}_2=(\Sigma^2/2)(1-\kappa_{io}^2)$, $f^{(0)}_3=-\kappa_{io}$, which gives contributions to conductivity 
\begin{equation}
	\delta \sigma_{DOS,1}^{(0)}=
	-\frac{\Sigma v_i^2\theta_{c,i}}{m_ex_q^2}\frac{(x_q^2+\Sigma^2(1-\kappa_{io}))^2}{(x_q^2+2\Sigma^2(1-\kappa_{io}))^2},
\end{equation}
\begin{equation}
	\delta\sigma_{DOS,2}^{(0)}=
	\frac{\Sigma^3 v_i^2\theta_{c,i}}{m_ex_q^2}(\kappa_{io}^2-1)\frac{(x_q^2+\Sigma^2(1-\kappa_{io}))}{(x_q^2+2\Sigma^2(1-\kappa_{io}))^2},
\end{equation}
\begin{equation}
	\delta\sigma_{DOS,3}^{(0)}=
	-\frac{\Sigma^5 v_i^2\theta_{c,i}}{m_ex_q^2}\frac{\kappa_{io}(1-\kappa_{io})^2}{(x_q^2+2\Sigma^2(1-\kappa_{io}))^2},
\end{equation}
which in $\Sigma/x_q\to \infty$ limit takes the form 
\begin{equation}
	\delta \sigma^{(0)}_{DOS,i\to o}(\Sigma/x_q\to \infty)=-\frac{v_i^2\theta_{c,i}\Sigma}{2m_ex_q^2}(1+\kappa_{io}),
\end{equation}
and $m_e\Sigma^{-1}$ term is absent from the expansion. The opposite limit reads
\begin{equation}
	\delta \sigma^{(0)}_{DOS,i\to o}(x_q/\Sigma\to \infty)=-\frac{\Sigma v_i^2\theta_{c,i}}{m_ex_q^4}.
\end{equation}
Given in small $\Sigma \ll Q_{max}$ limit the correction has a form of $\propto 2^2 \pi (x_q/\Sigma)(1/\theta_{c,i})p_i/q_c$, sign change of the i-o part of the conductance happens close to the point 
\begin{equation}
	\Sigma_{crit,i-o}\approx (x_q^{3/2}/m_e^{1/2})(p_i\theta_{c,i})^{-1},
\end{equation}
which is close to $10^{-5}$ in the range of the interest. 
\item Homo-contribution
Without loss of generality, we take an integral for the inner-inner part of the correction: 
\begin{equation}
	g_{i,DOS,i\to i}=\frac{1}{\Sigma m_e^5}\int 
	\frac{v_p^2 f_i}
	{\(\frac{i\Sigma}{m_e}(\kappa_{ii}\frac{i\Sigma}{m_e}+Q_{ii})-\frac{\Sigma^2+x_q^2}{m_e^2}\)^2+\(\frac{\Sigma\alpha}{m_e\gamma}\)^2\(\frac{i\Sigma}{m_e}(\kappa_{ii}+1)+Q\)^2}
\end{equation}
with auxiliary function being $f_1=1$, $f_2=m_e^2/2\((\kappa_{ii}-1)\frac{i\Sigma}{m_e}+Q\)\((\kappa_{ii}+1)\frac{i\Sigma}{m_e}+Q\)$, $f_3=-(im_e/\Sigma)\(\frac{i\Sigma}{m_e}\kappa_{ii}+Q\)$. In small $Q_c/\Sigma$ limit (for region 1) or $Q/\Sigma$ for all other regions, we expand into powers of $Q\equiv p_i^2+2p_iq \cos(\theta)$ to observe the sign change of conductivity correction. Additionally, for we are interested in regime $q/p_i<1$, we may expand in $\kappa_{ii}-1$. In the 0-th order then I have:
\begin{equation}
	g^{(0)}_{i,DOS,i\to i}=\frac{2\pi}{\Sigma m_e}
	\frac{v_p^2 \<f^{(0)}_i\>}
	{\(2\gamma\Sigma^2+\gamma x_q^2\)^2-\Sigma^4\(2\alpha\)^2},
\end{equation}
where averaging is performed over the angle $ \<f^{(0)}_i\>=1/(2\pi)\int d\theta f^{(0)}_i(\theta)$. Then its contributions conductivity are: 
\begin{equation}
		\delta\sigma^{(0)}_{1,DOS,i\to i}=-\frac{2\pi \Sigma }{ m_e x_q^2}
	\frac{v_p^2 \(\Sigma^22+ x_q^2\)^2}
	{\(2^2\Sigma^2+ x_q^2\)^2}
\end{equation}
\begin{equation}
		\delta\sigma^{(0)}_{2,DOS,i\to i}=0
\end{equation}
\begin{equation}
		\delta\sigma^{(0)}_{3,DOS,i\to i}=\frac{2\pi\Sigma}{m_e x_q^2}
	\frac{v_p^2 \Sigma^42^2}
	{\(2^2\Sigma^2+ x_q^2\)^2}
\end{equation}
which gives zero upon resummation in $\Sigma\to \infty$ limit. Expanding both expressions in $x_q/\Sigma$:
\begin{equation}
	\delta \sigma^{(0)}_{DOS,i\to i}=-\frac{\pi v_i^2}{m_e x_q^2}\frac{\Sigma}{2}
	\(\(1+\frac{x_q^2}{2\Sigma^2}-\frac{x_q^2}{2^2\Sigma^2}\)^2-(1-\frac{x_q^2}{2^2\Sigma^2})^2\)=-\frac{\pi v_i^2}{2m_e \Sigma},
\end{equation}
suggesting that there is no sign change of the homo-part of conductance as a function of $\Sigma/x_q$ \footnote{May happen because of additional terms in $g_i$ we neglected.}. Sign change point is
\begin{equation}
	x_q\propto 2m_e qp_i\theta_{i,i}/\pi,
\end{equation}
above which DOS-part of the homo-part of the conductance correction becomes positive.

\end{enumerate}

\subsubsection{Doubly corrected DOS} \label{appendix:conductivity:higher:xq4-term}
The conductivity correction corresponding to both Green function dressed, is:
\begin{equation}
	\sigma_{v,2}=-\frac{|x_q|^4}{2\pi}\int v_p^2 \text{Im}(G_p((G_pG_{p+q})^{-1}-x_q^2)^{-1}) \text{Im}(G_p((G_pG_{p+q})^{-1}-x_q^2)^{-1}).
\end{equation}
When written explicitly,
\begin{equation}
	\sigma_{v,2}=-\frac{\Sigma^2|x_q|^4}{2\pi}\int v_p^2
	\(\frac{\xi_p(\xi_p+\xi_{p+q})+(\xi_p\xi_{p+q}-x_q^2-\Sigma^2)}{\xi_p^2+\Sigma^2}\frac{1}{\gamma^2(\xi_p\xi_{p+q}-x_q^2-\Sigma^2)^2+\alpha^2\Sigma^2(\xi_p+\xi_{p+q})^2}\)^2.
\end{equation}
We can rewrite it in terms of generating function $g_{v,2}$ as three derivatives:
\begin{multline}
	\sigma_{v,2}
	=
	\frac{x_q^4\Sigma^2}{2^2\pi \gamma}\frac{\d }{\d \gamma}
	\int  \frac{v_p^2}{(\xi_p^2+\Sigma^2)^2}\frac{1}{\gamma^2(\xi_p\xi_{p+q}-x_q^2-\Sigma^2)^2+\alpha^2\Sigma^2(\xi_p+\xi_{p+q})^2}
	\\+
	\frac{x_q^4}{2^2\pi \alpha}\frac{\d }{\d \alpha}
	\int  \frac{v_p^2}{(\xi_p^2+\Sigma^2)^2}\frac{\xi_p^2}{\gamma^2(\xi_p\xi_{p+q}-x_q^2-\Sigma^2)^2+\alpha^2\Sigma^2(\xi_p+\xi_{p+q})^2}
	\\-
	\frac{x_q^3\Sigma^2}{2^2\pi \gamma^2}\frac{\d }{\d x_q}
	\int v_p^2 \frac{\xi_p(\xi_p+\xi_{p+q})}{(\xi_p^2+\Sigma^2)^2}\frac{1}{\gamma^2(\xi_p\xi_{p+q}-x_q^2-\Sigma^2)^2+\alpha^2\Sigma^2(\xi_p+\xi_{p+q})^2}.
\end{multline}
Since term goes has a multiplier $(x_q/Q_{max})^2\propto(Q_c/Q_{max})^2 $, it is of no importance for small value of this parameter (regime of Fermi arcs). In this range, two first terms are about of the same value for $(Q_c/Q_{max})^4\approx \Sigma/m_e$ the second goes at worst as $\propto x_q^4/(m_e^2Q_{max}^4\Sigma^2)$, and the first is $\propto m_e/\Sigma$ for small $\Sigma$. For large value of $(Q_c/Q_{max})^2$, the term dominates over DOS- and vertex-corrections. In this regime, pole at $\delta p=i\Sigma/m_e$ gives the largest contribution, and as a result, the first term with a derivative over $\gamma$ should dominate. Note, however, that generating function can be written (for all corrections) as
\begin{equation}
	g_{v,2}(x_q,\alpha,\gamma)=\gamma^{-2}G_{v,2}(x_q,\alpha/\gamma),
\end{equation}
therefore 
\begin{equation}
	\frac{\d }{\d \alpha}g_{v,2}(x_q,\alpha,\gamma)=\gamma^{-2}G'_{v,2}(x_q,\alpha/\gamma)(1/\gamma),
\end{equation}
\begin{equation}
	\frac{\d }{\d \gamma}g_{v,2}(x_q,\alpha,\gamma)=-2\gamma^{-3}G_{v,2}(x_q,\alpha/\gamma)-\gamma^{-4}G'_{v,2}(x_q,\alpha/\gamma)\alpha ,
\end{equation}
and so
\begin{equation}
		\frac{\d }{\d \gamma}g_{v,2}(x_q,\alpha,\gamma)=-2\gamma^{-3}G_{v,2}(x_q,\alpha/\gamma)-\frac{\d }{\d \alpha}g_{v,2}(x_q,\alpha,\gamma)\alpha\gamma^{-2}. \end{equation}
We now proceed to calculation of hetero-part.
\\\\

\paragraph{Hetero-contribution}
In variables $\theta$, $\delta p$, the integral reads:
\begin{equation}
	\sigma_{v,2,het,2}
	=
	\frac{x_q^4\Sigma^2}{2^3\pi \gamma }\frac{\d }{\d \gamma}
	\int  d\theta d\delta p\frac{v_p^2}{(\delta p^2+\frac{\Sigma^2}{m_e^2})^2}\frac{1}{\gamma^2(\delta p(\kappa_{io}\delta p+Q)+\frac{x_q^2}{m_e^2}+\frac{\Sigma^2}{m_e^2})^2+\frac{(\alpha\Sigma)^2}{m_e^2}(\delta p-\kappa_{io} \delta p-Q)^2},
\end{equation}
where $Q=p_o^2-p_i^2+q^2-2p_i q \cos(\theta)$. Up to a factor, contribution of quasiparticle poles (eigenstates of mean-field Hamiltonian) should be the same as in DOS. In large $Q_c/Q_{max}$ limit, it is $\delta p_{1,2}=i\Sigma/m_e$ pole that dominates, hence we write:  
\begin{equation}
	\delta \sigma_{v,2,het,3}
	=
	\frac{x_q^4}{2^4 \gamma m_e^5 \Sigma}\frac{\d }{\d \gamma}
	\int   \frac{d\theta v_i^2}{\gamma^2(\frac{i\Sigma}{m_e}(\kappa_{io}\frac{i\Sigma}{m_e}+Q)+\frac{x_q^2}{m_e^2}+\frac{\Sigma^2}{m_e^2})^2+\frac{(\alpha\Sigma)^2}{m_e^2}((1-\kappa_{io})\frac{i\Sigma}{m_e}-Q)^2}.
\end{equation}
The denominator has roots at:
\begin{equation}
	Q_{4,+}
	=
	-\frac{x_q^2}{m_e(1+\frac{\alpha}{\gamma})i\Sigma}
	-
	\frac{\Sigma}{i m_e}(1-\kappa_{io}),
	\end{equation}
\begin{equation}
	Q_{4,-}
	=
	-\frac{x_q^2}{m_e(1-\frac{\alpha}{\gamma})i\Sigma}
	-
	\frac{\Sigma}{i m_e}(1-\kappa_{io}).
	\end{equation}
$Q_{4,-}(1-\alpha/\gamma)\to_{\alpha\to\gamma} -x_q^2/(im_e\Sigma)$.
The whole thing can be rewritten in the form 
\begin{multline}
	\delta \sigma_{v,2,het,3}
	=
	-\frac{x_q^2 i }{2^3(\Sigma m_e)^2}\int \frac{d\theta v_p^2 (Q-Q_{4,+}^*)}{Q^2+|Q_{4,+}|^2}
	+\frac{x_q^4}{2^7(\Sigma m_e)^3}\int \frac{d\theta v_p^2 (Q-Q^*_{4,+})^2 }{(Q^2-|Q_{4,+}|^2)^2}
	-
	\frac{1}{2^5\Sigma m_e }\int \frac{d\theta v_p^2 Q(Q-Q_{4,+}^*)^2 }{(Q^2-|Q_{4,+}|^2)}
	\end{multline}
which for small $Q_{max}/|Q_{4,+}|$ goes as
\begin{equation}
	\delta \sigma_{v,2,het,3}\approx-\frac{(x_q v_i)^2Q_{max}}{2^3 J \Sigma m_e (x_q^2+2\Sigma^2(\kappa_{io}-1))}-\frac{x_q^4 v_i^2}{2^5\Sigma m_e \gamma^2 J}\frac{Q_{max}}{(x_q^2+2\Sigma^2(\kappa_{io}-1))^2},
\end{equation}
which changes effective power from $\Sigma^{-1}$ to $\Sigma^{-3}$. In the opposite limit  $|Q_{4,+}|/Q_{max}\ll 1$ there is a term independent on the order parameter:
\begin{equation}
	\delta \sigma_{v,2,het,3}
	\approx
	 -\frac{\pi}{2J}\(\frac{x_q v_i}{2^2 \Sigma m_e}\)^2-\frac{v_i^2}{2^5 \Sigma m_e}\frac{Q_{max}}{J}.
\end{equation}

\paragraph{Homo-contribution}
\begin{equation}
	\sigma_{v,2,hom}
	=
	\frac{x_q^4\Sigma^2}{2^3\pi \gamma m_e^6}\frac{\d }{\d \gamma}
	\int  d\theta d\delta p\frac{v_i^2}{(\delta p^2+\frac{\Sigma^2}{m_e^2})^2}\frac{1}{\gamma^2(\delta p(\kappa_{ii}\delta p+Q)-\frac{x_q^2}{m_e^2}-\frac{\Sigma^2}{m_e^2})^2+\frac{(\alpha\Sigma)^2}{m_e^2}((1+\kappa_{ii})\delta p+Q)^2},
\end{equation}
$Q=2p_i q\cos(\theta)+q^2$.
As in the hetero-contribution, it is plausible that the dominant contribution comes from $\delta p=i\Sigma/m_e$ pole. Hence we  write: 
\begin{equation}
	\sigma_{v,2,hom}
	\approx
	\frac{x_q^4}{ 2^4\gamma m_e^3\Sigma }\frac{\d }{\d \gamma}
	\int  d\theta \frac{v_i^2}{\gamma^2\(\frac{i\Sigma}{m_e}(\kappa_{ii}\frac{i\Sigma}{m_e}+Q_{ii})-\frac{x_q^2}{m_e^2}-\frac{\Sigma^2}{m_e^2}\)^2+\(\frac{\alpha \Sigma}{m_e}\)^2\(\frac{i\Sigma}{m_e}(1+\kappa_{ii})+Q_{ii}\)^2},
\end{equation}
which can be rewritten in the form:
\begin{equation}
	\sigma_{v,2,hom}
	=
	-\frac{x_q^2 i}{2^3 \Sigma^2 m_e^2}\int d\theta \frac{v_p^2 }{Q-Q_{4,+}}
	+\frac{x_q^4 }{2^7 \Sigma^3 m_e^3}\int d\theta \frac{v_p^2 }{(Q-Q_{4,+})^2}
	-\frac{1 }{2^5 \Sigma^3m_e}\int d\theta \frac{v_p^2Q}{Q-Q_{4,+}},
\end{equation}
that goes in the $Q_{max}/|Q_{4,+}|\ll 1$ limit as
\begin{equation}
	\sigma_{v,2,hom,3}
	\propto-\frac{x_q^2}{2^3 J m_e \Sigma}\frac{v_i^2 Q_{max}}{(x_q^2+\Sigma^2(1+\kappa_{ii})2)}, 
\end{equation}
while in the large $Q_{max}/|Q_{4,+}|\gg 1$ limit it follows the law
\begin{equation}
	\sigma_{v,2,hom,3}
	\propto-\frac{x_q^2 v_i^2\pi}{2^5\Sigma^2 m_e^2 J}-\frac{v_i^2 Q_{max}}{2^5\Sigma m_e J},
\end{equation}
so that the whole difference between homo- and hetero-corrections comes from the Jacobian and $Q_{max}$.

\subsubsection{Vertex correction} \label{appendix:conductivity:higher:vertex}
Based on \ref{appendix:conductivity:higher:4-th} and \ref{appendix:conductivity:higher:6-th} it is clear that the perturbative expansion for the correction with vertices having different momentum consists of terms of the form $\int v_p v_{p+q}\Im((G_p G_{p+q})^m)\Im((G_p G_{p+q})^n)$ to allow for the momentum change in the vertex. The sum 
\begin{multline}
	\sigma_{v}=
	\frac{|x_q|^2}{2\pi}\int v_p v_{p+q} \Im(G_pG_{p+q})(
	\Im(G_pG_{p+q})
	+
	|x_q|^2\Im((G_pG_{p+q})^2
	+
	|x_q|^4\Im((G_pG_{p+q})^3+...)\\+
	\frac{|x_q|^4}{2\pi}\int v_p v_{p+q} \Im((G_pG_{p+q})^2)
	(
	|x_q|^2\Im((G_pG_{p+q})^2
	+
	|x_q|^4\Im((G_pG_{p+q})^3+...)+...
\end{multline}
can be rewritten as
\begin{equation}\label{appendix:conductivity:higher:total:vertex:conductance:general}
	\sigma_{v}=
	\frac{|x_q|^2}{2\pi}\int v_p v_{p+q} \Im\(\frac{1}{G^{-1}_p G^{-1}_{p+q}-|x_q|^2}\) \Im\(\frac{1}{G_p^{-1}G_{p+q}^{-1}-|x_q|^2}\).
\end{equation}
Evaluation can be done along the same steps: we divide the integration into 4 regions and use approximate values for the poles. Note, however, that up to a factor and additional poles the generating function is the same. We also saw that additional poles for DOS correction, as well as additional terms (with derivatives of the generating) are not important at least in the limit of large mean-free path. We then simply use expressions from the previous section with corrected multipliers. \\\\
Using now general expressions for Green's functions we can rewrite \eqref{appendix:conductivity:higher:total:vertex:conductance:general} in the form:
\begin{equation}
	\sigma_v=-\frac{x_q^2}{2\pi}\frac{1}{2\alpha \gamma^2}\frac{\d }{\d \alpha}\int \frac{v_p v_{p+q}}{(\xi_p\xi_{p+q}-\Sigma^2-x_q^2)^2+\frac{\alpha^2}{\gamma^2}\Sigma^2(\xi_p+\xi_{p+q})^2}=\frac{(x_q\Sigma)^2}{2\alpha}g_{v},
\end{equation}
where generating function is nearly identical to that we used in the DOS-section:
\begin{equation}
	g_{v}=-\frac{1}{2\pi(\gamma\Sigma)^2}\int \frac{v_p v_{p+q}}{(\xi_p\xi_{p+q}-\Sigma^2-x_q^2)^2+\frac{\alpha^2}{\gamma^2}\Sigma^2(\xi_p+\xi_{p+q})^2},
\end{equation}
following the notation of the previous section, that would correspond to a term with numerator $\<f\>\<s\>=-m_e^2/\Sigma^2$.

Within the constant velocity approximation, we can simply write answers using results of the previous section. 

\paragraph{Hetero-contribution}
	We then immediately fill the gaps for the hetero-contribution.
	For concreteness, start with i-o proccess. In the hetero case,  $\xi_{o,p+q}=-m_e(p^2+2pq\cos(\theta)+q^2-p_o^2)$ and $\xi_{i,p}=m_e(p^2-p_i^2)$. In terms of a variable $\delta p=p^2-p_i^2$ it can be written $\xi_{o,p+q}=-m_e(\delta p(1+\frac{q}{p_i}\cos(\theta))+2p_iq\cos(\theta)+p_i^2+q^2-p_o^2)\equiv-m_e( \kappa_{io} \delta p+Q_{io})$ and $\xi_{i,p}=m_e\delta p$. Since $\theta_{c,i}\approx 0$ we approximate $v_{p+q}\approx p_i+q$. Now we use results of the previous section.\\\\ Also, for  $x_q/m_e<Q(0)/(2\kappa_{io})$  exhibits all 3 possible behaviors, we consider only that.
\begin{enumerate}[label={\alph*)}]
	\item  In the far-left region, $Q>Q_c+\frac{\Sigma \kappa_{io}}{4m_e}$
\begin{equation}
		\sigma_{v,het,1}\approx\frac{ x_q  }{\Sigma m_e^2}\frac{\kappa_{io}^{5/2}}{\kappa_{io}+1}\frac{v_i v_o}{p_i q\sin(\theta_{c,i})}\tan^{-1}\(\frac{\sqrt{Q_0-Q_c}}{Q_c^{1/2}}\),
\end{equation}
  	\item In the middle-region, $Q_c-\frac{\Sigma \kappa_{io}}{4m_e}<Q<Q_c+\frac{\Sigma \kappa_{io}}{4m_e}$, within the constant poles approximation
  	 \begin{multline}
	\sigma_{v,het,2}\approx -\frac{x_q\tau_D^2\kappa_{io}^{9/2}(p_o/q)}{2^2\gamma^2\sin(\theta_{c,i})}\frac{(2\tau_C/\tau_D)^{1/2}}{-\(\frac{\alpha q}{\gamma p_i}\cos(\theta_{c,i})\)^2+\(\frac{\tau_D}{2\tau_C}\)}
	\(\frac{1}{2\tau_C}\frac{\d }{\d \alpha}\tau_C+\frac{\(2\alpha\(\frac{ q}{\gamma p_i}\cos(\theta_{c,i})\)^2+\(\frac{\tau_D}{2\tau_C^2}\frac{\d }{\d \alpha}\tau_C\)\)}{\(\frac{\tau_D}{2\tau_C}\)-\(\frac{\alpha q}{\gamma p_i}\cos(\theta_{c,i})\)^2}\)
\end{multline}
where again  $\(\(\frac{\Sigma}{m_e}\)^2\kappa_{io,1}^2+Q_c^2\(\frac{\alpha}{\gamma}\)^2\kappa_{io,2}^2\)^{1/4}=(m_e\tau_{C})^{-1/2}$, and $\kappa_{io,1}=2^2\kappa_{io}+\(\frac{\alpha}{\gamma}\frac{q}{p_i}\)^2\cos(\theta_{c,i})^2$, $\kappa_{io,2}=2^2\kappa_{io}+\frac{2q}{p_i}\cos(\theta_{c,i})$. And the derivative $\d \tau_C/\d \alpha=-\tau_C^3\alpha m_e^2/\gamma^2\(Q_c^2\kappa_{io,2}^2+(q/p_i)^2\cos(\theta_{c,i}^2)/(m_e\tau_D)^2\)$
\item For the far-right region, $Q_c-\frac{\Sigma \kappa_{io}}{4m_e}>Q$:
\begin{equation}
	\sigma_{v,het,3}\approx 2 \pi x_q^{1/2}\tau_D^{1/2}\frac{\kappa_{io}^{13/4}(p_o/q)}{\sin(\theta_{c,i})}\frac{1}{(\kappa_{io}+1)^{1/2}}.
\end{equation}
  	\end{enumerate}
  	
 \paragraph{Homo-contribution}
Changes in the story of homo-contribution are more intricate, since the angle where the thing is peaked is not small generically.  Using result of the previous section \eqref{appendix:conductivity:higher:total:DOS:homo:generating:small-qpi}, we write:
 \begin{equation}\label{appendix:conductivity:higher:total:vertex:homo:generating:small-qpi}
	g_i\approx
	 -\frac{1 }{2\gamma \alpha  m_e^5 \Sigma}\frac{\kappa_{ii}}{\kappa_{ii}+1}\int\frac{ v_p v_{p+q} \<f_v\> }{Q^2+Q_c^2}.
\end{equation}.
For $Q_c\ll Q_{max}$ we still should be able to use linear angle approximation. Then we immediately get:
\begin{multline}
	\sigma_{ii}\approx
	 -\frac{x_q^2}{2\gamma \alpha  m_e^3 \Sigma}\frac{\kappa_{ii}}{\kappa_{ii}+1}\int\frac{ v_{p_F} v_{p_F+q} }{Q^2+\tilde{Q_c}^2}=-\frac{x_q^2}{2\gamma \alpha  m_e^3 \Sigma}\frac{\kappa_{ii}}{\kappa_{ii}+1}\frac{v^i_p v^i_{p+q}}{\tilde{Q_c}J_{ii}} \tan^{-1}\(Q_{max}/\tilde{Q}_c\)\\\approx-\frac{x_q}{2^2\gamma \alpha  m_e^2 \Sigma J_{ii}}\frac{\kappa_{ii}^{1/2}}{\kappa_{ii}+1}v^i_p v^i_{p+q}\tan^{-1}\(Q_{max}/\tilde{Q}_c\).
\end{multline}
where $\tilde{Q}_c=2(\kappa_{ii}x_q^2/m_e^2+\kappa_{ii}\Sigma^2/m_e^2)^{1/2}$.
Clearly, this correction is going to be $\propto \tau_D$. Or, without the approximation, using an explicit form for velocities, I get:
 \begin{equation}
	g_i\approx
	 -\frac{2}{\gamma^2\alpha m_e \Sigma^3}\frac{\kappa_{ii}}{\kappa_{ii}+1}\oint \frac{dz}{zi}
	 \(
	\frac{(z^2+1)^2/q^2}{(\frac{qz}{p_i}+(z^2+1)^2)^2+\frac{Q_c^2z^2}{(p_i q)^2}}
	+
	\frac{2(z^2+1)z/(p_i q)}{(\frac{qz}{p_i}+(z^2+1)^2)^2+\frac{Q_c^2z^2}{(p_i q)^2}}
	\)
\end{equation}.
where the poles the same as in DOS section (\eqref{appendix:conductivity:higher:total:DOS:homo:generating:poles}). After substitution, I get:
\begin{multline*}
	g_V=-\frac{2^2\pi}{\gamma \alpha m_e \Sigma^3}\frac{\kappa_{ii}}{\kappa_{ii}+1}\frac{1}{q^2}
	-\frac{\pi}{2q^2\gamma \alpha m_e \Sigma^3}\frac{\kappa_{ii}}{\kappa_{ii}+1}\frac{1}{-iQ_c+(D-D^*)/2}
	\\\Bigg(\(\frac{((q^2+iQ_c-D)^2+(2p_i q)^2)^2}{D(-iQ_c+\frac{1}{2}(D+D^*))(-q^2-iQ_c+D)}
	-\frac{((q^2-iQ_c-D^*)^2+(2p_i q)^2)^2}{D^*(iQ_c+\frac{1}{2}(D+D^*))(-q^2+iQ_c+D^*)}\)
	\\+(2q)^2\(\frac{(q^2+i Q_c-D)^2+(2p_iq)^2}{D(-iQ_c+(D+D^*)/2)}-\frac{(q^2-i Q_c-D)^2+(2p_iq)^2}{D(iQ_c+(D+D^*)/2)}\)\Bigg),
\end{multline*}
so that the contribution to the conductance is
\begin{multline*}
	\sigma_{hom,v}=\frac{2\pi x_q^2}{\gamma \alpha m_e \Sigma}\frac{\kappa_{ii}}{\kappa_{ii}+1}\frac{1}{q^2}
	+\frac{\pi x_q^2}{(2q)^2\gamma \alpha m_e \Sigma}\frac{\kappa_{ii}}{\kappa_{ii}+1}\frac{1}{-iQ_c+(D-D^*)/2}
	\\\Bigg(\(\frac{((q^2+iQ_c-D)^2+(2p_i q)^2)^2}{D(-iQ_c+\frac{1}{2}(D+D^*))(-q^2-iQ_c+D)}
	-\frac{((q^2-iQ_c-D^*)^2+(2p_i q)^2)^2}{D^*(iQ_c+\frac{1}{2}(D+D^*))(-q^2+iQ_c+D^*)}\)
	\\+(2q)^2\(\frac{(q^2+i Q_c-D)^2+(2p_iq)^2}{D(-iQ_c+(D+D^*)/2)}-\frac{(q^2-i Q_c-D)^2+(2p_iq)^2}{D(iQ_c+(D+D^*)/2)}\)\Bigg).
\end{multline*}

\end{document}